%% file: main.tex
\documentclass[
9pt, % The default document font size, options: 10pt, 11pt, 12pt
oneside, % Two side (alternating margins) for binding by default, uncomment to switch to one side
english, % ngerman for German
singlespacing, % Single line spacing, alternatives: onehalfspacing or doublespacing
nohyperref, % Uncomment to not load the hyperref package
headsepline, % Uncomment to get a line under the header
chapterinoneline, % Uncomment to place the chapter title next to the number on one line
]{MastersDoctoralThesis} % The class file specifying the document structure

\usepackage[utf8]{inputenc} % Required for inputting international characters
\usepackage[T1]{fontenc} % Output font encoding for international characters
\usepackage{palatino} % Use the Palatino font by default
\usepackage{amsmath}
\usepackage{graphicx}
\usepackage{color}
\usepackage[hang,flushmargin]{footmisc}
\usepackage{upgreek}
\usepackage{microtype}
\usepackage{pdfpages}
\usepackage{caption}
\usepackage[titles]{tocloft}
\usepackage[small]{titlesec}
\titleformat{\chapter}[hang]{\normalfont\LARGE\bfseries}{\thechapter}{0.5em}{} 
\titlespacing*{\chapter}{0pt}{0pt}{3pt}
\titlespacing*{\section}{0pt}{9pt}{3pt}
\titlespacing*{\subsection}{0pt}{9pt}{1pt}
\titlespacing*{\subsubsection}{0pt}{9pt}{1pt}

\usepackage{etoolbox}
\newcommand{\zerodisplayskips}{%
  \setlength{\abovedisplayskip}{10pt}%
  \setlength{\belowdisplayskip}{10pt}%
  \setlength{\abovedisplayshortskip}{5pt}%
  \setlength{\belowdisplayshortskip}{5pt}}
\appto{\normalsize}{\zerodisplayskips}
\appto{\small}{\zerodisplayskips}
\appto{\footnotesize}{\zerodisplayskips}

\usepackage[style=numeric-comp,sorting=none,firstinits=true,doi=false]{biblatex}
\usepackage{hyperref}
\hypersetup{colorlinks,linkcolor=black,citecolor=black}

\thesistitle{Experimental exploration of gold semi-\\ continuous films in the near- and far-field} % Your thesis title, this is used in the title and abstract, print it elsewhere with \ttitle
\supervisor{Assoc. Prof. Nicolas \textsc{Stenger}\\Assoc. Prof. Andrei V. \textsc{Lavrinenko}\\Prof. N. Asger \textsc{Mortensen}} % Your supervisor's name, this is used in the title page, print it elsewhere with \supname
\examiner{} % Your examiner's name, this is not currently used anywhere in the template, print it elsewhere with \examname
\degree{Doctor of Philosophy} % Your degree name, this is used in the title page and abstract, print it elsewhere with \degreename
\author{Christian \textsc{Frydendahl}} % Your name, this is used in the title page and abstract, print it elsewhere with \authorname
\addresses{} % Your address, this is not currently used anywhere in the template, print it elsewhere with \addressname

\subject{} % Your subject area, this is not currently used anywhere in the template, print it elsewhere with \subjectname
 % Keywords for your thesis, this is not currently used anywhere in the template, print it elsewhere with \keywordnames
\university{Technical University of Denmark} % Your university's name and URL, this is used in the title page and abstract, print it elsewhere with \univname
\department{Department of Photonics Engineering.} % Your department's name and URL, this is used in the title page and abstract, print it elsewhere with \deptname
\group{\textit{Structured Electromagnetic Materials}, and\\
\textit{Plasmonics and Metamaterials} groups,} % Your research group's name and URL, this is used in the title page, print it elsewhere with \groupname
\faculty{Faculty Name} % Your faculty's name and URL, this is used in the title page and abstract, print it elsewhere with \facname

\AtBeginDocument{
\hypersetup{pdftitle=\ttitle} % Set the PDF's title to your title
\hypersetup{pdfauthor=\authorname} % Set the PDF's author to your name
\hypersetup{pdfkeywords=\keywordnames} % Set the PDF's keywords to your keywords
}

\begin{document}

\frontmatter % Use roman page numbering style (i, ii, iii, iv...) for the pre-content pages

\pagestyle{plain} % Default to the plain heading style until the thesis style is called for the body content

%----------------------------------------------------------------------------------------
%	TITLE PAGE
%----------------------------------------------------------------------------------------

\includepdf[pages=-]{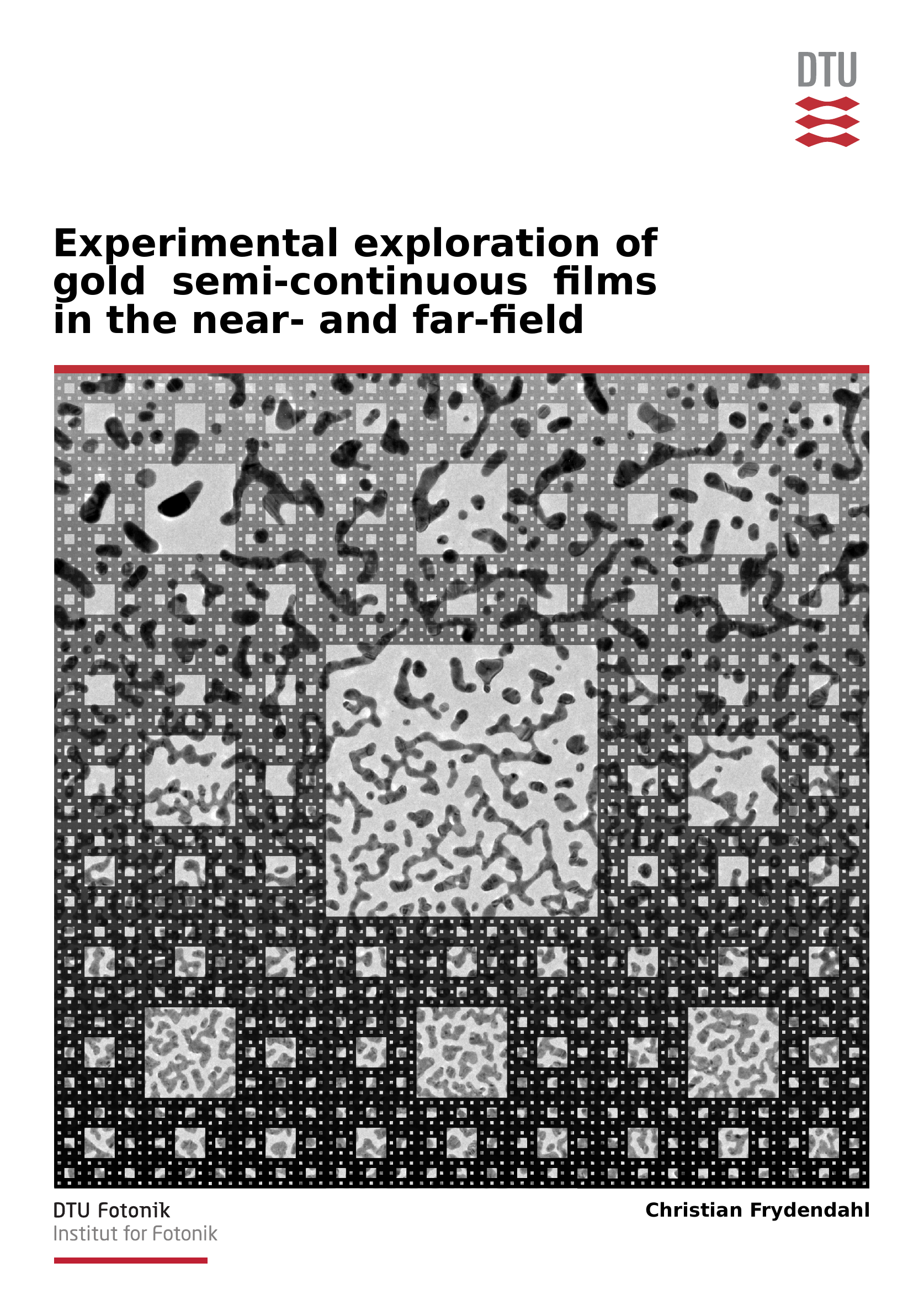}
\newpage\null\thispagestyle{empty}\newpage

\begin{titlepage}
\begin{center}

\vspace*{.01\textheight}
{\scshape\LARGE \univname\par}\vspace{0.5cm} % University name
\textsc{\large Doctoral Thesis}\\[0.9cm] % Thesis type

\HRule \\[0.4cm] % Horizontal line
{\large\bfseries \ttitle\par}\vspace{0.4cm} % Thesis title
\HRule \\[1.3cm] % Horizontal line
 
\begin{minipage}[t]{0.4\textwidth}
\begin{flushleft}
\emph{\textbf{Author:}}\\
\authorname % Author name - remove the \href bracket to remove the link
\end{flushleft}
\end{minipage}
\begin{minipage}[t]{0.5\textwidth}
\begin{flushright} 
\emph{\textbf{Supervisors:}} \\
\supname % Supervisor name - remove the \href bracket to remove the link  
\end{flushright}
\end{minipage}\\[1.3cm]

\textit{A thesis submitted in fulfilment of the requirements\\ for the degree of \degreename}\\[0.3cm] % University requirement text
in the\\[0.4cm]
\groupname\\\deptname\\ % Research group name and department name

\vspace{0.65cm}
\includegraphics[width=1.3cm]{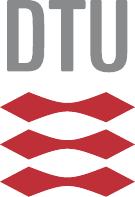}\\
\vspace{0.5cm}
{\large November, 2017} % Date

\end{center}
\end{titlepage}

\newpage\null\thispagestyle{empty}\newpage

\vspace*{0.2\textheight}

\hfill\noindent{\itshape{\Large ``You live and learn. At any rate, you live.''}}\bigbreak

\hfill \large-- Douglas Adams

\setcounter{page}{1}
\newpage\null\thispagestyle{empty}\newpage

%----------------------------------------------------------------------------------------
%	ABSTRACT PAGE
%----------------------------------------------------------------------------------------

\chapter*{Abstract (English)}
\addchaptertocentry{Abstract (English)} % Add the abstract to the table of contents
\normalsize
This thesis is the final work of my PhD-study performed in the groups of '\textit{Structured Electromagnetic Materials}' and '\textit{Plasmonics and Metamaterials}' at the Technical University of Denmark (DTU). The thesis is a compilation of my research activities during that three year period, with a particular focus on experimental investigations of optical and plasmonic properties of self-organized gold percolation films, and how these can be modified by fs-laser pulses.

The thesis will start with a short introduction to metal optics and plasmonics in general. We will then introduce percolation systems and the concept of the percolation threshold. We detail how different percolation film morphologies can be fabricated by simple metal deposition techniques, and also discuss some of the physics behind such films' self-organization. We finally discuss how 'hotspots' of strong plasmonic excitation exists in such fractal structures.

We will then move on to introduce the concepts of gold photoluminescence and two-photon luminescence (TPL). We will show how it is possible to observe greatly enhanced TPL from gold percolation films, as compared to bulk gold. We will also prove that the greatest enhancement of TPL is found for gold films close to the percolation threshold. We will also briefly discuss how it is possible to observe strong white light continuum generation from such percolation films, but the physics responsible are still poorly understood.

To continue our discussion of results, we will briefly have to introduce the techniques of scanning near-field optical microscopy (SNOM) and electron energy-loss spectros\-copy (EELS). We also show some partial SNOM results from experiments done at DTU.

Using high power fs-laser pulses, we demonstrate that it is possible to locally extinguish the TPL-signal from gold percolation films on silicon substrates. This effect proves to be strongly polarization sensitive, only extinguishing the TPL-signal from polarizations parallel to that of the writing laser. We explain the origin of this writing effect to be the photothermal melting and reshaping of plasmonic hotspots in the percolation films. As only those hotspots that match the polarization and wavelength of the laser are reshaped, the result is a polarization and wavelength sensitive extinction of TPL.

When we repeat this laser writing experiment on percolation films on thin glass substrates, we find a very different kind of laser writing effect. Due to the poorer heat conductivity of glass compared to silicon, the heat energy is no longer confined to just the local geometry that supports the hotspots. The result is a total reshaping of the film. We do however still report a polarization sensitive plasmonic response, this time showing plasmon resonances aligned parallel to the writing laser's polarization. We record EELS maps of our structures and compare them to finite-element simulations of the same morphologies, explaining the origin of the resonances to be from nanorod-like particles that have been shaped by the laser writing.

Finally a summary and outlook is provided, detailing how these results could be used for plasmonic laser writing applications, like ultra-dense data storage.

%\newpage\null\thispagestyle{empty}\newpage

\chapter*{Resum\'e (Danish)}
\addchaptertocentry{Resum\'e (Danish)} 
\normalsize
Denne afhandling er det sidste værk i mit PhD-studie udført i '\textit{Structured Electromagnetic Materials}' og '\textit{Plasmonics and Metamaterials}' grupperne på Danmarks Tekniske Universitet (DTU). Af\-handlingen er en samling af mine forskningsaktiviteter igennem denne 3-års periode, med et særligt fokus på eksperimentel ud\-forskning af optiske og plasmoniske egenskaber af selvorganiserede guldperkolationsfilm, samt hvordan disse kan ændres med fs-laserpulser.

Afhandlingen starter med en kort introduktion til metaloptik og plasmonik generelt. Vi vil så introducere perkolationssystemer og konceptet om perkolationsgrænsen. Vi detaljere hvordan forskellige perkolationsfilmsmorfologier kan fremstilles ved simple metaldeponeringsteknikker, og diskutere også nogle af de fysiske mekanismer der står bag sådanne films selvorganisation. Endeligt diskutere vi hvordan 'hotspots' af stærk plasmonisk excitation findes i sådanne fraktale strukturer.

Vi fortsætter så ved at introducere koncepterne guldfotoluminescens og to-foton luminescens (TPL). Vi viser hvordan det er muligt at observere skarpt forøget TPL fra guldperkolationsfilm, sammenlignet med solidt guld. Vi beviser også at den største forøgelse af TPL findes i guldfilm nær perkolationsgrænsen. Vi vil også kort diskutere hvordan det er muligt at observere stærk hvid-lys kontinuumgeneration fra sådanne perkolationsfilm,\linebreak men den fysiske mekanisme bag er stadig dårligt forstået.

For at fortsætte vores diskussion af resultater, må vi kort introducere teknikkerne skanning nærfelts optisk mikroskopi (SNOM) og elektron energitabs-spektroskopi (EELS). Vi viser også nogle delvise resultater fra SNOM-eksperimenter udført på DTU.

Ved brug af fs-laserpulser med høj effekt, demonstrere vi at det er muligt at lokalt slukke for TPL-signalet fra guldperkolationsfilm på siliciumsubstrater. Denne effekt viser sig at være stærkt polarisationsfølsom, da kun TPL-signalet fra polarisationer parallel med polarisationen af skrivelaseren slukkes. Vi forklarer årsagen af denne skriveeffekt som værende smeltningen og omformningen af plasmoniske hotspots i perkolationsfilmene. Da kun de hot\-spots som matcher polarisationen og bølgelængden af laseren er omformet, bliver resultatet en polarisations- og bølgelængdefølsom slukning af TPL.

Når vi gentager dette laserskrivningseksperiment på perkolationsfilm på tynde glassubstrater, finder vi en meget anderledes laserskrivningseffekt. På grund af den ringere varmeledningsevne af glas sammenlignet med silicium, er varmeenergien ikke længere begrænset til den lokale geometri som understøtter hotspots'ene. Resultatet er en totalomformning af filmen. Vi finder dog stadigt et polarisationsfølsomt plasmonisk respons, denne gang i form af plasmonresonanser der er alignet parallelt til skrivelaserens polarisation. Vi optager EELS-kort af vores strukturer og sammenligner disse med endeligt-element simuleringer af de samme morfologier. Fra disse forklarer vi at årsagen til resonanserne er nanostavlignende partikler der er formet under laserskrivningen.

Endeligt gives et resum\'e og et outlook. Heri detaljeres hvordan disse resultater kunne anvendes i plasmoniske laserskrivningsanvendelser, så som datalagring med ultrahøj tæthed.

\newpage\null\thispagestyle{empty}\newpage
\newpage\null\thispagestyle{empty}\newpage

%----------------------------------------------------------------------------------------
%	ACKNOWLEDGEMENTS
%----------------------------------------------------------------------------------------

\begin{acknowledgements}
\addchaptertocentry{\acknowledgementname} % Add the acknowledgements to the table of contents
\normalsize
\noindent I would like first to thank my former supervisor, Prof. Philip Hofmann at Aarhus University. If not for his words of encouragement during my Master's studies, I would not have had the confidence or ambition to pursue a further career in research.

I would also like to sincerely thank my three supervisors: Nicolas, Asger, and Andrei. Before starting this PhD-study I had very limited knowledge or experience with experimental optics, and next to no knowledge of plasmonics. I thank them for giving me this opportunity, introducing me to so many interesting new topics in physics, and for providing me with counsel and guidance through these three years. I can barely believe now, how much I have learned in this short amount of time, and I certainly could not have done so without their help.

I also extend my thanks to all current and former members of the '\textit{Structured Electromagnetic Materials}' and '\textit{Plasmonics and Metamaterials}' groups. You have all contributed to an amazing environment in which to perform a PhD-study, with amble time for stimulating scientific discussions. Special thanks goes to my office-mates and fellow PhD-students, Kåre, Johan, Mathias, and Andr\'e. Even if I have not had the opportunity to explicitly collaborate and work closely with all of you, I have still learned many valuable things from our open discussions in the office about science, life, politics, and everything else in the observable universe.

Most of my work presented here, has been the product of a strong collaboration with the group of Prof. Sergey I. Bozhevolnyi at the University of Southern Denmark. I am deeply grateful to everyone that has participated in this collaboration, as well as to all my other co-authors and collaborators in my projects during the PhD. Even though several of my ideas and projects did not come to full fruition during the three years, I have still learned many skills and lessons for the future from them. An extra special thanks goes to Taavi, for assisting me with additional simulations for one of the figures in this thesis.

I will also take this opportunity to offer my most sincere gratitude to the members of my PhD-evaluation committee: Prof. Mark I. Stockman, Prof. Uriel Levy and Prof. Peter Uhd Jepsen. Thank you for taking time out of your busy schedules to not only read this thesis, but also to participate in the evaluation of its contents.

Finally, I wish to thank my friends, family and girlfriend -- for always offering me support, love, kindness, and innumerable distractions from my work.

\end{acknowledgements}

\newpage\null\thispagestyle{empty}\newpage

%----------------------------------------------------------------------------------------
%	List of publications
%----------------------------------------------------------------------------------------

\chapter*{List of Publications}
\addchaptertocentry{List of Publications}

\subsection*{Journal articles}

\normalsize
\begin{itemize}
	\item[A:] S.~M.~Novikov, J.~Beermann, \textbf{C.~Frydendahl}, N.~Stenger, V.~Coello, N.~A.~Morten\-sen, and S.~I.~Bozhevolnyi. Enhancement of two-photon photoluminescence and SERS for low-coverage gold films. Optics express \textbf{24}, 16743-16751, 2016.
    \item[B:] S.~M.~Novikov, \textbf{C.~Frydendahl}, J.~Beermann, V.~A.~Zenin, N.~Stenger, V.~Coello, N.~A.~Mortensen, and S.~I.~Bozhevolnyi. White light generation and anisotropic damage in gold films near percolation threshold. ACS Photonics \textbf{4}, 1207-1215, 2017.
    \item[C:] \textbf{C.~Frydendahl}, T.~Rep\"an, M.~Geisler, S.~M.~Novikov, J.~Beermann, A.~V.~Lavrinenko, S.~Xiao, S.~I.~Bozhevolnyi, N.~A.~Mortensen, and N.~Stenger. Optical reconfiguration and polarization control in semi-continuous gold films close to the percolation threshold. Nanoscale, \textbf{9}, 12014-12024, 2017.
\end{itemize}

\subsection*{Conference contributions}

\begin{itemize}
	\item[I:] \textbf{C.~Frydendahl}, S.~M.~Novikov, J.~Beermann, N.~Stenger, N.~A.~Morten\-sen, and S.~I.~Bozhevolnyi. Enhancement of two-photon photoluminescence for low coverage gold films. NFO \textbf{14}, Tu-9P-7, 2016.
    \item[II:] S.~M.~Novikov, \textbf{C.~Frydendahl}, J.~Beermann, V.~A.~Zenin, N.~Stenger, V.~Coello, N.~A.~Mortensen, and S.~I.~Bozhevolnyi. White light generation and anisotropic damage in gold films near percolation Threshold. SPP \textbf{8}, P-07-23, 2017.
    \item[III:] \textbf{C.~Frydendahl}, T.~Rep\"an, M.~Geisler, S.~M.~Novikov, J.~Beerman, A.~Lavrinenko, S.~Xiao, S.~I.~Bozhevolnyi, N.~A.~Mortensen, and N.~Stenger. Near- and far field spectroscopy of semi-continuous gold films with optically induced anisotropy. SPP \textbf{8}, P-07-22, 2017.
\end{itemize}
%

%----------------------------------------------------------------------------------------
%	LIST OF CONTENTS/FIGURES/TABLES PAGES
%----------------------------------------------------------------------------------------

\tableofcontents % Prints the main table of contents

\newpage\null\thispagestyle{empty}\newpage

\dedicatory{To my parents, Ole and Elsebeth.} 

\newpage\null\thispagestyle{empty}\newpage

%----------------------------------------------------------------------------------------
%	THESIS CONTENT - CHAPTERS
%----------------------------------------------------------------------------------------

\mainmatter % Begin numeric (1,2,3...) page numbering

\pagestyle{thesis} % Return the page headers back to the "thesis" style

% Include the chapters of the thesis as separate files from the Chapters folder
% Uncomment the lines as you write the chapters

\input{chapters/chap_intro.tex}

\newpage\null\thispagestyle{empty}\newpage
\input{chapters/chap_opt.tex}

\input{chapters/chap_plas.tex}

\input{chapters/chap_metal.tex}

\input{chapters/chap_tpl.tex}

\input{chapters/chap_snom.tex}

\input{chapters/chap_eels.tex}

\newpage\null\thispagestyle{empty}\newpage
\input{chapters/chap_recon.tex}

\newpage\null\thispagestyle{empty}\newpage
\input{chapters/chap_con.tex}

\newpage\null\thispagestyle{empty}\newpage

%----------------------------------------------------------------------------------------
%	THESIS CONTENT - APPENDICES
%----------------------------------------------------------------------------------------

\appendix % Cue to tell LaTeX that the following "chapters" are Appendices

% Include the appendices of the thesis as separate files from the Appendices folder
% Uncomment the lines as you write the Appendices

%\include{Appendices/AppendixA}
%\include{Appendices/AppendixB}
%\include{Appendices/AppendixC}

%----------------------------------------------------------------------------------------
%	BIBLIOGRAPHY
%----------------------------------------------------------------------------------------

\printbibliography[heading=bibintoc]
\newpage\null\thispagestyle{empty}\newpage


\section*{Supplementary material}

Christian Frydendahl received a BSc (2011) and a MSc (2013) in physics from the Department of Physics and Astronomy (IFA), Aarhus University (AU), Denmark.

In 2014 he started his PhD-study at The Technical University of Denmark (DTU), at the Department of Photonics Engineering (DTU Fotonik). The study has been performed in the groups of "*Structured Electromagnetic Materials*" and "*Plasmonics and Metamaterials*", and was supervised by Assoc. Prof. Nicolas Stenger, Assoc. Prof. Andrei V. Lavrinenko, and Prof. N. Asger Mortensen (now of University of Southern Denmark).
The thesis was successfully defended on November 2nd, 2017. The evaluation committee consisted of Prof. Mark I. Stockman, Prof. Uriel Levy, and Prof. Peter Uhd Jepsen.

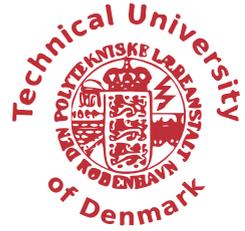

Metallic nanostructures can support so-called plasma oscillations (plasmons). Plasmons allow for the concentration of the energy from light, down to sizes well below the conventional diffraction limit known from optics. Plasmonics thus allows for a plethora of new optical applications at the nanoscale.

In this thesis, we have investigated the optical and plasmonic properties of semi-continuous gold films (also called percolation films). These films consist of complex tortuous fractal patterns on the nanoscale. They are an easy, fast, and scalable method to fabricate metallic nanostructures.

We show, that despite their very complex overall geometry, a large part of the films' properties can be understood alone based on their strongly localized plasmonic 'hotspots' – areas of just a few nanometres in the films, wherein optical fields are enhanced several thousand times.

Additionally, we show that such films can be used for the enhancement of gold two-photon photoluminescence and white light continuum generation. We have also shown that it is possible, using femtosecond-laser pulses, to inscribe information into the films, *via* phototheral processes. This has potential applications for ultra-dense information storage, and plasmonic colour laser printing.



%\newpage\null\thispagestyle{empty}\newpage

%\addchaptertocentry{Paper A - Enhancement of two-photon photoluminescence and SERS for low-coverage gold films}
%\includepdf[pages=-]{OpticsExpress.pdf}
%\newpage\null\thispagestyle{empty}\newpage

%\addchaptertocentry{Paper B - White Light Generation and Anisotropic Damage in Gold Films near Percolation Threshold}
%\includepdf[pages=-,trim=-6mm -0mm -6mm -0mm, clip]{ACSPhotonics.pdf}
%\newpage\null\thispagestyle{empty}\newpage

%\addchaptertocentry{Paper C - Optical reconfiguration and polarization control in semi-continuous gold films close to the percolation threshold}
%\includepdf[pages=-,trim=-10mm -0mm -10mm -0mm, clip]{Nanoscale.pdf}

%----------------------------------------------------------------------------------------

\end{document}

%% file: chapters/chap_intro.tex
\chapter{Introduction}

Metallic nanostructures posses the ability to localize and enhance optical fields down to the nanoscale via collective excitations of their conduction electrons. These excitations are known by their quantized quasi-particle representation as plasmons, and have been the subject of intense study in the recent decades\cite{Schuller:2010}. The ability to concentrate and manipulate light on the nanoscale offers many new promising applications in the field of optics. By carefully engineering metallic nanostructures, their optical properties can be fine-tuned to allow for sub-diffraction light focusing for spectroscopy and sensing\cite{Jeon:2016}, enhancing light-matter interactions in new emerging 2D materials\cite{Li:2017,Low:2016}, and quantum information processing technologies\cite{Tame:2013,Pelton:2015,Bozhevolnyi:2017b}. In the last few years in particular, interest has also been growing in the field of plasmonic colours\cite{Kristensen:2016}. By tuning the spectral reflectivity of a surface due to plasmonic structures on it, it is possible to create colour printed images with sub-diffraction pixel resolution\cite{Zijlstra:2009,Kumar:2012,Zhu:2016}.

Nanometre sized structures are needed to realize these applications, and defining features on the nanoscale often requires elaborate fabrication methods like electron-beam lithography (EBL) and focused ion beam (FIB) milling\cite{Fang:2013}. EBL allows for very precise and reproducible definition of nanostructures, but at the cost of several time consuming process steps to produce a mask pattern in a polymer resist\cite{Chen:2015}. In addition, the spatial resolution limit is still in the order of 10\,nm for state of the art systems\cite{Chen:2015}. FIB milling can offer an alternative mask-less fabrication method of similar resolution to EBL. However, the method still requires long pattern writing times, and also comes with the potential risk of contaminating the structure materials with the type of ions used for milling\cite{Kim:2012}, potentially altering their optical properties.

However, another alternative exists in the realm of self-assembled and self-organized materials. By letting already nanoscale particles or elements structure themselves, sub-nanometre gaps and protrusions can routinely be achieved. Such small features are perfect for hosting plasmons that enable high levels of field-enhancement and localization. The big advantage of such self-assembled structures is their fast and easy, bottom-up, fabrication methods\cite{Colson:2013,Klinkova:2014,Xi:2015}. Such structures are naturally limited in what structures can be achieved, based on their component sizes and method of their self-assembly. It is however possible to alter their assembly processes directly to somewhat expand the range of possible structures\cite{Gwo:2016}. Another method to expand the range of achievable structures, is by laser induced photothermal reshaping post-assembly\cite{Shalaev:2007,Hu:2012,Garnett:2012,Herrmann:2014,Mertens:2016,Zhu:2016,Zhu:2017}.

In this thesis, we will study the optical properties of self-organized gold percolation films, and how it is possible to modify these properties by fs-laser induced photothermal reshaping. The intrinsic nanostructured morphology of percolation metal films comes from the Volmer--Weber growth process for metals on dielectric substrates\cite{Greene:2010}. During a deposition process, the metal atoms have a mutually strong interaction, while they will bond less with the substrate. During the earliest part of the deposition, this leads to the formation of isolated clusters. With continuing deposition these initial clusters tend to grow in the substrate plane, eventually reaching a 'percolation threshold' where they merge to form a connected system. The total metal film is now dominated by one large cluster with several gaps in it. Any further metal deposition will close up these remaining gaps in the film morphology, and eventually the system will transition into a continuous metal film. Any additional deposition will then simply be metal on metal growth. 

Using standard metal evaporation systems, it is possible to fabricate large-scale areas of such metal percolation films, wherein the smallest feature sizes can be at the sub-nanometre scale\cite{Greene:2010}. Semi-continuous metal films thus make an ideal alternative platform for many plasmonic technologies, and have been the subject of much theoretical and experimental study already\cite{Shalaev:2007,Stockman:1996,Stockman:1997,Genov:2003b,Ducourtieux:2000,Losquin:2015}.

\section{Outline}

The first two chapters will serve to introduce much of the underlying physics for understanding light's interaction with metals, both at the macro- and nanoscale. Chapter 2 will briefly introduce the simple and intuitive Drude model for a metal's dielectric function, as well as some of its results and limitations. Chapter 3 will introduce the concept of collective plasma oscillations (plasmons) in metals. The purpose of these two chapters is to (hopefully) help establish a simple intuition of plasmonics for the unfamiliar reader. For readers already familiar with the topics discussed, these chapters can be skipped without much consequence. Further details on these concepts are available in references \cite{Novotny:2012,Fox:2002,Ashcroft:1976,Maier:2007}.

In chapter 4 we will look at how these discussed concepts can be realized in the random metal structures known as percolation films. We will discuss how such films can be fabricated, the underlying physics behind their self-organization, and how they can be modelled theoretically. We will then look at a specific application of such film geometries to enhance two-photon luminescence (TPL) from gold, and how it is possible to generate a white light continuum from these films in chapter 5.

In chapter 6 we will detail a practical limitation of experimental optics known as the diffraction limit. We will discuss how it is possible to still perform optics in the near-field to circumvent this limitation. We will in this same chapter detail the basic working principles of aperture- and aperture-less scanning near-field optical microscopes (SNOMs), and look at some example images recorded from gold percolation films. To go even further beyond the spatial resolution of modern SNOM devices, we will also cover how it is possible to investigate plasmon resonances by electron-energy loss spectroscopy (EELS) in a transmission electron microscope (TEM) in chapter 7.

In chapter 8 we will discuss how it is possible to alter the morphology of plasmonic nanostructures by photothermal processes from fs-laser illumination. We will show how it is possible change the TPL-enhancement from gold percolation films on a silicon substrate locally by a fs-laser writing process. Using TEM and EELS measurements, We will also detail the morphology and plasmon resonance changes from a similar laser writing process for percolation films on thin glass substrates.

Finally, we will conclude with a summary of many of the results presented throughout the thesis, and present a short outlook of further work that could be done to expand on these results for future applications.

%% file: chapters/chap_opt.tex
\chapter{Optical properties of metals}
\label{chap:opt}

Perhaps the most universally striking optical feature of metals, when polished to a smooth surface, is their high reflectivity. Early human civilizations around the world have utilized mirrors of polished copper as early as 4000 B.C.\cite{Enoch:2006}, so the ability of metals to reflect light has been well known through a majority of human civilization. Despite this, our current theoretical framework for understanding the properties of metals only begins around the 20th century with P.~Drude's proposal of a kinetic model for the free charge carriers in metals in 1900\cite{Drude:1900a,Drude:1900b}. The Drude model is a classical model, and we know it only offers an incomplete picture of the physics as we understand them today. However, the model still provides good agreement with many observations and offers a simple intuitive framework, and as such remains popular.

In the following we will discuss how from simple kinetic considerations we can explain much of the broadband reflective properties of metals, and we will also look at the initial breakdown of this simple model when dealing with imperfect metals that show absorption due to interband transitions in their band structure.

\section{Free electrons}

When collecting individual atoms together to form a solid material, the individual atomic energy levels of the solid's constituent atoms will start to shift and merge together to form broad bands of electronic states, instead of discrete levels like in an atom. This is also the basis for the very successful electronic band structure theory of condensed matter physics\cite{Fox:2002}.

\begin{figure}[h]
	\center
	\includegraphics[width=3.8cm]{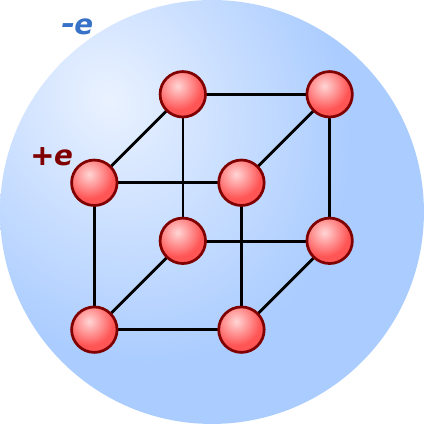}
    \caption{Schematic of a metal. The positively charged atomic nuclei make up a rigid lattice, and the conduction electrons form a charged plasma surrounding the lattice.}
    \label{fig:free_elec}
\end{figure}
\clearpage
\noindent Because of the types of bonds and crystal structures that become most energetically favourable when bonding metal atoms together, the resulting solid from this collection of atoms will not have any energy gap between its valence and conduction bands in its band structure. This means that for a metal, electrons are always populating states in the conduction band, no matter their energy. Here they will approximately behave like free electrons, as the electrons in lower energy states are tightly bound to their atomic cores in the crystal lattice, and will help screen off their host atomic nuclei's positive charge. The end result is that we can treat the metal as a static lattice of ions, with a plasma of electrons weakly bound to it\footnote{Especially since the ions are many orders of magnitude more massive than the electrons.}. A schematic drawing of this can be seen on Fig.~\ref{fig:free_elec}.

\section{The Drude model}
\begin{figure}[h]
	\center
	\includegraphics[width=0.7\columnwidth]{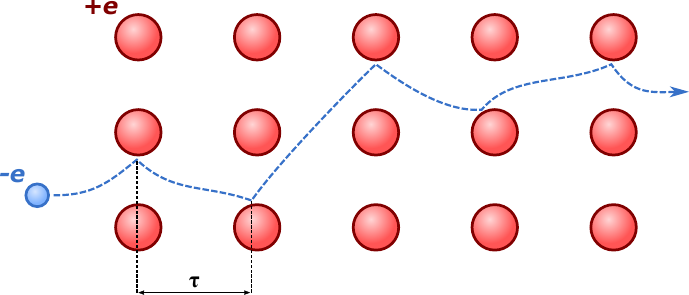}
    \caption{Schematic of a Drude metal. A semi-free electron propagates and experiences scattering events with the atomic lattice, with a mean time of $\tau$ between events.}
    \label{fig:drude}
\end{figure}
\noindent In the framework as described above, we find the Drude model for metal conductivity. The model operates with simple classical physics, and assumes that the conduction electrons in the metal can propagate through the metal if an electric field is applied. During their propagation through the metal the electrons will scatter off of the positive ionic lattice, with some mean time, $\tau$, between scattering events (see Fig.~\ref{fig:drude}). Different metals can now be understood by having different scattering times, explaining their differences in electrical conductivity.

We can now consider how the electric field of external optical radiation would couple to the conduction electrons in a metal. The field of propagating light is an oscillating transverse wave, so if we consider an electron via an oscillator model. We can write an equation of motion for the displacement, $x$, of an electron with mass $m_e$, by a field, $E(t)$, as:
\begin{align}
	m_e \frac{d^2x}{dt^2} + m_e \gamma \frac{dx}{dt} = -e E(t) = -e E_0\mathrm{e}^{-i \omega t}.
    \label{eqn:DL_motion}
\end{align}
\noindent The first left-hand term is the acceleration of the electron, while the second left-hand term is the frictional damping of the electron from its scattering with the lattice ions. We can correlate the damping coefficient, $\gamma$, to the scattering time in the Drude conductivity model, $\tau$, as simply $\gamma = 1/\tau$. The right-hand side of eqn.~\ref{eqn:DL_motion} is simply the force of the oscillating optical field on the electron, with $E_0$ the amplitude of the electric field, and $\omega$ its angular frequency. Notice the absence of a restoring force term, as we are treating the electrons as only weakly coupled to the lattice, we neglect their interaction with the ions. 

By solving eqn.~\ref{eqn:DL_motion} by substitution of $x=x_0 \mathrm{e}^{i \omega t}$, and regarding the polarizability of a system of $N$ such electrons as $P = Nex$, it is possible to arrive at the following equation for the relative dielectric function, $\epsilon_r(\omega)$, of a Drude metal:
\begin{align}
	\epsilon_r \left( \omega \right) = 1 - \frac{\omega_p^2}{\omega^2+i\gamma\omega} \, ,
    \label{eq:drude}
\end{align}
\noindent with $\omega_p$ the plasma frequency, defined as:
\begin{align}
	\omega_p = \left( \frac{Ne^2}{\epsilon_0 m_e} \right)^{1/2}.
\end{align}
\noindent We will see later how the plasma frequency is an important parameter in understanding metals' optical and plasmonic properties. It's worth noting the only 'free' parameter in the definition of $\omega_p$ is the electron count $N$. This means metals will have plasma frequencies in the ultraviolet range, while ceramics and doped semiconductors can have infrared plasma frequencies\cite{Naik:2013}.

\clearpage

\begin{figure}[h]
	\center
	\includegraphics[width=0.65\columnwidth]{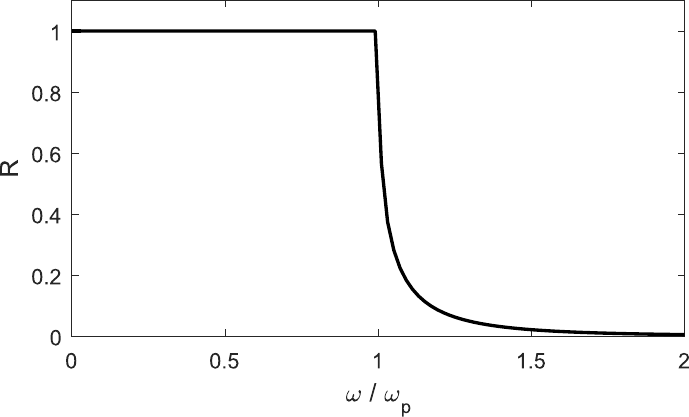}
    \caption{Reflectivity spectrum, $R$, of a lossless Drude metal, showing the transition to transparency for frequencies above $\omega_p$.}
    \label{fig:R_plot}
\end{figure}

\noindent If we consider the case of low damping ($\gamma=0$), we can write the refractive index as $n = \sqrt{\epsilon_r}$, and from the refractive index we can calculate the reflectivity as:
\begin{align}
	R = \left| \frac{n - 1}{n + 1} \right|^2.
\end{align}
\noindent From plotting this expression in units of $\omega_p$ in Fig. \ref{fig:R_plot}, we see how for frequencies below the plasma frequency the metal will act as a perfect reflector, while for frequencies above $\omega_p$ the metal becomes transparent.

\subsection{Thin metal films}

In the previous section we have only treated the case for bulk metals, but it is worth briefly mentioning how very thin ($1-15$\,nm) metallic films will differ from the bulk properties quite dramatically due to the finite skin depth of electric fields in metals. If we consider the absorption of the electric field strength through the metal by a simple Beer's absorption law, the skin depth, $\delta$, at which the field has decayed by a factor of $\mathrm{e}^{-1}$, can be defined as:
\begin{align}
	\delta = \left( \frac{2}{\sigma_0 \omega \mu_0} \right)^{1/2},
\end{align}
\noindent with $\sigma_0$ the DC conductivity of the metal. From this we can estimate the typical\footnote{A typical metal has $\sigma_0$\textasciitilde$10^7\,\mathrm{S/m}$ and visible light has $\omega$\textasciitilde$10^{15}\,\mathrm{Hz}$.} skin depth of metals in the visible spectrum to be around \textasciitilde$10$\,nm. As a result we can expect metal films below and around this thickness to be partially transparent, as the light is able to partially pierce through the metal.

\begin{figure}[h]
	\center
	\includegraphics[width=0.65\columnwidth]{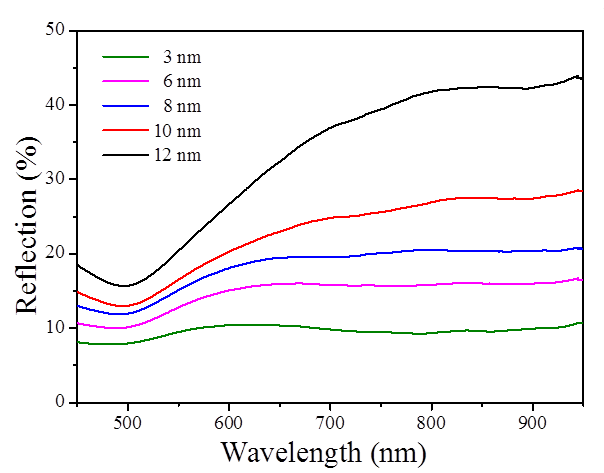}
    \caption{Reflectivity spectra of thin semi-continuous gold films of various thicknesses, deposited on glass substrates.}
    \label{fig:spec}
\end{figure}

\noindent In Fig.~\ref{fig:spec}.a we see the reflection spectra of several thin semi-continuous gold films. We see that for increasing film thickness the reflectivity increases, as expected. Likewise we see that the reflectivity is larger for longer wavelengths. However, around the $520$\,nm range we see a strong decrease in reflectivity. We could expect this to be explained by the plasma frequency of gold, but gold's $\omega_p$ corresponds to a wavelength of \textasciitilde$145$\,nm.

The essence of the problem here, is that our simple lossless model for metals does not explain the different colours of metals. It is thus only suitable for describing certain metals in certain wavelength ranges\footnote{For example: The Drude model is quite accurate for aluminium and silver in the the visible range, and gold and copper in the infrared range.}. In the Drude framework, our main way of adding losses would be by electron scattering, either off of the lattice or lattice defects. The net effect of this would be a general broadband reduction in reflection, as this kind of scattering just equally removes energy from the electrons. It is possible to introduce this kind of wavelength selective reduction by adding resonant Lorentzian oscillator terms to eqn.~\ref{eq:drude} to make a Drude-Lorentz model\cite{Rakic:1998}. However, to understand the physical cause of the absorption we will need a more detailed framework.

\clearpage
\section{Interband transitions}

In this section, we will discuss how band structure theory can explain the different colours of metals through specific energy absorption by so-called interband transitions. We will also expand our Drude model to account for the different colours of metals.

\begin{figure}[h]
	\center
	\includegraphics[width=0.7\columnwidth]{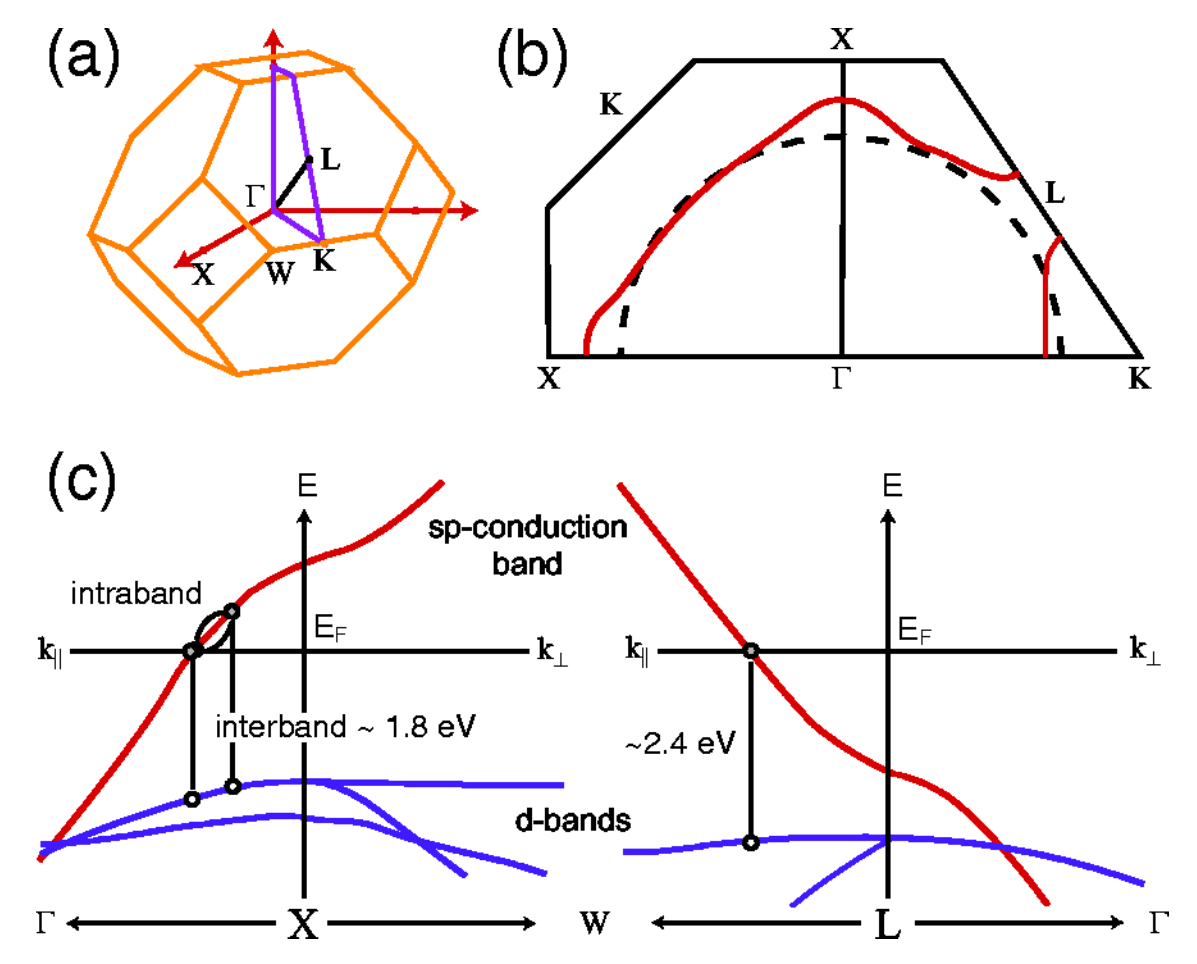}
    \caption{\textbf{(a)} Symmetry points in gold's first Brillouin zone. \textbf{(b)} The Fermi surface as mapped out along a path between the
high-symmetry points \textit{X} and \textit{L}. Near these points the Fermi surface is deformed, compared with that of free electrons (dashed circle). This is caused by Van Hove singularities in the density of states. \textbf{(c)} Shows the band structure near \textit{X} and \textit{L}, close to the Fermi level. The interband edges are marked. The many parallel $d$-bands give rise to large density of states that permits interband transitions. Reproduced from reference \cite{Beversluis:2003}.}
    \label{fig:gold_bs}
\end{figure}

\noindent Gold has the electron configuration $[\mathrm{Xe}]4f^{14}5d^{10}6s^1$, and a face-centred cubic crystal structure. This results in the partial band structure seen on Fig.~\ref{fig:gold_bs}. We see that the outer $d$- and $sp$-electron states of gold form two very different classes of bands. The $d$-bands are all very tightly stacked horizontally in a narrow energy range below the Fermi level, while the more free-electron-like $sp$-bands criss-cross each other and go far below and far above the Fermi level. It's exactly around these points where the $sp$-states cross through the Fermi level and the region of $d$-states that interband transitions can occur. The smallest gap between the bands sets the edge for where interband transitions can start to happen. For gold this is around 2.4\,eV (\textasciitilde520\,nm, which is where we start to see the strong decrease in reflection in Fig.~\ref{fig:spec}). Essentially, gold becomes a strong absorber for wavelengths below this, and as these wavelengths of visible light are not reflected but absorbed, it causes the characteristic colour of gold. In chapter~\ref{chap:tpl} we will study how it is possible to stimulate photoluminescence from this 'band gap', in a process somewhat reminiscent of that of photoluminescence in semiconductors.

\begin{figure}[h]
	\center
	\includegraphics[width=0.7\columnwidth]{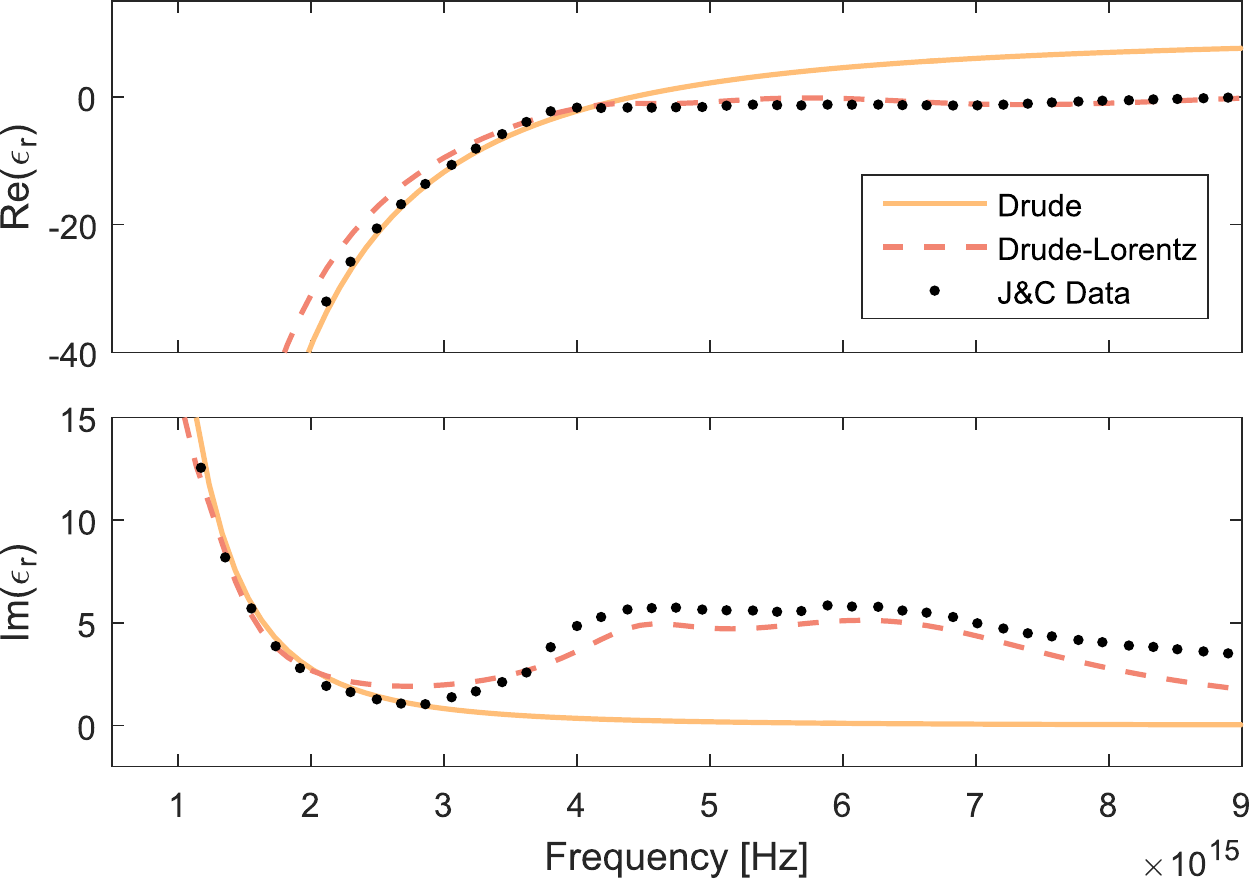}
    \caption{Real and imaginary parts of gold's dielectric function. A simple Drude model (details in text) and a Drude-Lorentz model, as given in reference \cite{Rakic:1998}. Experimental data from P.~B.~Johnson and R.~W.~Christy\cite{Johnson:1972}}
    \label{fig:au_dielec}
\end{figure}

\noindent To account for the interband transitions in our Drude framework, we need to perform either of two modifications to eqn.~\ref{eq:drude}. First, we can add a constant term to account generally for the material response at high frequency, $\epsilon_\infty$, where the $\omega$-dependent term of eqn.~\ref{eq:drude} otherwise goes to zero:
\begin{align}
	\epsilon_r \left( \omega \right) = \epsilon_\infty - \frac{\omega_p^2}{\omega^2+i\gamma\omega} \, .
    \label{eq:drude2}
\end{align}
\noindent However, as mentioned before, if we truly want to capture the wavelength dependent nature of the interband transitions, we need to add additional resonant Lorentz oscillator terms\cite{Rakic:1998}. A Drude-Lorentz model can look like:
\begin{align}
	\epsilon_r \left( \omega \right) = 1 - \frac{f_0\omega_p^2}{\omega^2-i\gamma_0\omega} + \sum_{n=1}^N \frac{f_n\omega_p^2}{\omega_n^2-\omega^2-i \gamma_n \omega} \, ,
    \label{eq:drude-lorentz}
\end{align}
\noindent where $f_n$ is the oscillator strength, $\omega_n$ the resonance frequency, and $\gamma_n$ the damping rate of the $n$'th oscillator. Fig.~\ref{fig:au_dielec} shows comparisons between experimental data and a simple Drude model with $\epsilon_\infty = 10$, $\omega_p = 1.4 \cdot 10^{16}$\,Hz and $\gamma = 1.13\cdot10^{14}$. The Drude-Lorentz model with $N=5$ of reference \cite{Rakic:1998} is also shown. 

From Fig.~\ref{fig:au_dielec} we clearly see how the simple Drude model breaks down at the onset of interband transitions around $4 \cdot 10^{15}$\,Hz (roughly \textasciitilde550\,nm), while the Drude-Lorentz model much more closely resembles the data. We do however still see that the simple Drude model perfectly represents the data for the low frequency range, where gold behaves very close to an ideal metal. The conclusion is thus simply: If working with a metal in a frequency range where no interband transistions can occur in the metal, then a simple Drude model like eqn.~\ref{eq:drude2} is perfectly sufficient, otherwise a more advanced Drude-Lorentz model like eqn.~\ref{eq:drude-lorentz} will be required.

In the following chapters we will see how nanostructuring of metals can introduce new optical properties, based on collective electron oscillations near a metal/dielectric interface. The general bulk properties as discussed here remain important for understanding the optical response of metal nanostructures, and especially the losses and optical absorption associated with interband transitions in gold will be important for later chapters.

%% file: chapters/chap_plas.tex
\chapter{A brief introduction to plasmonics}
\label{chap:plas}

In this chapter we will introduce the concept of collective electron oscillations, also better known by their quantization as quasi-particles called plasmons\cite{Schuller:2010}. We will focus our attention on plasmons in metals using the previously discussed free electron picture of chapter~\ref{chap:opt}. We will start with discussing bulk plasmons, and then moving on to surface plasmon polaritons (propagating surface waves) and finish with localized surface plasmons.

\section{Bulk plasmons}

Bulk plasmons are generally of little interest in optics (for reasons that will become clear later). They do however make a suitable starting point for discussing plasmons, as they are a simple pedagogical demonstration of the concepts. They are also important historically, as they were the first plasmon phenomenon to be realized and understood experimentally\cite{Ruthemann:1948,Lang:1948,Watanabe:1956}.

Bulk plasmons were initially observed independently by G.~Ruthemann\cite{Ruthemann:1948} and W.~Lang\cite{Lang:1948} in 1948 as straight lines in electron-energy loss spectra through thin metal films (We will discuss this technique in broader detail in chapter~\ref{chap:eels}). Y.~Cauchois initially proposed in 1952 that excitations of electrons via interband transitions could be responsible\cite{Cauchois:1952} (we will discuss a very interesting consequence of this kind of excitation in chapter~\ref{chap:tpl}). However, the shapes of the observed spectra were found to be incompatible with interband transitions. D.~Pines and D.~Bohm would later propose a model for plasmons being responsible\cite{Pines:1952,Bohm:1953}. H.~Watanabe would go on to experimentally verify the plasmon explanation in 1956 by carrying out a series of measurements across several materials, testing the spectra's dependence on the angle of the electron beam and the temperature of the sample\cite{Watanabe:1956}.

\begin{figure}[h]
	\center
	\includegraphics[width=0.4\columnwidth]{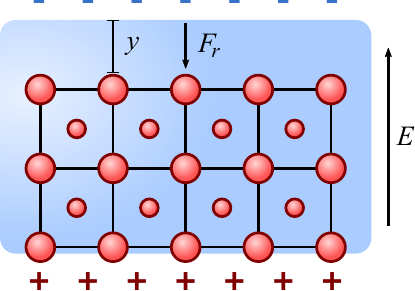}
    \caption{The displacement, $y$, of the electron plasma surrounding a metal crystal. The offset in charges causes an electric field, $E$, along the displacement. This results in a restoring force, $F_r$, pulling the electrons back towards the lattice.}
    \label{fig:plasma_osc}
\end{figure}

We can look at a plasma oscillation for a metal crystal in a very simple classical oscillator picture. If we consider a displacement, $y$, of the free electron plasma of $N$ electrons surrounding a positive ion lattice in a metal thin film, like in Fig.~\ref{fig:plasma_osc}\footnote{This initial displacement could be caused by an electron beam passed through the metal film.}.

\clearpage

\noindent The displacement of charges will result in a net charge distribution $+Ney$ on the bottom of the lattice, and $-Ney$ on the top. This then results in the electric field $E=Ney /\epsilon_0$ from the bottom to the top of the lattice. The charged electrons will now feel a restoring force, $F_r = -e E$, giving the following equation of motion:
\begin{align}
	m_e \frac{d^2 y}{dt^2} = -e E = - y \frac{Ne^2}{\epsilon_0},
\end{align}
\noindent which has the well-known solution of the form: $y(t)=y_0 \cos(\omega_p t)$, with the plasma frequency $\omega_p = \sqrt{Ne^2/\epsilon_0 m_e}$ as we also found it in chapter~\ref{chap:opt}. The bulk plasmons observed is electron-energy loss experiments are in integer multiples of the energy $\hbar \omega_p$\cite{Novotny:2012}. From our initial assumptions of the oscillation mode, we see that the bulk plasmon is exclusively a longitudinal oscillation mode\footnote{This fact can also be shown in greater detail, see reference \cite{Fox:2002}}. This means that it is impossible to excite directly with light (outside some extreme circumstances), as the electric field of light is always transverse to its propagation direction\cite{Fox:2002}. We will thus not discuss bulk/volume plasmons in more detail here, as they are outside the scope of this thesis.

\section{Surface plasmon polaritons}

Following the developments on bulk plamons, R.~H.~Ritchie would go on to predict the existence of surface plasmons in 1957\cite{Ritchie:1957}. Ritchie's original motivation for further developing the theory on plasmons, was to explain the presence of lower energy losses present in the past experiments, which were not accounted for in the bulk plasmon picture\cite{Ritchie:1962}.

\begin{figure}[h]
	\center
	\includegraphics[width=0.65\columnwidth]{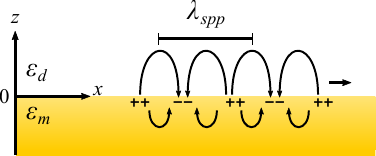}
    \caption{Geometry of a surface plasmon polariton at the interface of a metal and dielectric. The electrons near the interface oscillate in a longitudinal wave along the interface, creating a propagating transverse electric field along the interface.}
    \label{fig:spp_geo}
\end{figure}

\noindent As we saw above, bulk/volume plasmons are longitudinal modes of oscillation. If we now focus on just the interface between a dielectric (like air) and a metal, we can examine surface plasmons. Surface plasmons are a mix of transverse and longitudinal oscillations, and as such can couple to external radiation, forming a surface plasmon polariton. Surface plasmon polaritons were first demonstrated by optical excitation by A.~Otto in 1968\cite{Otto:1968}, and only a few months later independently verified by E.~Kretschmann and H.~Raether\cite{Kretschmann:1968}\footnote{We will discuss their experiments in more detail later in this section.}.

If we choose our coordinates such that the interface sits at the point $z=0$ (see Fig.~\ref{fig:spp_geo}), then we can write the dielectric function as:
\begin{align}
	\epsilon_j (\omega) = \begin{cases} \epsilon_d (\omega) &  \text{for $z>0$}\\
    									\epsilon_m (\omega) & \text{for $z<0$}	
                          \end{cases},
\end{align}
\noindent with $j = d,m$ for either the metal or dielectric region. We have written no explicit $x$-dependence, as we assume the interface to be continuous along this axis. From Maxwell's equations we can obtain the equation for the electric field, $\mathbf{E}(\mathbf{r},\omega)$\cite{Novotny:2012}:
\begin{align}
	\nabla \times \nabla \times \mathbf{E}(\mathbf{r},\omega) - \frac{\omega^2}{c^2}\epsilon_j(\omega)\mathbf{E}(\mathbf{r},\omega) = 0,
    \label{eq:WE}
\end{align}
\noindent with $\mathbf{r}=[x,y,z]$ a 3D-position vector. We are now looking for solutions of eqn.~\ref{eq:WE} along the metal/dielectric interface. To achieve this, we can essentially make a set of ansatzes about the form of the electric fields at either half of the interface, and then find the correct boundary conditions to unite the fields across the interface in $z=0$.

The first ansatz we can make is that the surface plasmons should be confined to the interface, and as such should exponentially decay along the $z$-axis\footnote{This kind of exponential surface confinement is seen for many other types of surface modes across physics.}. Secondly, the modes can propagate along the interface, i.e. the $x$-axis. As we want our surface mode to decay along the $z$-axis, we start by investigating strictly $p$-polarized waves ($\mathbf{E}$-field lying in the $xz$-plane). From this we can construct our plane wave electric field vector for either material half, $j$, as:
\begin{align}
	\mathbf{E}_j = \left(\begin{array}{c} E_{j,x}\\0\\E_{j,z} \end{array}\right) \text{e}^{i k_x x-i \omega t} \text{e}^{i k_{j,z} z}.
    \label{eq:field1}
\end{align}
\noindent We know that across the interface the $x$-component of the wave vector, $k$, is conserved, so:
\begin{align}
	k_x^2+k_{j,z}^2 = \epsilon_j k^2
    \label{eq:kx}
\end{align}
\noindent Utilizing that the displacement field in either half of the material is divergence free ($\nabla \cdot \mathbf{D} = 0$), it is possible to write: $k_x E_{j,x}+k_{j,z} E_{j,z} = 0$, with which eqn.~\ref{eq:field1} can be rewritten as:
\begin{align}
	\mathbf{E}_j = E_{j,x} \left(\begin{array}{c} 1\\0\\-k_x/k_{j,z} \end{array}\right) \text{e}^{i k_x x-i \omega t} \text{e}^{i k_{j,z} z}.
\end{align}
\noindent From requiring continuity of the field components across the metal/dielectric interface, we can write:
\begin{equation}
	\begin{aligned}
		E_{d,x}-E_{m,x} = 0,\\
    	\epsilon_d E_{d,z} - \epsilon_m E_{m,z} = 0.
	\end{aligned}
\end{equation}
\noindent The above equations can now either be solved for $k_x = 0$, which is the trivial solution, or for:
\begin{align}
	\epsilon_d k_{m,z} - \epsilon_m k_{d,z} = 0.
    \label{eq:kz}
\end{align}
\noindent Combined with eqn.~\ref{eq:kx}, eqn.~\ref{eq:kz} gives us the dispersion relation for surface plasmon polaritons:
\begin{align}
	k_x^2 = \beta^2 = \frac{\epsilon_d \epsilon_m}{\epsilon_d + \epsilon_m} \frac{\omega^2}{c^2},
    \label{eq:spp_disp}
\end{align}
\noindent with $k^2 = \omega^2/c^2$. In many contexts, $k_x$ will also be referred to as the propagation constant of the plasmon, $\beta$.

\begin{figure}[h]
	\center
	\includegraphics[width=0.7\columnwidth]{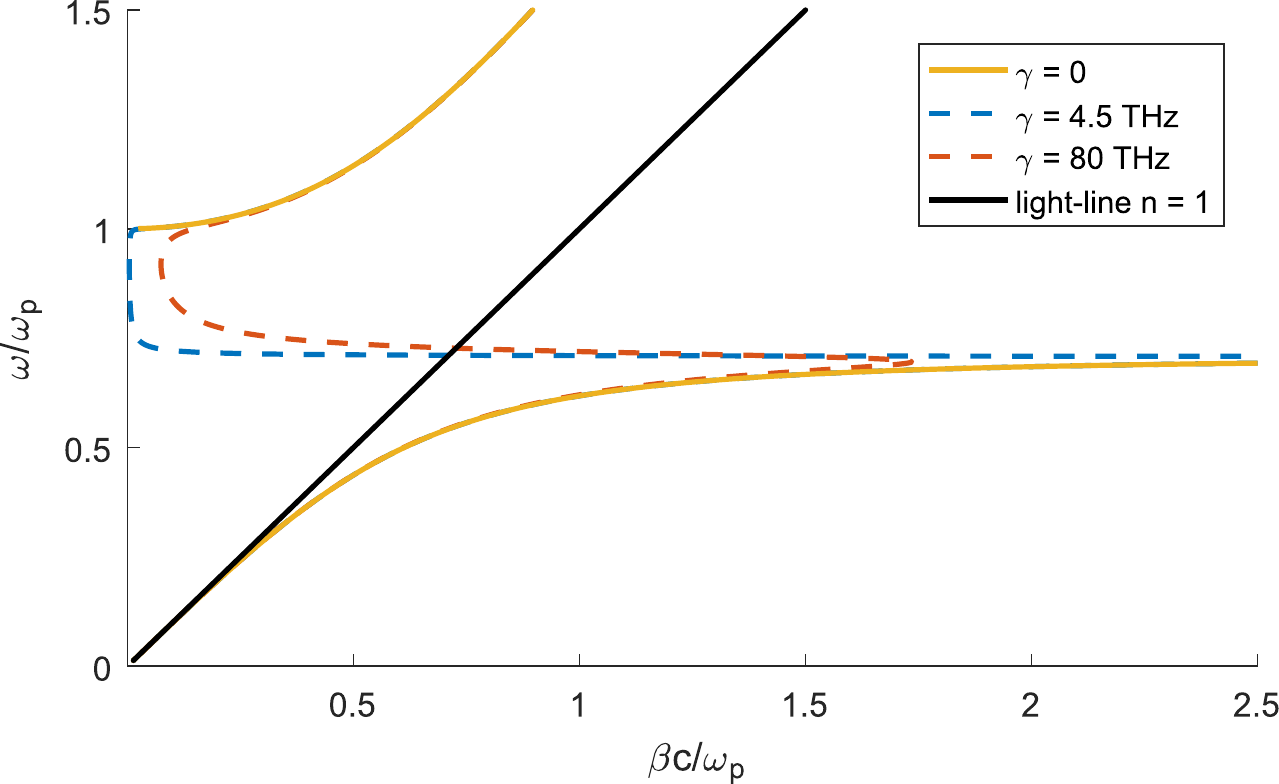}
    \caption{Dispersion relation for surface plasmons between a metal and air interface. The dispersion of light is also plotted, showing the momentum mismatch between surface plasmon polaritons and free-space light. For increasing losses in the metal, earlier linking of the high energy Brewster mode to the surface plasmon is observed.}
    \label{fig:spp_dispersion}
\end{figure}

On Fig.~\ref{fig:spp_dispersion} a plot is shown of eqn.~\ref{eq:spp_disp} for $\epsilon_d = 1$ and a Drude metal with different losses ($\gamma=0;4.5;80$\,THz). The dispersion of light ($\omega=ck$) in free-space is also shown. Two distinct branches are immediately apparent: A high energy branch to the left of the light-line, and a low energy branch to the right of the light-line. The high energy branch is not related to true surface modes, and is known as the Brewster mode. It is related to modes where $k_{m,z}$ is not purely imaginary, and thus allows the mode to propagate into the metal\cite{Novotny:2012}.

The low energy branch however relates to true propagating surface modes. It is worth noting that the low energy branch is solidly to the right of the dispersion of light. This means that surface plasmons have larger wave vectors, and i.e. shorter wavelengths than free space light ($\lambda_{spp}=2\pi/\beta$). Surface plasmon polaritons are thus a way to confine light below its normal free-space mode volume. Without accounting for losses, the lower energy branch just continues off asymptotically, but once losses are introduced a kind of 'back bending' will occur where the Brewster and surface plasmon modes become linked. Essentially, this sets a limit for how confined the plasmon can be at the surface of the metal\cite{Novotny:2012}. We see for increasing losses, that this 'back bending' happens earlier. We can see from this that the losses in metals intrinsically limits their plasmonic properties\cite{Khurgin:2015}.

\noindent Another consequence of the plasmon dispersion being to the right of the light-line, is that free-space light does not carry sufficient momentum on its own to excite the plasma oscillations. We can rectify this limitation in various ways, such as scattering the light off of a grating or other surface structure to confer an additional in-plane wave vector component, excite the plasmons by evanescent near-fields (more on this later), or even excite the surface plasmons by an electron beam (see chapter~\ref{chap:eels}).

One way to achieve an evanescent field for plasmon excitation is through total internal reflection in a prism. If light in the prism at an angle of $\theta$ is reflected at the prism's facet (and $\theta$ is greater than the critical angle), no propagating component of the light will be transmitted and the light is totally reflected. However, it can be shown that an evanescent field will emerge on the other side of the prism's facet\cite{Griffith:1981}. Due to the requirement of continuity of the fields across the interface, the in-plane component of this evanescent wave must then be:
\begin{align}
	k_x = k\sqrt{\epsilon_d}\sin\theta ,
    \label{eq:OK}
\end{align}
\noindent with $k$ the free space wave vector and $\epsilon_d$ the dielectric constant for the prism. From eqn.~\ref{eq:OK} we see that we should expect a strong angle dependence for the plasmon coupling, as by changing the angle we can directly tune the size of the $k_x$/$\beta$ of the light we are trying to couple to the plasmon. Likewise, we see that the refractive index ($n_d=\sqrt{\epsilon_d}$) of the prism material is important, as higher refractive indices will allow for larger in-plane wave vectors for a lower input angle.

\begin{figure}[h]
	\center
	\includegraphics[width=0.8\columnwidth]{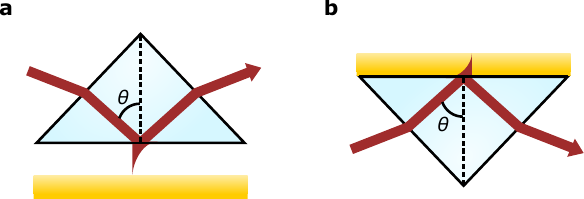}
    \caption{Coupling to surface plasmon polaritons in a metal film by evanescent fields from a prism, in the \textbf{(a)} Otto- and \textbf{(b)} Kretschmann configuration.}
    \label{fig:prism}
\end{figure}

\noindent Two configurations for prism coupling exists, following the experiments by A.~Otto, E.~Kretschmann and H.~Raether in 1968\cite{Maier:2007,Novotny:2012}. They can be seen on Fig.~\ref{fig:prism}. The Otto-configuration\cite{Otto:1968} has a small air-gap between the metal film and the prism, and relies on the 'tunneling' of the photons through the air to couple to the plasmons. This is very reminiscent of the way an optical fiber can be used for plasmon excitation and measurement in a scanning near-field optical microscope (we will discuss such devices in more detail in chapter~\ref{chap:snom}). The Kretschmann-configuration\cite{Kretschmann:1968} instead has no air-gap, and the film is deposited directly onto the prism. For both configurations the plasmon will be observed as a strong drop in the recorded reflectivity of the input laser beam, as energy is lost in the plasmon excitation. For sensitive samples where the surface should not be touched directly the Otto-configuration is preferred. However, the Kretschmann-configuration is generally more commonly used, as it is easier to implement in practice\cite{Maier:2007}. It is worth noting that using a prism to couple light into the plasmon, equally allows for the light to couple out again. As such, a higher degree of radiative losses are associated with prism coupled plasmons. This can be exploited for imaging of plasmons in so-called leakage radiation microscopy\cite{Maier:2007}.

\begin{figure}[h]
	\center
	\includegraphics[width=0.95\columnwidth]{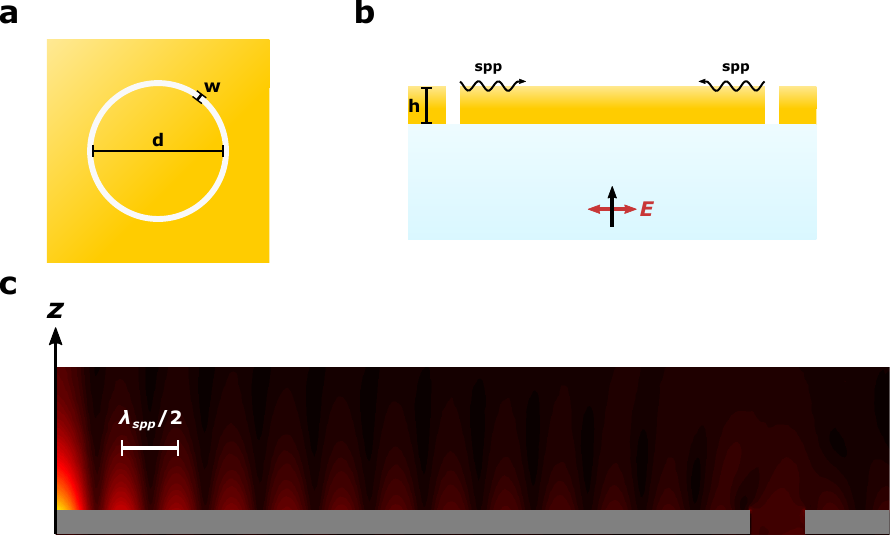}
    \caption{\textbf{(a)} Top view of the plasmonic lens geometry, defining diameter $d=16$\,$\upmu$m and slit width, $w=300$\,nm. \textbf{(b)} Profile view of the lens, defining the metal film thickness, $h=150$\,nm as well as showing the glass substrate and how the lens is illuminated from below by a 633\,nm plane wave. \textbf{(c)} $\left| \mathbf{E} \right|^2$ plotted for half of the plasmonic lens. The geometry is rotationally symmetric around the $z$-axis. Unpublished.}
    \label{fig:plasmon_lens}
\end{figure}

\noindent Another common technique to couple external light to plasmon polaritons (or manipulate their propagation) is by scattering from a grating. This technique is especially convenient for on-chip plasmonics and waveguiding applications\cite{Maier:2005,Maier:2007,Lerman:2009,Yanai:2012,Lerman:2013,Grajower:2013}. In a similar fashion, even a simple slit in a metal film can be used to excite plasmon polaritons\cite{Mehfuz:2010}. This has been exploited for focusing applications in plasmonic lenses\cite{Fu:2010,Liu:2005,Lerman:2009}. By milling a circular slit in a thin metal film, plasmon polaritons will be launched towards the centre of the circle from all directions, creating a focused spot of field enhancement. Because each point of the circle also has an opposite point launching similar plasmons, an interference pattern of constructive/destructive interference will be observed as concentric circles. The period of these interference fringes will be $\lambda_{spp}/2$. Fig.~\ref{fig:plasmon_lens} shows the geometry and a cross section of a simulation of the excited fields for such a lens with a diameter of 16\,$\upmu$m. The evanescent nature of the fields is clearly visible in Fig.~\ref{fig:plasmon_lens}.c, as the field intensity rapidly decays with distance from the metal film. The fields were obtained by a finite element simulation performed with COMSOL Multiphysics 4.4, with gold's dielectric function modelled using values from reference \cite{Johnson:1972}. From a rough estimate, using eqn.~\ref{eq:spp_disp}: $\lambda_{spp}/2 \simeq 300$\,nm, which fits the scale in Fig.~\ref{fig:plasmon_lens} very well.

\subsection{Localized surface plasmons}

We have discussed the two cases for when light interacts with macroscale metal structures, but in this final section we will deal with the special case of exciting plasmons in sub-wavelength metal structures\footnote{For visible light this being the nanoscale.}. Limiting the physical size of the particle in which the plasmon is excited allows for further confinement of the plasmon, resulting in localized surface plasmons.

If we consider the case of a metal sphere of radius, $r$, which is significantly smaller than the exciting wavelength, and also smaller than the skin depth in the material it is made from, $\delta$ (see chapter~\ref{chap:opt}). What will then occur upon illumination, is that the external field will fully penetrate into the sphere and uniformly perturb the sphere's electron plasma. Similar to the case for the bulk plasmon, this will then cause a collective oscillation of the electrons. However, in this case we can specifically drive the oscillation with the frequency of light used, $\omega$, and the limited size of the metal particle will help determine the resonance frequency of the surface plasmon. 

\begin{figure}[h]
	\center
	\includegraphics[width=0.6\columnwidth]{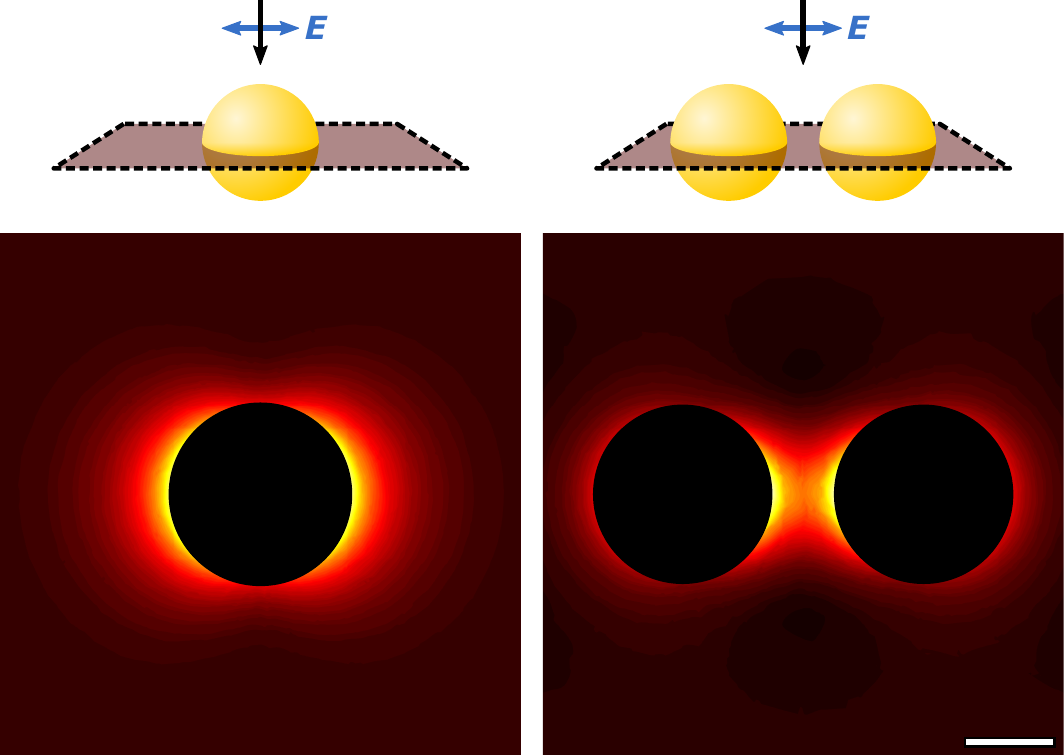}
    \caption{Excitation of plasmons in metal nanospheres from plane wave illumination. $\left|\mathbf{E}\right|^2$ is plotted from the highlighted planes. The left images show a simple dipole resonance for a single sphere, aligned with the polarization of the exciting light. The right images show a gap plasmon between two such spheres. The scale bar is 40\,nm. Simulations performed by T.~Rep\"an.}
    \label{fig:monomer-dimer}
\end{figure}

An analogy to this could be the vibrations in a fixed string, where the length of the string plays a critical role in its fundamental vibration modes. In a more generalized description of localized surface modes, we can relax the criteria of the structures being of similar size to the skin depth and still get similar results\cite{Maier:2007}. The general point however remains: For localized surface plasmons, we do not need any momentum matching to couple to external light, but the structures will also only have a strong plasmonic response when we illuminate them with a wavelength of light sufficiently close to the eigenmodes of the specific geometry. The left-hand side of Fig.~\ref{fig:monomer-dimer} shows a simple dipole resonance for a metal sphere.

We can now also consider what happens if we take two such spheres and put them within a short distance of each other. As the two particles approach each other, eventually their plasmon modes will start to influence each other and hybridize into a coupled gap mode for the particle pair. We can expect two cases, one where both spheres' plasmon oscillate in the same phase, or the other case where they oscillate out of phase. We refer to the in phase plasmons as a 'bonding' mode, and the out of phase as an 'anti-bonding' mode, due to the similarity to hybridization seen in molecular bonding\cite{Nordlander:2004}.

\noindent For the bonding mode, its resonance energy will tend to redshift the closer we move the two particles, while for the anti-bonding mode its energy will blueshift. Generally the anti-bonding mode is considered a 'dark' plasmon mode, as it is difficult to excite by light due to the geometry (as the structure is sub-wavelength, it will be uniformly engulfed in the electric field of the light, favouring in-phase oscillations\cite{Nordlander:2004}).

Localizing the excitation mode in the smaller gap between two structures generally leads to a smaller mode volume for the plasmon, which allows for a higher degree of electric field enhancement and localization, as well as a greater degree of fine-tuning of resonance frequency\cite{Novotny:2011,Fischer:2008}. This is especially exploited for applications in various optical antenna geometries\cite{Novotny:2011,Fischer:2008}. See right-hand side of Fig.~\ref{fig:monomer-dimer} for a gap plasmon mode between two metal nanospheres.

\clearpage

\begin{figure}[h]
	\center
	\includegraphics[width=0.55\columnwidth]{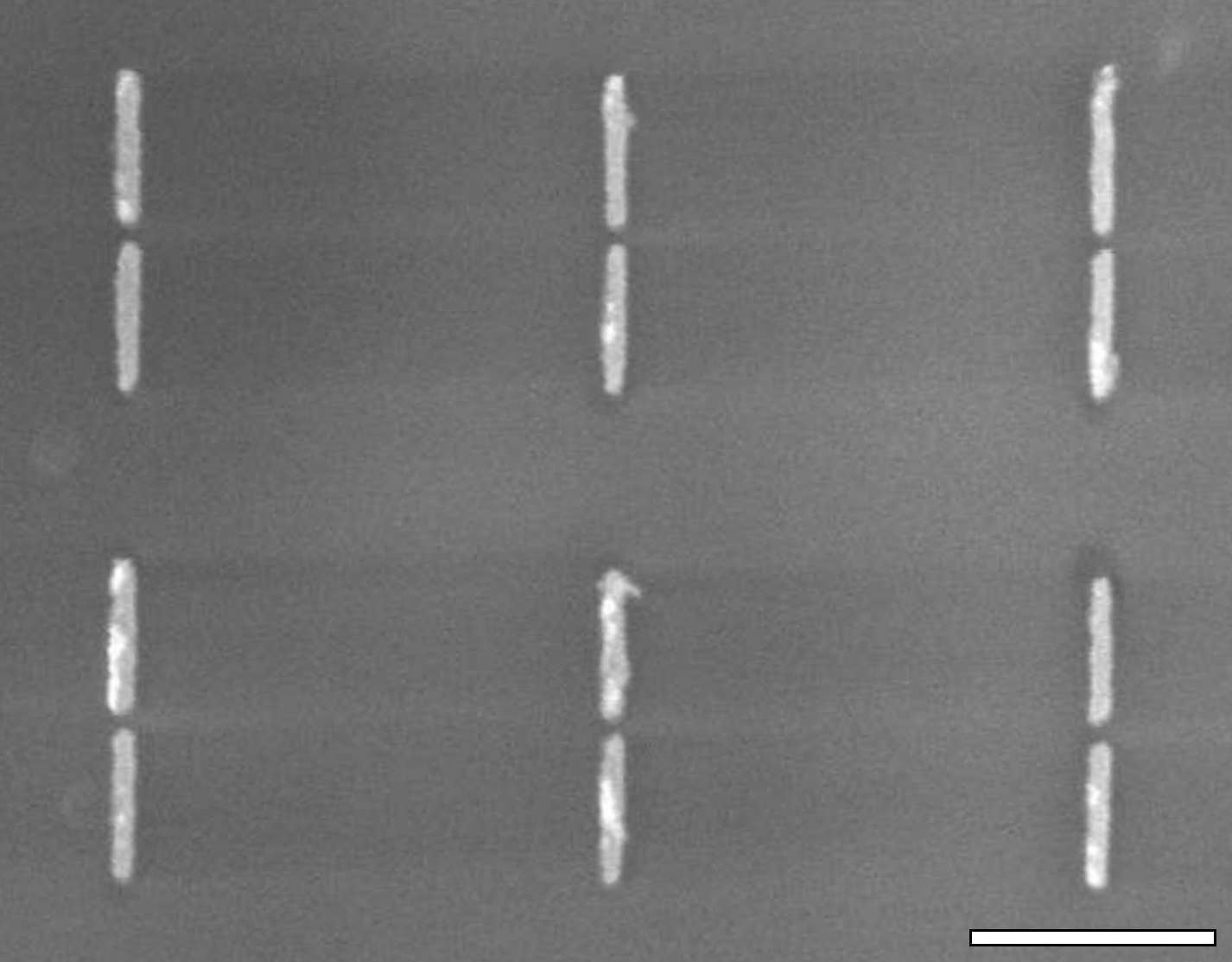}
    \caption{SEM-image of plasmonic gap antennas made with EBL. 40\,nm gold with 10\,nm titanium adhesion layer on a glass substrate. Scale bar is 500\,nm. Unpublished.}
    \label{fig:antenna}
\end{figure}

\noindent For practical applications, nanoantennas will typically be defined using electron-beam lithography (EBL). Fig.~\ref{fig:antenna} shows an example of an array of gap antennas fabricated during my studies in collaboration with N.~Stenger. By varying the length of the individual rods making up the gap, it is possible to change their individual plasmon resonance, and as such change the hybridized gap plasmon as well\cite{Fischer:2008}.

%% file: chapters/chap_metal.tex
\chapter{Metal percolation films}
\label{chap:metal}

As we have discussed previously, metals are able to host collective electron excitations known as plasmons. These excitations can in the right circumstances couple to external electrical fields, such as light. We also touched on the fact that structuring metal at the nanoscale changes the available modes of excitation. In this chapter, we will look at an example of self-organized random nanostructures in metals, achieved by a simple one-step fabrication process. These structures are known as metal percolation films\footnote{Also often referred to as semi-continuous metal films, or metal/dielectric composite films.}.

\section{Percolation theory}

Perhaps an ideal place to start our discussion is with the mathematical discipline of percolation theory, where many of the concepts and terminology originates. The verb "percolate" refers to the windy and tortuous path a liquid will take as it passes through a porous medium (a well-known example for most people being hot water passed through coffee grounds to brew coffee in a regular drip coffee maker). Percolation theory thus deals with this kind of random motion/growth of a system, where at each time interval the system can branch/evolve randomly in a direction, based on the points it has already reached\cite{Saberi:2015}. This results in a system of initially isolated self-similar structures, which will experience dramatic changes in properties at the point of percolation (when the system becomes dominated by one connected region\cite{Saberi:2015}).

\clearpage

\begin{figure}[h]
	\center
	\includegraphics[width=0.8\columnwidth]{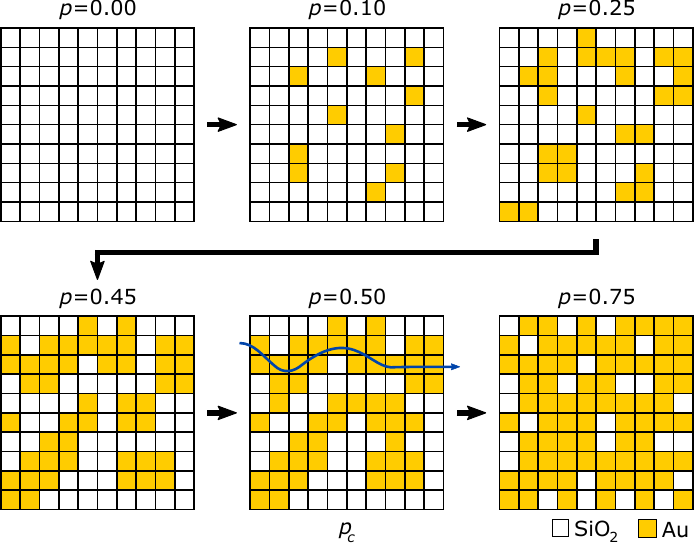}
    \caption{Simple top-view cartoon of the growth process of a gold film on a SiO$_2$ substrate during constant metal deposition. A potential route from left to right is marked in the fifth frame were the system has reached its percolation threshold, $p_c$, and becomes connected.}
    \label{fig:perco}
\end{figure}

\noindent There are multiple ways to describe and define percolation systems (as indeed many different types of systems also exists.)\cite{Shalaev:2007,Saberi:2015}. We will here limit our attention to what is typically referred to as 'site percolation' in a 2D geometry. An example of such a system is shown on Fig.~\ref{fig:perco}. The system consists of a 2D lattice of either vacant (SiO$_2$ substrate) or occupied (Au) 'sites'. As time progresses more and more lattice elements become occupied, and the result is isolated clusters that eventually 'percolate' together, forming a connected system. we can for each frame designate a probability $p$ of a site being occupied, and a probability $1-p$ of a site being unoccupied. In the fifth frame the system reaches its percolation threshold ($p=p_c$), and an electrical current could be passed from the left to the right of the lattice through the connected gold elements. As a straight path is not available, the resistivity of the percolation film will generally be a lot larger than for a fully formed film\footnote{In our Drude model terms from chapter~\ref{chap:opt}: A longer path in the metal means more scattering events, which limits conductivity.}. As additional material is deposited the resistivity will also decrease as a more and more straight path becomes available. The percolation threshold thus marks a phase transition of the system as a whole, as it marks the point where the collective film goes from being an insulator to a conductor\footnote{The individual clusters that make up the film before percolation are of course still conductive, the percolation threshold simply marks the transition from isolated to connected clusters.}.

This style of growth pattern is observed in many self-assembled structures\cite{Shalaev:2007}, and overall results in a high degree of self-similarity. This means that a small subset of the structure is representative of the whole, even across different size scales\cite{Shalaev:2007}. 

\begin{figure}[h]
	\center
	\includegraphics[width=0.9\columnwidth]{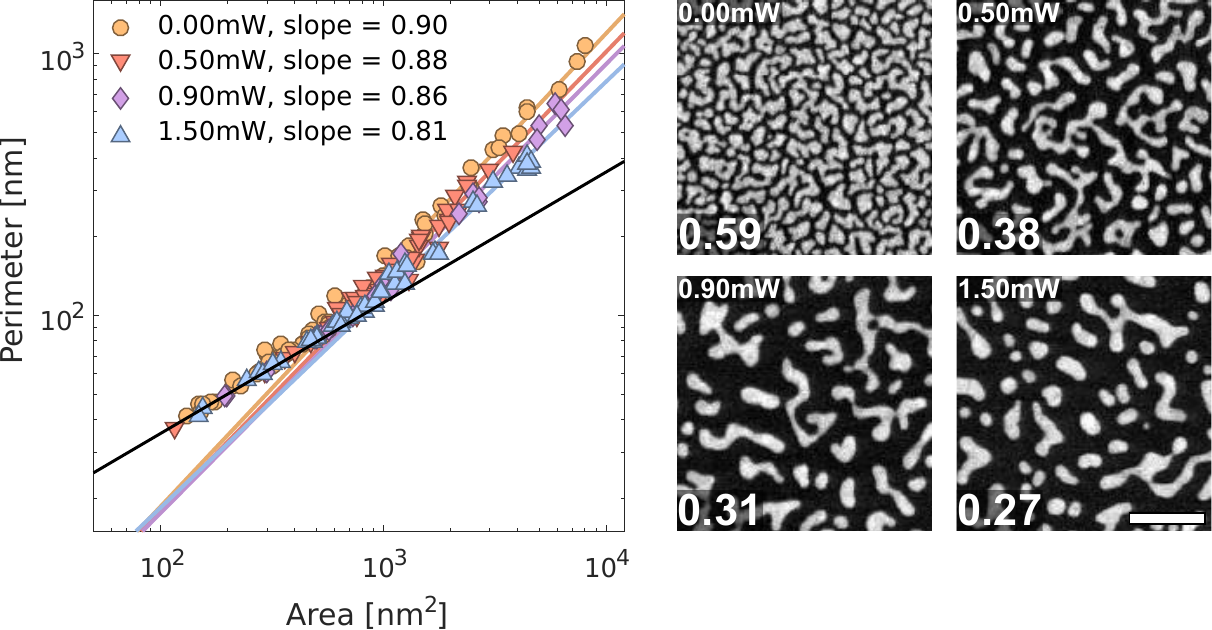}
    \caption{Perimeter vs. area plot of four 5\,nm gold percolation film morphologies. The solid black line is the theoretical expression for perfectly circular particles. STEM dark-field images of the samples are labelled with the laser power used to alter the morphology in the top left corner, and in the bottom left the filling fraction of gold in the image is noted. The scale bar is 150\,nm, all images at same scale.}
    \label{fig:fractal}
\end{figure}

\noindent A simple way to see the self-similar/fractal nature of the 2D percolation geometry is in a perimeter vs. area plot of its constituent particles. On Fig.~\ref{fig:fractal} four different morphologies can be seen: A pristine 5\,nm percolation film, and three examples of morphologies obtained by partially melting similar films with three different laser powers. By recording scanning transmission electron microscope (STEM) dark-field images of the morphologies, it is possible to gain high quality greyscale images of the sample structures. From these we can extract the perimeter and areas of each particle present in the image. If we plot the perimeter vs. particle area of each particle in the images, we get the set of curves on the log-log plot in the left of Fig.~\ref{fig:fractal}. The fact that the points fall on a straight line in the log-log plot indicates that the perimeter/area ratio ($P/A$) of the particles is preserved across size scales (with the ratio given by the slope of the linear fit). For the case of perfectly circular particles of random sizes, we would expect the black curve with slope $0.5$ also shown in Fig.~\ref{fig:fractal}. As expected the $P/A$-ratio decreases for the samples that have been partially melted, as more and more of the particles contract to more spherical shapes for higher laser powers. The scale invariance of the $P/A$-ratio is a strong indicator of self-similarity, and a common feature of fractal systems\cite{Shalaev:2007}. We will see in a later section how this fractal/self-similar morphology of the films plays a large impact in their optical properties, and in later chapters we will also see how the percolation threshold is critical for maximizing plasmonic field enhancement in these kinds of films.

\section{Fabrication}

The gold percolation films studied here have all been fabricated by metal evaporation onto a dielectric substrate, using an electron-beam physical vapor deposition (e-beam) system in the cleanroom facilities at DTU Danchip. We will here discuss the physics responsible for the nanostructuring that happens inherently during this kind of deposition, as well as how metal evaporation is achieved and controlled in an e-beam set-up.

\subsection{The Volmer-Weber growth process}

\begin{figure}[h]
	\center
	\includegraphics[width=0.8\columnwidth]{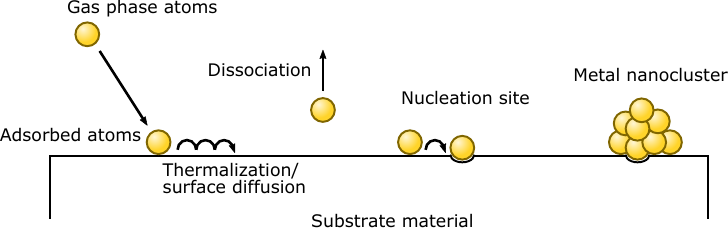}
    \caption{Microscale schematic of the deposition process. Gas phase atoms adsorb on the substrate surface, and then thermalize/diffuse until they either dissociate back into the gas phase, or bond to a point of lower potential on the substrate, or another metal atom.}
    \label{fig:VW_growth}
\end{figure}

When evaporating metals onto a substrate for thin film growth, we are dealing with a phase transition of gas/liquid phase metal atoms to a solid phase metal film on the substrate material. We refer to the first stable points where the metal atoms can coalesce together into a solid as nucleation sites\cite{Greene:2010}. These points are where the metal atoms have a higher probability of growing in time during deposition, rather than dissociate. The substrate temperature and evaporation temperature at the metal source will naturally influence the required size for such a nucleation site to be stable, as more energetic atoms will require larger clusters to become stably bound\cite{Greene:2010}. During a deposition the gas phase atoms will adsorb unto the substrate material. Here they thermalize with the substrate, and start to diffuse on it. During diffusion they will either dissociate back into the gas phase, or form nucleation sites/bond to nucleation sites. A schematic picture of the different processes that can occur when the gas phase atoms hit the substrate material can be seen on Fig.~\ref{fig:VW_growth}.

\clearpage

\noindent In the case of metal deposition onto a dielectric substrate, the metal atoms will have a much stronger mutual bonding than to the substrate. This means that the metal atoms as they are deposited onto the substrate are more likely to attach to already existing metal clusters than to form new nucleation sites. This type of film growth is referred to as 3D-island or Volmer-Weber growth. In the case of equally strong interaction between the deposited material and the substrate, the atoms will instead quickly form flat layers, leading to 2D-layer or Frank–van der Merwe growth. An example of 2D-growth could be metal on metal deposition. Finally, a third kind of growth exists that is a mix of 2D-layer and 3D-island growth, called Stranski-Krastanow growth. Here the deposition initially results in 2D-layers, but due to a lattice mismatch between substrate and deposited material eventually the strain and energy cost associated with additional 2D-layers becomes too great. This results in 3D-islands forming on top of the 2D-layer, typically after 1-3 monolayers. An example of Stranski-Krastanow growth is Ge/Si(001), where germanium quantum dots can be grown on a silicon (001) surface simply from the \textasciitilde4\% mismatch in lattice constants between Ge and Si(001)\cite{Greene:2010,Voigtlander:1999,Vailionis:2000}.

If we reconsider the Volmer-Weber process, we can understand how a percolation film will naturally self-organize. As the initial metal clusters grow during continuous metal deposition, they will eventually become connected in increasingly larger and larger structures, that also become more twisted and tortuous due to the randomness of where additional material is added. Eventually the film will reach a percolation threshold where it becomes possible to find a fully connected path through the system. For additional deposition after this point, the film will slowly become more and more interconnected until it forms a rough fully connected film with large holes/gaps in it. Further deposition will close the film together completely, and the growth will shift to a 2D layer by layer growth, as the gold is now being deposited onto a gold film, with more or less equal likelihood of bonding to any point\cite{Greene:2010,Singh:2005}.

\begin{figure}[h]
	\center
	\includegraphics[width=0.6\columnwidth]{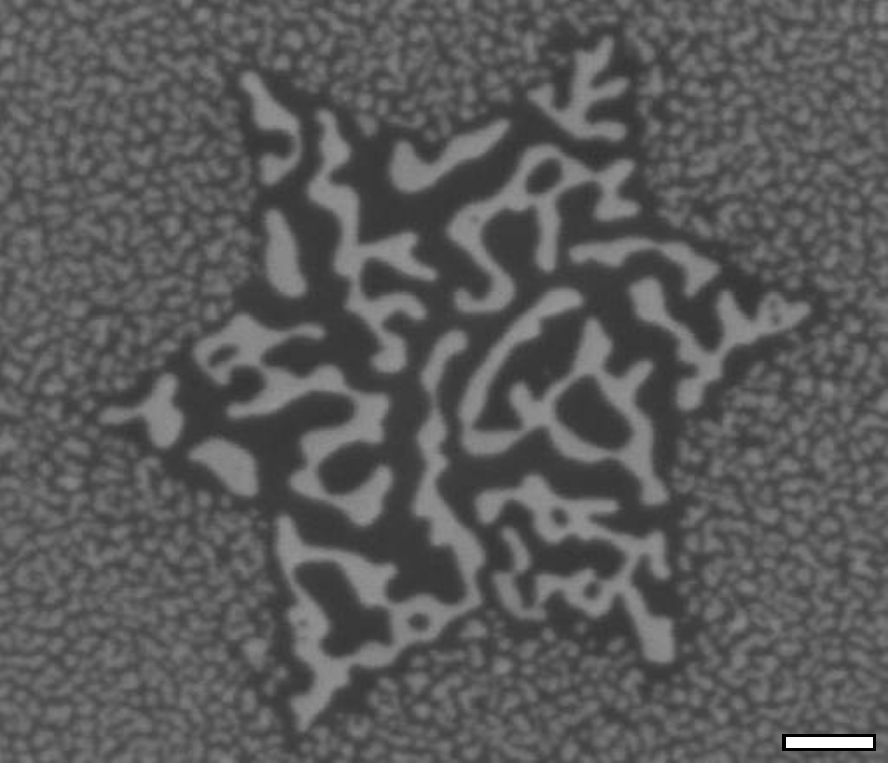}
    \caption{SEM-image of a multi-pointed MoS$_2$ monolayer flake on a SiO$_2$/Si-substrate. 4\,nm of gold has been deposited. Scale bar is 250\,nm.}
    \label{fig:mos2}
\end{figure}

\noindent An example of the strong influence of substrate material for film growth can be seen on Fig.~\ref{fig:mos2}. Here we see an electron micrograph of a monolayer\footnote{MoS$_2$ is an indirect band gap semiconductor, but transitions to a direct band gap for monolayers\cite{Manzeli:2017}} MoS$_2$ flake with 4\,nm gold deposited. It is very clear that the gold structures formed on top of the flake differ drastically from the ones on the surrounding SiO$_2$/Si-substrate. The reason for this can be explained by the surface energy between gold and MoS$_2$ versus that of gold and silicon. The stronger interaction between gold and MoS$_2$ allows for much more efficient nucleation and film growth, resulting in bigger clusters at the same deposition thickness.

\subsection{E-beam metal deposition}

One method of metal deposition is that of electron beam (e-beam) evaporation. A high energy beam of electrons is used to heat part of a metal ingot enough to evaporate atoms off of it. As only a small part of the ingot is evaporated, it is possible to conserve deposition material, and the evaporated material is also projected into a directional jet of gas phase atoms\cite{Singh:2005}.

\clearpage

\begin{figure}[h]
	\center
	\includegraphics[width=0.5\columnwidth]{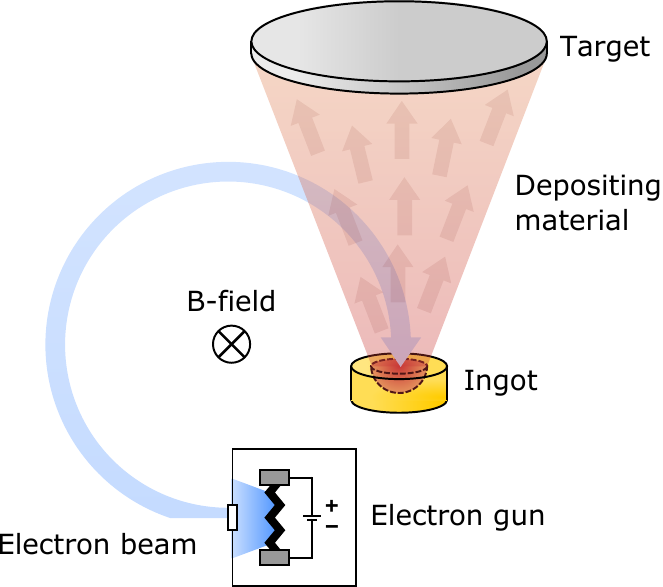}
    \caption{Cartoon of the e-beam evaporation and deposition process. In a high-vacuum chamber, an electron gun generates a high current of electrons that are diverted into a metal ingot by a magnetic field. The kinetic energy of the electrons locally heat the ingot and cause a section of it to evaporate, creating a directed jet of metal atoms towards a target.}
    \label{fig:e_beam}
\end{figure}

\noindent The electron beam is typically generated by a tungsten\footnote{Tungsten is used due to its exceptionally high melting point of \textasciitilde$3400\,^\circ$C} filament that is Ohmically heated by applying a large voltage to it. By heating the filament enough the electrons in it eventually receive enough kinetic energy that they are able to escape to the vacuum level. At this point the filament will eject electrons in many different directions, and a pinhole is usually placed in front of it to limit the angular distribution of electrons. The beam passing the pinhole is thus relatively collimated and can be accelerated by electric fields and directed by magnetic fields. As an electron beam cannot propagate in air (due to collisions with the air molecules), the entire deposition process has to happen in high-vacuum\footnote{Pressures of about \textasciitilde$10^{-4}$-$10^{-8}$\,torr.}. As low air-pressure is generally desired for the deposition process to be more uniform and well-controlled, this is not a major disadvantage to the e-beam method. As both the amount of electrons and their energy can be controlled well electrically in the system, it is possible to vary the deposition rate across a broad range of values, and likewise it is possible to evaporate many different materials, despite their differences in melting point\cite{Singh:2005}.

During a deposition process, a series of steps will be performed before the actual material deposition. This is to ensure consistent material evaporation during the process, removing any potential contaminants from the source material surface, and also to protect the source metal ingot and maximize its lifetime. Before actual deposition, the source ingot will first go through a soaking process where it is slowly raised to temperature by scanning the e-beam across it. This helps ensure a uniform evaporation from the ingot, and by letting the initial outer layers evaporate with the shutter between the ingot and target staying closed, any surface contaminants on the source ingot can be removed without contaminating the target substrate.

\begin{figure}[h]
	\center
	\includegraphics[width=0.65\columnwidth]{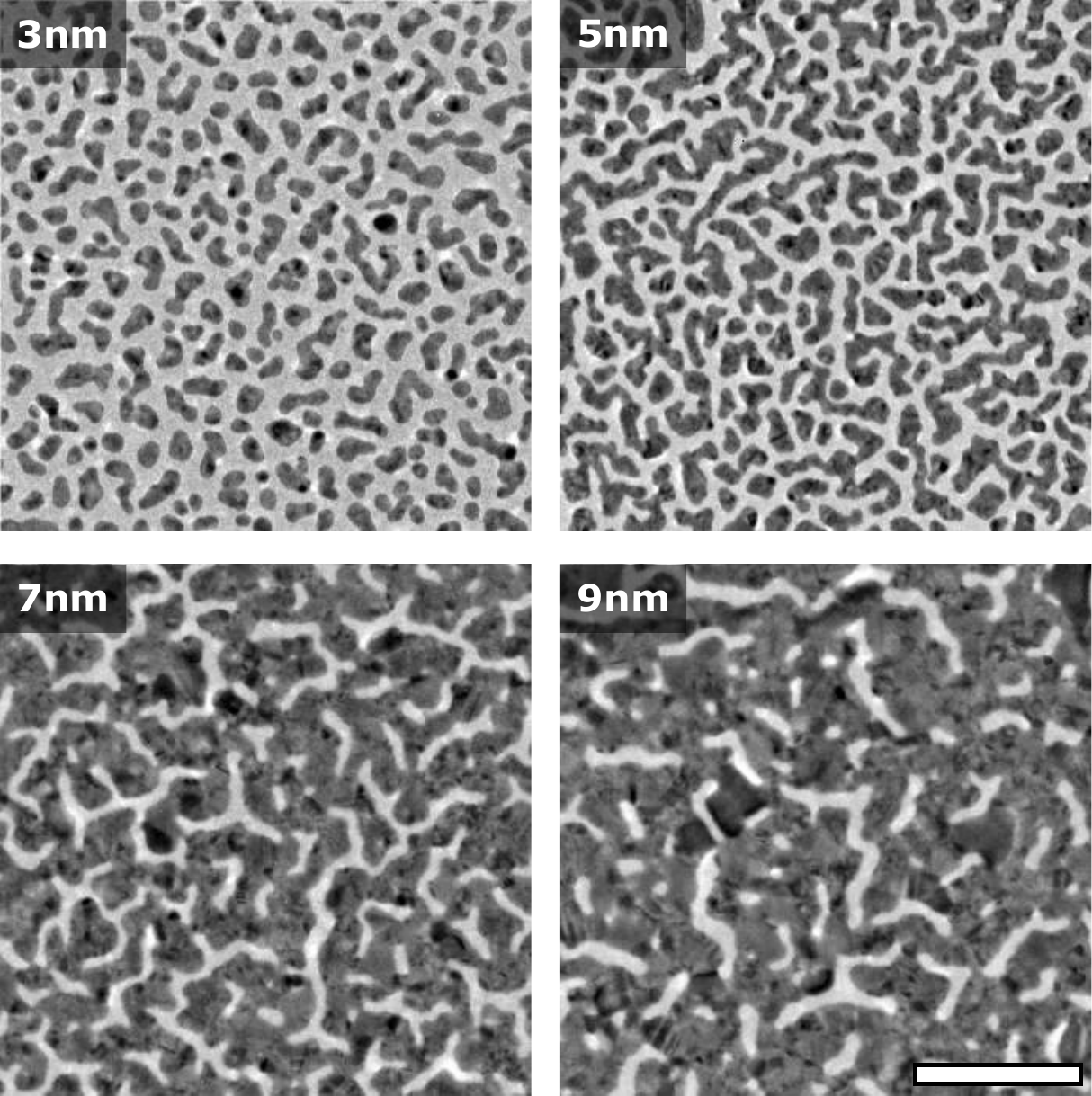}
    \caption{TEM-images of four different gold morphologies achieved by varying the deposition time at a constant deposition rate of 2\,\AA/s. The listed thickness is the total deposited amount, which is not necessarily the actual thickness of the clusters. The dark contrast is gold, the lighter substrate is SiO$_2$. Scale bar is 50\,nm, all images at same scale.}
    \label{fig:TEM_images}
\end{figure}

\noindent The samples investigated in this thesis were all fabricated by maintaining a constant gold deposition rate of $2.0$\,\AA/s (within experimental limits of roughly $\pm 0.1$\,\AA/s). High-resolution transmission electron microscope (TEM) images of four different film morphologies achieved by varying the deposition time can be seen on Fig.~\ref{fig:TEM_images}. The contrast in the TEM-images also offers us some additional insight into the structure of the gold films, as the contrast is somewhat proportional to the thickness of the material (thicker material appears darker). From this we can see that the majority of the film is actually of a similar thickness, with the exception of sparsely distributed 'grains' of significantly darker contrast. For the thicker sample depositions these grains also seem to grow significantly in size, while  the surrounding film uniformly seems to become thicker. Essentially, what we are seeing here is the initial nucleation sites of the film deposition, slowly growing into larger and larger crystal domains of what will eventually become a highly polycrystalline/amorphous gold film if depositing beyond the 9\,nm case.

\section{Optical properties}

The optical properties of metal percolation films (and random media in general), has already been studied in detail\cite{Shalaev:2007,Stockman:1998}. We will here try to summarize some of these results, and also discuss how the fractal and self similar morphology of metal percolation films naturally results in very broadband optical properties.

\subsection{Theoretical approaches}

An obvious first approach to treat the optical properties of semi-continuous media in general is that of effective-medium and mean-field theories\cite{Garnett:1906,Shalaev:2007}. Such approaches are able to predict the linear properties of semi-continuous media quite well\cite{Shalaev:2007}. However, because such models do not account for the exact local field distributions in the structures, they miss out on some of the crucial phenomena\cite{Shalaev:2007}.

Several models have been developed to calculate the exact optical response of random metal clusters when illuminated with external radiation\cite{Bergman:1992,Shalaev:1998,Stockman:1996,Stockman:1997,Genov:2003b}. We will here focus on the theoretical model as outlined in references \cite{Shalaev:2007,Brouers:1997a,Brouers:1997b,Brouers:1998,Gadenne:1997,Shalaev:1998}. The premise is quite straight-forward: We consider a lattice of $L\times L$ elements in $x,y$-coordinates. Each site member then consists either of a dielectric or metal element. External radiation is directed perpendicular to the $xy$-plane (along the $z$-axis), and each lattice element is much smaller than the wavelength of the light exciting the structure\footnote{I.e. we can neglect field variation across the structures, also known as the quasistatic approximation.}. We model the metal's dielectric function as a Drude metal (see chapter~\ref{chap:opt}), and the dielectric as the constant $\epsilon_d$. For treating the film's response to an external field, $\mathbf{E}_0(\mathbf{r})$, we consider the local AC conductivity of the system, $\sigma(\mathbf{r})$. For metal elements this becomes $\sigma_m = -i\omega\epsilon_0(\epsilon_m-1)$ and for dielectric elements it is $\sigma_d = -i\omega\epsilon_0(\epsilon_d-1)$. We can now express the total electric field in terms of the local potential, $\phi(\mathbf{r})$, and the external field:
\begin{align}
	\mathbf{E}(\mathbf{r}) = -\nabla \phi(\mathbf{r}) + \mathbf{E}_0 (\mathbf{r}).
\end{align}
\noindent The local current, $\mathbf{j}(\mathbf{r})$, is given by Ohm's law\footnote{$\mathbf{j}(\mathbf{r})=\sigma(\mathbf{r})\mathbf{E}(\mathbf{r})$}, so we can write:
\begin{align}
	\mathbf{j}(\mathbf{r}) = \sigma(\mathbf{r}) \left[-\nabla\phi(\mathbf{r}) + \mathbf{E}_0\right].
\end{align}
\noindent From current conservation, $\nabla \cdot \mathbf{j}(\mathbf{r}) = 0$, we finally get the Laplace equation for the local potential:
\begin{align}
	\nabla\cdot\left(\sigma(\mathbf{r}) \left[-\nabla\phi(\mathbf{r}) + \mathbf{E}_0\right]\right)=0.
    \label{eq:laplace}
\end{align}
\noindent To solve eq.~\ref{eq:laplace}, we can regard the conductivity between the bonds of our lattice elements. For the $i$'th site this becomes:
\begin{align}
	\sum_j \sigma_{ij} (\phi_i - \phi_j + E_{ij}) = 0,
    \label{eq:kirchoff}
\end{align}
\noindent where the summation over $j$ is for the 4-point connected nearest neighbours to lattice site $i$ (i.e., $j=i-1, i+1, i+L, i-L$). $\sigma_{ij}$ is the conductivity between site $i$ and $j$, given as $\sigma_{ij}=2\sigma_i\sigma_j/(\sigma_i+\sigma_j)$. $E_{ij}$ is the electromotive force between sites $i$ and $j$. If $a$ is the physical size of each lattice element, and the external radiation is $y$-polarized, we can define $E_{ij}= \pm a E_0$, for bonds in the $\pm y$-directions, respectively, and bonds along the $x$-direction $E_{ij}=0$ (for $x$-polarized light, this would just be reversed).

Eq.~\ref{eq:kirchoff} is essentially a Kirchoff equation, and we can use a matrix representation for the full set of equations of the system:
\begin{align}
	\hat{H} \mathbf{\Phi} = \mathbf{F},
    \label{eq:HF}
\end{align}
\noindent where $\hat{H}$ is the $L^2 \times L^2$ sparse block matrix Kirchoff Hamiltonian of the system. It has diagonal elements $\hat{H}_{ii}=\sum_{j}\sigma_{ij}$ (the summation here is like eq.~\ref{eq:kirchoff}) and off-diagonal elements $\hat{H}_{ij}=-\sigma_{ij}$. $\mathbf{\Phi} = [\phi_i]$ and $\mathbf{F} = [-\sum_j \sigma_{ij} E_{ij}]$ are simply two $L^2$-length column vectors.

By solving eq.~\ref{eq:HF} for $\mathbf{\Phi}$, it is possible to get the explicit local electric field distribution for a given percolation film geometry\footnote{The actual direct solving of eq.~\ref{eq:HF} is incredibly computationally intensive, and a block elimination method is normally used\cite{Genov:2003a,Genov:2003b}. Alternatively, the sparse nature of $\hat{H}$ allows for sparse matrix numerical methods, greatly reducing computational requirements.}. Eq.~\ref{eq:HF} can be solved with periodic boundary conditions -- i.e. the first row of lattice sites are connected to the last row, and the first column is connected to the last column. However, when opting for such boundary conditions, the system of equations represented by eq.~\ref{eq:HF} becomes inconsistent, as one of the equations will become linearly dependent on the others. This can be rectified by 'grounding' one of the lattice sites\cite{Genov:2003b} (for example, the corner $i,j=1,1$). In practice, this is achieved by simply assuming its potential $\phi_i=0$, and removing the corresponding row $i$ and column $j$ from $\hat{H}$, and rows $i$ from $\mathbf{\Phi}$ and $\mathbf{F}$ when doing the calculation.

\begin{figure}[h]
	\center
	\includegraphics[width=\columnwidth]{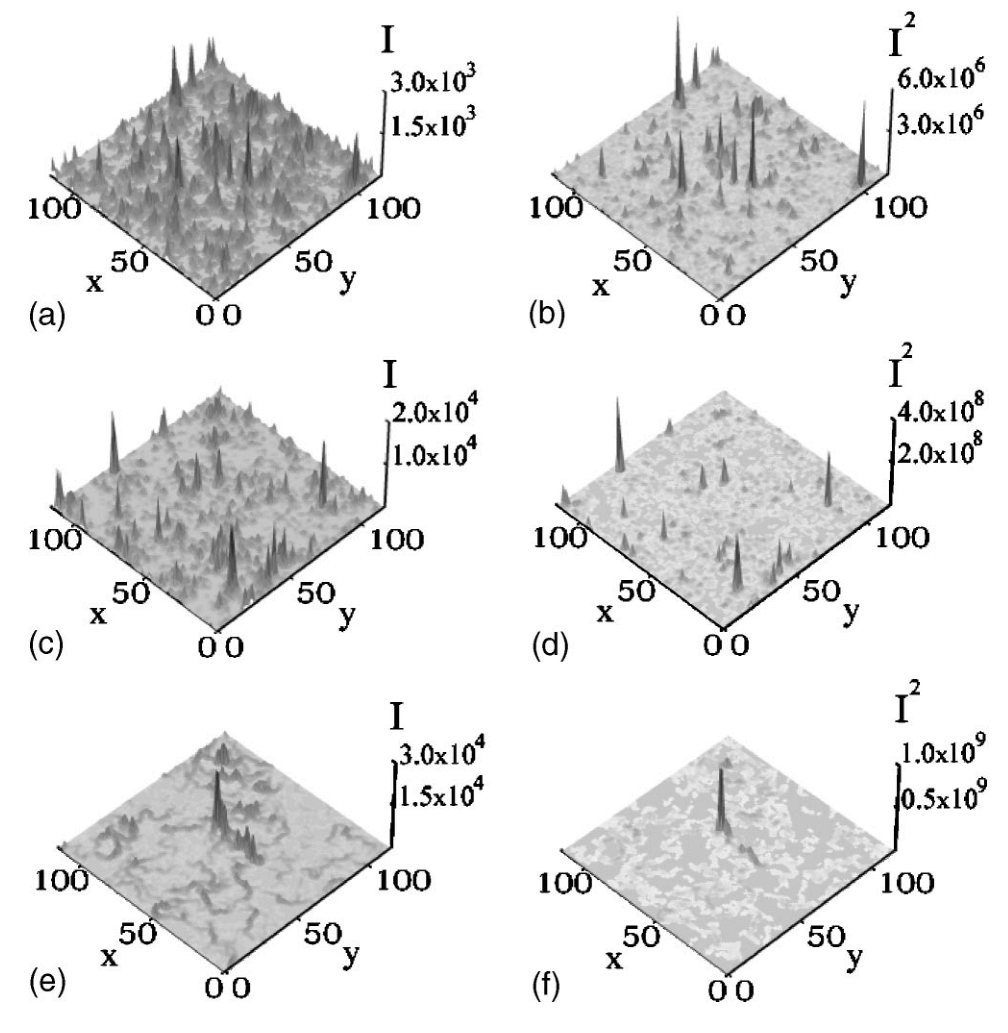}
    \caption{The spatial distributions for the normalized local intensity $I(x,y)$ and for the non-linear intensity $I^2(x,y)$ for silver on glass percolation films. The distributions are calculated for three different wavelengths: \textbf{(a-b)} $\lambda=0.370$\,$\upmu$m, \textbf{(c-d)} $\lambda=1$\,$\upmu$m, and \textbf{(e-d)} $\lambda=5$\,$\upmu$m. The metal filling factor is chosen as $p=p_c$ for all cases. Reproduced from reference \cite{Genov:2003b}.}
    \label{fig:perco_fields_theory}
\end{figure}

Fig.~\ref{fig:perco_fields_theory} shows the local field intensities $I(x,y)$ and squared intensities $I^2(x,y)$ for silver percolation films at three different excitation wavelengths, obtained from solving eq.~\ref{eq:HF} in reference \cite{Genov:2003b}. From these distributions we clearly see that the optical response of the film is dominated by very intense plasmonic 'hotspots'. We also see that these spots are strongly localized to small regions of the film, and that the spatial distributions of the hotspots vary for different excitation wavelengths. Finally, we see that the excitations from longer wavelengths of light also have a larger general intensity than for shorter wavelengths. The localization of the plasmonic hotspots is often attributed to the fact that fractal or chaotic systems, like percolation films, lack translation symmetry. Thus they do not support propagating waves\cite{Ducourtieux:2000} (the most extreme case of this being potential Anderson localization\cite{Stockman:2001}). The fact that several different resonance energies are supported can likewise be attributed to the fractal/self-similar geometry. The localized plasmon resonance of a metallic structure strongly depends on its size, and as the percolation films have similar shaped metal structures at different size scales, we likewise expect them to have plasmon resonances across different energies.

\clearpage

\subsection{Experimental investigations}

\begin{figure}[h]
	\center
	\includegraphics[width=0.75\columnwidth]{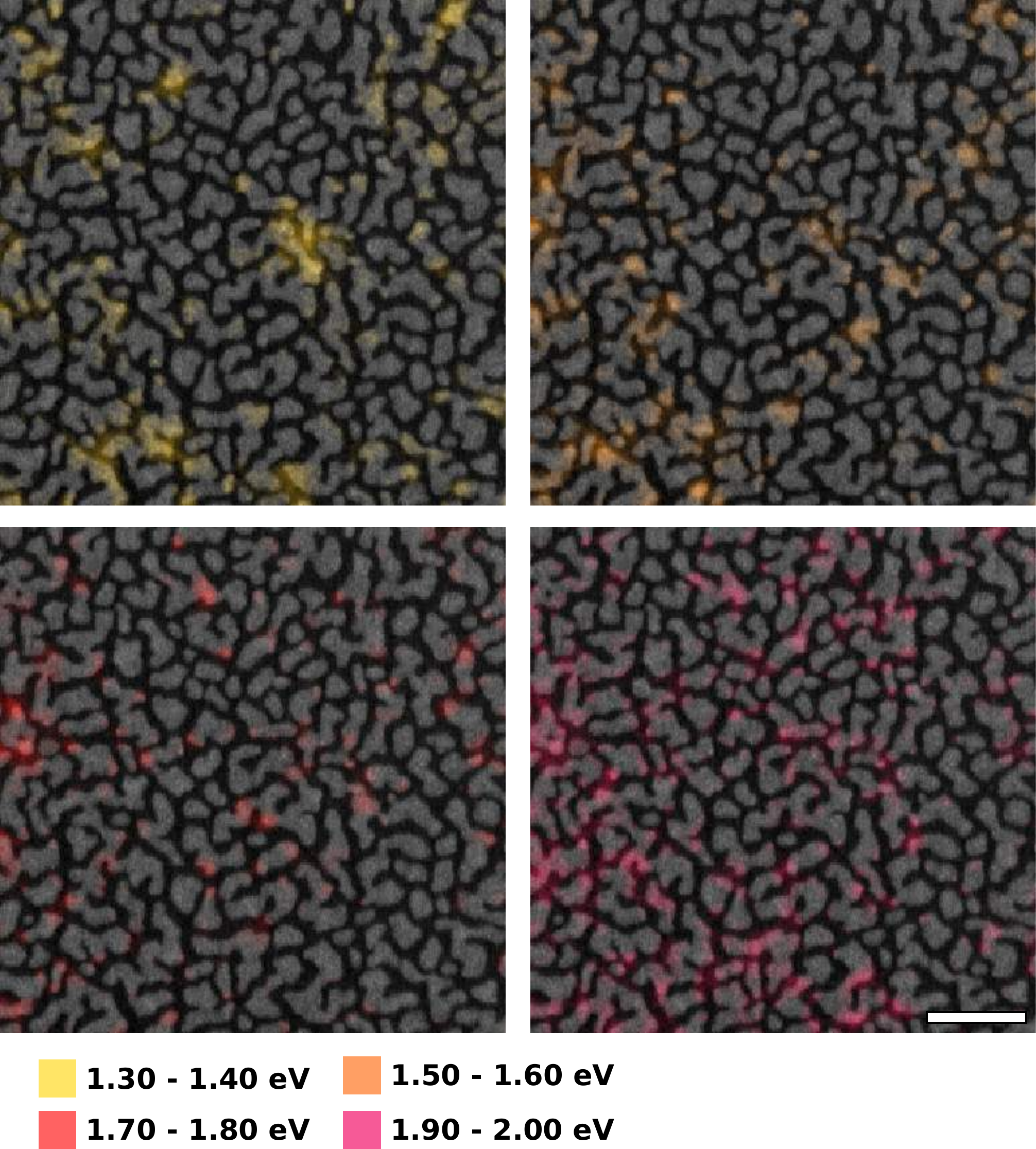}
    \caption{STEM dark-field images of the 5\,nm thin film gold sample with integrated EELS maps projected on top. The background corrected EELS data were integrated in different energy ranges to produce intensity maps of the recorded EELS signal. The coloured parts of the images generally show between 5-15 times as high EELS signal as the rest of the image. Scale bar is 100\,nm.}
    \label{fig:5nm_maps}
\end{figure}

We will here highlight some of our own experimental results for near-field measurements of gold percolation films, performed with electron-energy loss spectroscopy\footnote{See chapter~\ref{chap:eels} for an explanation of the method and equipment.}. On Fig.~\ref{fig:5nm_maps} are a set of examples of how the plasmons in a 5\,nm gold film can be spatially visualized by integrating the EELS data over various energy ranges\cite{Bosman:2011} and overlaying this on the sample morphology also recorded in the scanning transmission microscope (STEM).

\clearpage
\noindent From the images we see that the local field distribution in the gold film varies quite considerably for different energy ranges. For example, we see that the higher energy plasmons are typically quite well localized on the surface of the gold, while the lower energy modes can in some cases extend fairly significantly. In general, several hot spots are apparent where the local intensity is significantly larger than the immediate surrounding areas. We also see that the apparent position of these hot spots is not related to the geometry of the film in a simple way, as many similar looking regions appear dark in the EELS measurement, while others show large EELS intensity signal. This is in agreement with similar measurements for silver films, see reference \cite{Losquin:2013}.

\begin{figure}[h]
	\center
	\includegraphics[width=0.7\columnwidth]{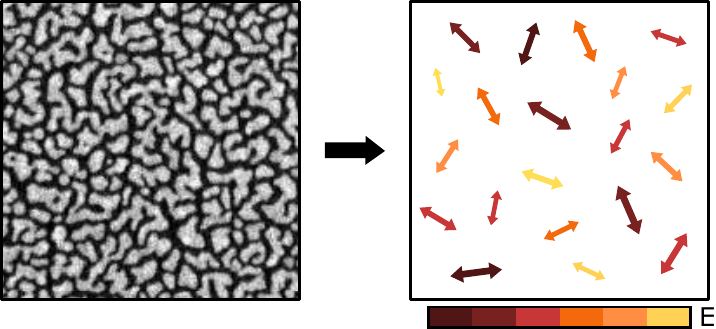}
    \caption{The complex geometry of the percolation film results in various plasmonic hotspots at different resonance energies and with different polarizations. We can thus think of an equivalent model for the percolation film that consists of a random distribution of linearly polarized plasmon resonances. We here represent these hotspots as double arrows aligned with their polarization. The size of the arrows indicates their relative field intensity, and the colours represents their resonance energy.}
    \label{fig:perco_model}
\end{figure}

\noindent Because of the random nature of their morphologies, the specific detailed spatial distribution of plasmon resonances can be difficult to reproduce and compare across percolation film samples. As a result, direct qualitative maps of the plasmons like in Fig.~\ref{fig:5nm_maps} best serve to give us a general impression of how the plasmons are distributed. Making detailed comparisons between several EELS maps like in Fig.~\ref{fig:5nm_maps} at different energies also quickly becomes very overwhelming to interpret. Lastly, the fact that many of the films' optical properties are dominated by strongly localized plasmonic hotspots\cite{Stockman:1994,Stockman:1996,Shalaev:2007,Genov:2003b} allows us to make an equivalent model of the films based on their local properties\cite{Gaio:2015}.

Fig.~\ref{fig:perco_model} shows a schematic of how a complex gold percolation geometry can be thought of equivalently as a random distribution of plasmonic hotspots with random polarizations\footnote{We will show the polarization dependence explicitly in chapter~\ref{chap:snom}.} and resonance energies. Such a random distribution of plasmonic hotspots can more or less be fully described in a purely quantitative statistical framework. Because the sizes of the particles and gaps in our images are all several orders of magnitude smaller than the investigated regions (500$\times$500\,nm$^2$), we assume that our EELS maps are recorded at the size scale where the films can be considered isotropic. In other words: Our measured regions, while only a small subset of the entire sample, are large enough that any probabilistic conclusions we draw from them can be said to apply to the films as a whole, due to their isotropy.

To make a statistical description of our samples, we take advantage of the methods of previous theoretical and experimental works on self-similar and fractal structures\cite{Stockman:1994,Bozhevolnyi:2001,Genov:2003a,Genov:2003b}. We can extract two parameters that describe the plasmon resonances in our EELS maps: The central position of the plasmon peak in the EELS data, and the amplitude of the resonances (its EELS intensity). We achieve this by a series of sequential Gaussian fits, that progressively become more and more detailed (full method in chapter~\ref{chap:eels}). After having identified the plasmon resonance energies and their intensities, we can now perform two rather simple treatments. First, a histogram can be constructed of the different resonances energies to show how the plasmons are distributed in terms of their energy in the sample. Secondly, we can construct a probability density function (PDF) of the EELS intensities of the plasmons. By then binning the intensities, for a certain range of resonance energies, we form a distribution of intensities in the spectral image (i.e. a histogram). We then normalize this histogram with the total amount of resonances detected to get a probability. By breaking this PDF analysis down by energy ranges, we can sort out which part of the energy spectrum that provides the largest EELS intensities in the percolation film (right-most points of the PDFs), and by that token which energy range would have the largest field enhancement.

\clearpage
\begin{figure}[h]
	\center
	\includegraphics[width=\columnwidth]{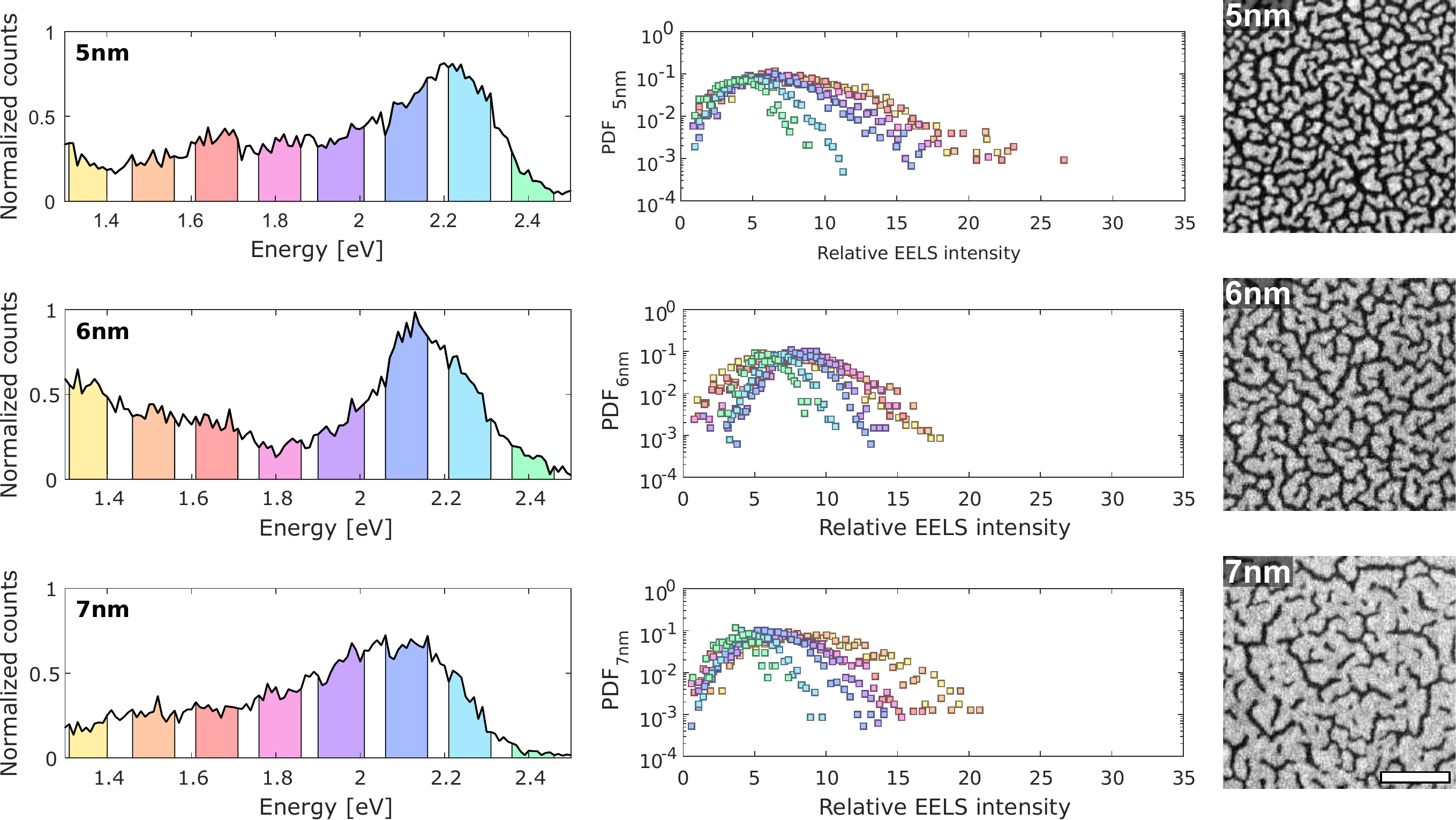}
    \caption{Quantitative analysis of the spectral images recorded from three different percolation film thicknesses: 5, 6, and 7\,nm. The left row shows histograms for how the plasmon resonances are distributed in terms of energy. The centre column shows constructed PDFs of how the EELS intensity is divided inside subset energy ranges identified with the colours in the resonance histograms on the left. On the right is shown a STEM dark-field image of the sample morphology. Scale bar is 150\,nm.}
    \label{fig:perc_stat}
\end{figure}

\noindent Fig.~\ref{fig:perc_stat} shows the results of such an analysis carried out on 5, 6, and 7\,nm gold samples. In the left column simple histograms are plotted showing the distribution in central peak energies for the plasmon modes identified. From these we can see a general peak in the amount of resonances near 2.1-2.2\,eV. For increasing deposition thickness this peak shifts towards lower energies. The centre column of Fig.~\ref{fig:5nm_maps} shows PDFs of the relative EELS intensities for these plasmons, based on taking data from the highlighted coloured energy ranges in the histograms. The EELS intensities are normalized to the median of the 50 lowest intensities seen in the individual maps, to get a relative measure of how much larger some resonances might be than others. From these PDFs we are able to see that across all three samples, the highest intensities are found in the lower energy ranges 1.3-1.7\,eV. This is also in agreement with prior work\cite{Shalaev:2007,Brouers:1997a,Brouers:1997b,Brouers:1998,Gadenne:1997,Shalaev:1998,Losquin:2015}.

%% file: chapters/chap_tpl.tex
\chapter{Photoluminescence in gold}
\label{chap:tpl}

The ability of noble metals to emit light after initial optical excitation has been an active area of study for the last several decades. Despite this, the underlying physical mechanisms are still poorly understood for some systems\cite{Mertens:2017,Borys:2013,Haug:2015}.

Metallic photoluminescence was first addressed by A.~Mooradian in 1969\cite{Mooradian:1969}. In this initial publication, Mooradian showed that it was possible to get a spectrum of photoluminescence from noble metals when exciting them with a powerful near-UV light source. He also showed that the shape of this spectrum was dependent on the temperature of the metal.

\begin{figure}[h]
	\center
	\includegraphics[width=0.55\columnwidth]{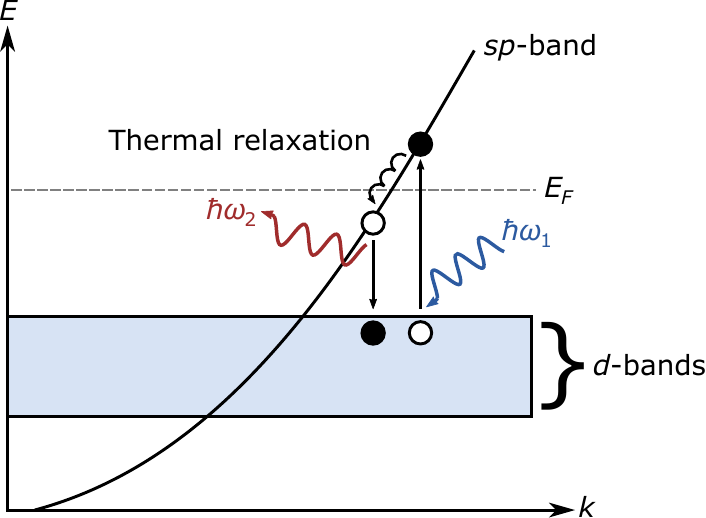}
    \caption{Schematic of the photoluminescence process in metals. $d$-band electrons are optically excited into a higher energy $sp$-band. After some set of thermal relaxation processes, the excited electron recombines with a $d$-band hole, emitting a photon of different wavelength.}
    \label{fig:gold_photo_lum}
\end{figure}

\noindent Mooradian's explanation for the observations was that by illuminating the metal close to the interband transition edge, it is possible to excite electrons from the $d$-states up into the $sp$-conduction band. Here the electrons will have time to thermally relax by phonon emission, before radiatively recombining with holes in the $d$-band (see Fig.~\ref{fig:gold_photo_lum}). Because there is no strict band gap, like in a semiconductor, the electrons can equally efficiently relax to various different energy states before re-emitting a photon. This creates a broad spectrum of luminescence\cite{Mooradian:1969}. By cooling the metal down to cryogenic temperatures, it is observed that this spectrum becomes more narrow. At low temperature, the electrons and holes inherently have a more narrow distribution of initial thermal energy, and they will be more likely to re-emit photons of similar energies as the exciting photon\cite{Mooradian:1969}.

Later in 1986 G.~T.~Boyd \textit{et al.} would investigate the enhancement of gold photoluminescence from rough surfaces\cite{Boyd:1986}. They attribute the higher luminescence signal observed from rough films to be due to localized surface plasmons hosted in the nanostructured surface roughness. Because of the presence of resonant structures, both the light absorption and emission processes are enhanced, as compared with the flat films\cite{Boyd:1986,Mertens:2017}. Like Mooradian, they find that the efficiency for the luminescence process from the smooth films is about $\eta$\textasciitilde$10^{-10}$, while in some cases the signals from rough films can be close to an order of magnitude larger. They also investigate two-photon excitation of luminescence, reporting that a luminescence signal is only observed from the rough films. They attribute this to the fact that a two-photon process is even more sensitive to the field enhancement from the plasmons in the rough films.

In 1998 J.~P.~Wilcoxon \textit{et al.} were the first to observe photoluminescence from gold nanoclusters\cite{Wilcoxon:1998}. They found a significant dependence between the metal cluster sizes and emission efficiency. For the smallest of their investigated clusters, they reported $\eta$\textasciitilde$10^{-4}$-$10^{-5}$. Later studies by several other groups would go on to improve this to $\eta$\textasciitilde$10^{-3}$, when investigating nanoparticles or nanostructures of different dimensions\cite{Mohamed:2000,Eustis:2005,Yorulmaz:2012,Huang:2015,Yin:2016,Fang:2012}. These later results also showed a very strong dependence of the luminescence on the plasmon resonances of the structures. This has mainly been attributed to the plasmons enhancing absorption and emission of certain wavelengths of light. However, from these types of resonant nanostructures it has become possible to observe photoluminescence from noble metals in infrared energy ranges that could not be supported by interband transitions\cite{Mertens:2017}. This has resulted in multiple alternative explanations for the observed luminescence, ranging from black body radiation\cite{Haug:2015,Andersen:2015,Fang:2012,Tcherniak:2011}, inelastic light scattering\cite{Otto:2006,Hugall:2015}, \textit{intraband} excitation emission processes\cite{Beversluis:2003,Haug:2015,Lin:2016}, to direct photoexcitation of electrons in plasmonic structures\cite{Mertens:2017}.

\section{Two-photon luminescence}

As discussed above, it is also possible to stimulate luminescence in metals by two-photon absorption. The basic principle is very simple: Instead of using one photon of energy $\hbar\omega_1$ (like in Fig.~\ref{fig:gold_photo_lum}) to excite the luminescence, two photons of energy $\frac{1}{2}\hbar\omega_1$ are used\cite{Boyd:1986}. Two-photon luminescence (TPL) is thus a three particle process, and as such the efficiency decreases. Two-photon excitation however also leads to new spectral features\cite{Andersen:2015,Boyd:1986}. Usually for single-photon luminescence prominent peaks are seen in the spectrum from radiative recombination near the interband edge of the metal. For TPL, the spectrum is generally more broad and evenly distributed, with a potential peak for second-harmonic generation (SHG). This is also how TPL was initially observed in 1981 by C.~K.~Chen \textit{et al.}, as a broad background spectrum in their experiment of enhancement of SHG from rough metal films\cite{Chen:1981}. Because of its much higher sensitivity to field enhancement effects, TPL can be spectrally engineered to a very high degree\cite{Andersen:2015,Tcherniak:2011,Yorulmaz:2012}.

\subsection{Measurement of TPL}
\begin{figure}[h]
	\center
	\includegraphics[width=0.65\columnwidth]{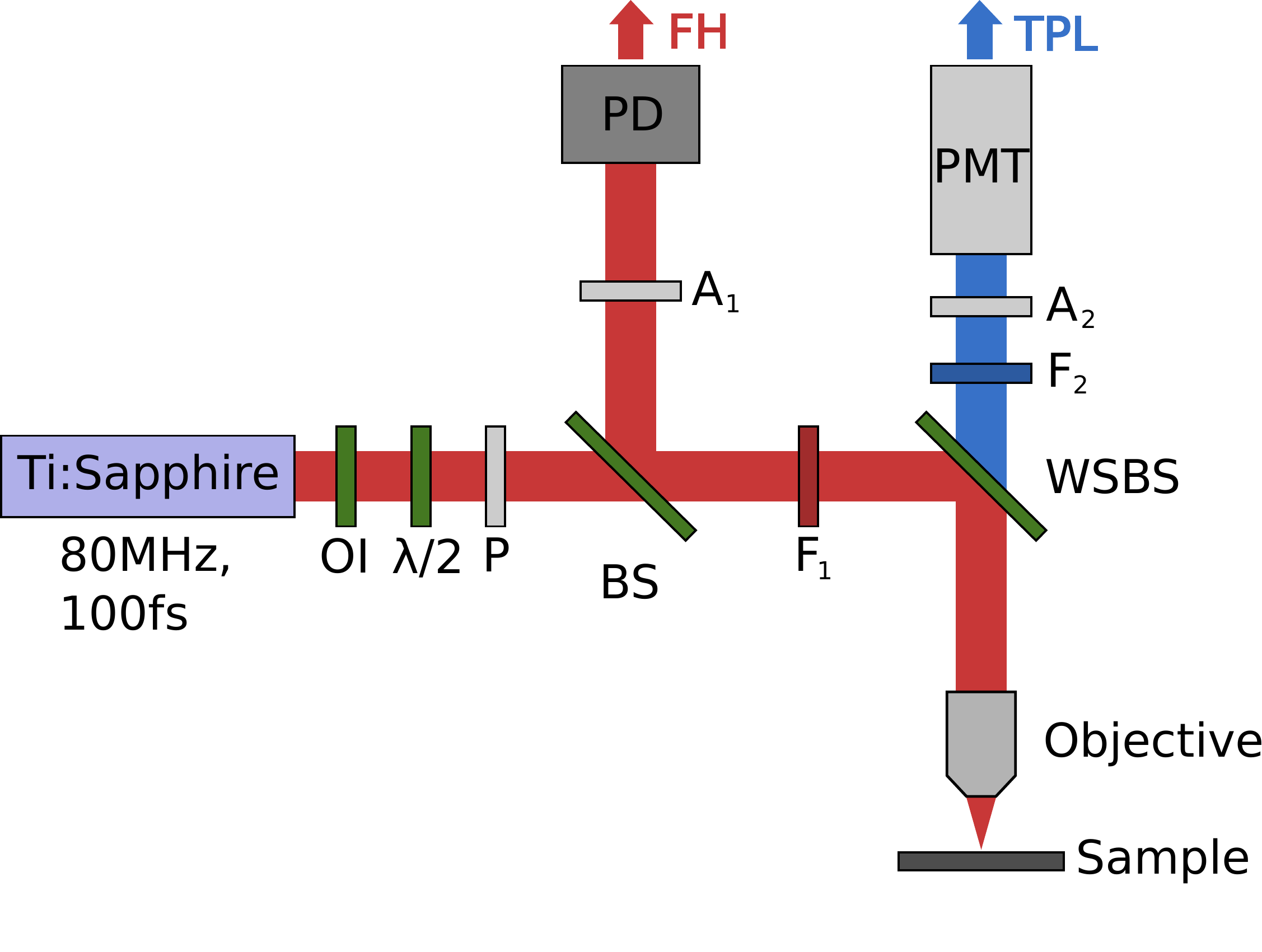}
    \caption{Overview of the TPL set-up utilized for all TPL-measurements presented here. The optics consists of a reflection microscope with a scanning sample stage, using a fs-laser as illumination source.}
    \label{fig:tpl_setup}
\end{figure}
\noindent All TPL-investigations presented here were done in collaboration with Prof. S.~I.~Bozhe\-volnyi's group at University of Southern Denmark, with the measurements carried out by either S.~M.~Novikov or J.~Beermann. On Fig.~\ref{fig:tpl_setup} is a schematic drawing of the TPL set-up that was used. The set-up is based on a reflection microscope, where the same objective is used for sample illumination and signal collection. The light source used is a fs-pulsed Ti:Sapphire laser ($730-860$\,nm, \textasciitilde$100$\,fs, $80$\,MHz repetition). The laser is typically operated at 740\,nm, which we will refer to as the fundamental harmonic (FH) frequency.

The light from the laser is passed through an optical isolator (to stop back-reflection from the sample into the laser cavity), a half-wave plate, polarizer, beam splitter, bandpass filter (matching FH wavelength), and a wavelength-selective beam splitter, before being focused onto the sample by a $100\times$ ($\text{NA}=0.70$) microscope objective. From the reflection back from the sample, the wavelength-selective beam splitter sorts the TPL-signal to go through a filter (350-550\,nm pass) before being collected in a photomultiplier tube. The FH-signal is then sent back through the first colour filter, and the signal is diverted to a photodiode by the first beam splitter. This allows for simultaneous collection of the sample's linear reflectivity (FH-signal) and its TPL-signal (the two filters serving to reduce mixing of signals).

\begin{figure}[h]
	\center
	\includegraphics[width=0.35\columnwidth]{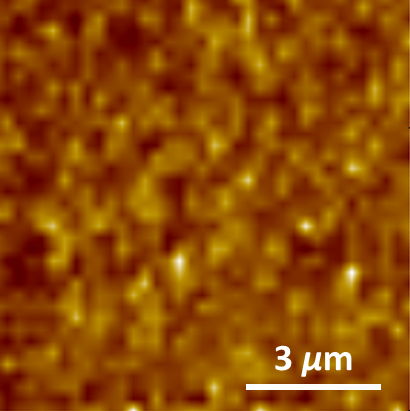}
    \caption{Example of a typical TPL-image recorded from a $6$\,nm percolation gold film. Isolated 'hot spots' show higher signal by about $20$-$30$\% than the rest of the film. As the resolution of the measurement is diffraction limited, it is not possible to truly resolve individual metal clusters.}
    \label{fig:tpl_img}
\end{figure}

By mounting the sample on a computer controlled scanning stage, it is possible to make scanned images of the TPL- and FH-signal across an area of the sample. See Fig.~\ref{fig:tpl_img} for an example of a typical TPL-image. By the half-wave plate and polarizer, it is possible to vary the incident power and polarization of light very accurately. As the initial light entering the setup is linearly polarized to the axis of the optical isolator, the half-wave plate and second polarizer allows the operator to easily scale the power of the beam by partially cross polarizing it at the desired polarization.
\clearpage
\subsection{Enhancement of TPL}

As discussed above, it is possible to see an enhancement of photoluminescence from a rough metal film as compared to a smooth metal film. This has been attributed to the electric field enhancement from the plasmons hosted in the nanoscale features in the rough film\cite{Boyd:1986}. In this section we will discuss how it is also possible to see large TPL-signal enhancement from semi-continuous gold films\cite{Iwaszczuk:2013}.

\begin{figure}[h]
	\center
	\includegraphics[width=0.95\columnwidth]{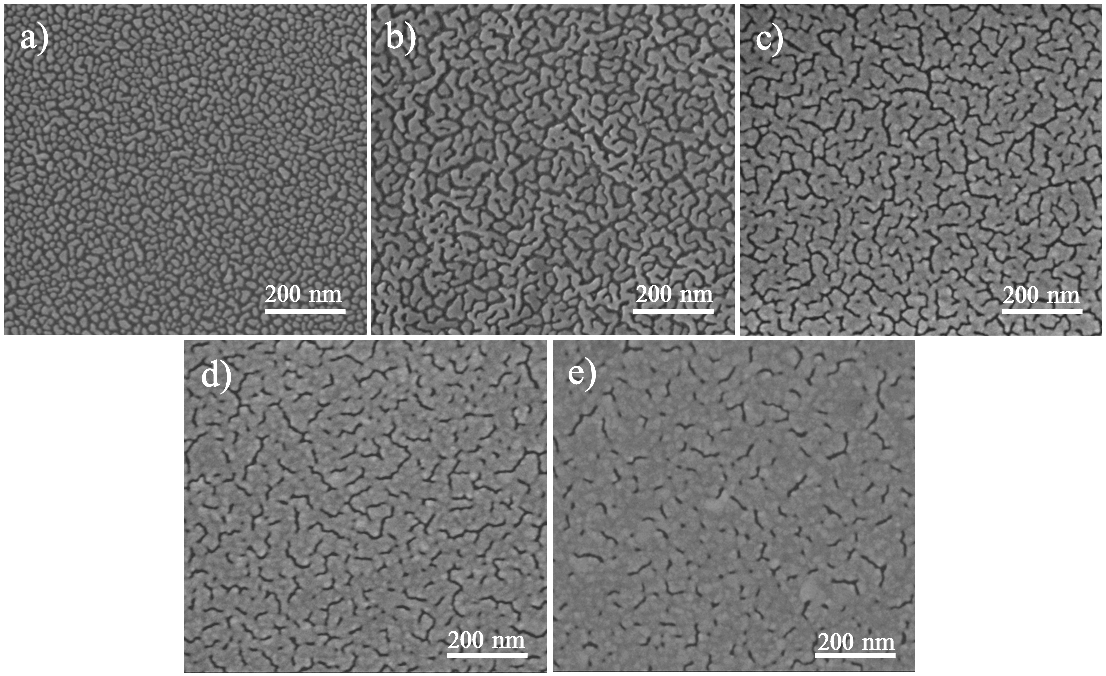}
    \caption{SEM-images of five different percolation gold film morphologies achieved by deposition onto a glass substrate. \textbf{(a)} $3$\,nm, \textbf{(b)} $6$\,nm, \textbf{(c)} $8$\,nm, \textbf{(d)} $10$\,nm, and \textbf{(e)} $12$\,nm.}
    \label{fig:tpl_morph}
\end{figure}

\noindent By varying the deposition time while depositing at a constant rate of $2$\,\AA/s, five gold films on glass substrates were fabricated. Scanning Electron Microscope (SEM) images of the resulting gold morphologies can be seen on Fig.~\ref{fig:tpl_morph}. From these images we see again that the gold follows the characteristic Volmer-Weber growth process, starting out as isolated clusters in Fig.~\ref{fig:tpl_morph}.a, then merging to large intertwined clusters in Fig.~\ref{fig:tpl_morph}.b, before finally fully percolating and becoming a progressively denser film through Fig.~\ref{fig:tpl_morph}.c-e.

\clearpage
\begin{figure}[h]
	\center
	\includegraphics[width=0.8\columnwidth]{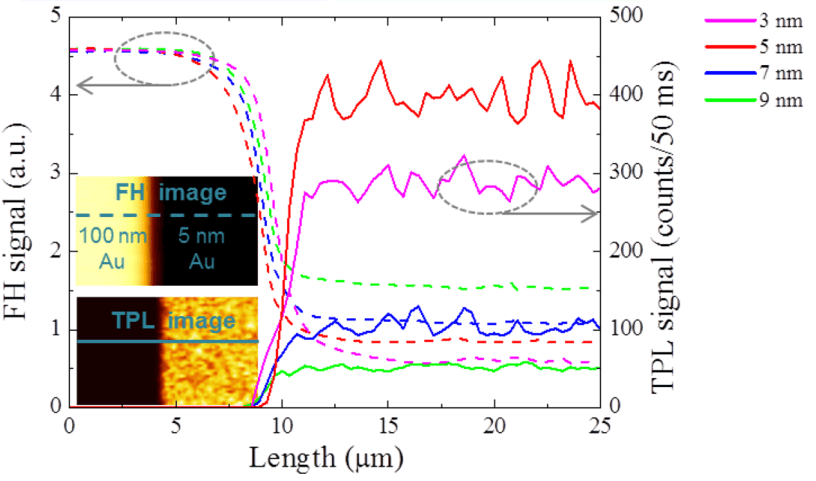}
    \caption{TPL (solid lines) and FH/reflectivity (dashed lines) from $100$\,nm bulk gold and thin gold film samples. One half of the scanned area of the sample was bulk gold, and the other a thin percolating film.}
    \label{fig:tpl_half}
\end{figure}

On Fig.~\ref{fig:tpl_half} we show the difference in TPL-signal between a bulk gold film and four different percolation films. To produce these samples, first a $100$\,nm bulk gold film was deposited on one half of the substrate by using a silicon shadowmask to cover the other half. Next the same sample had a thin gold film deposited across it without the shadowmask. This allows us to record the TPL- and FH-signals from both types of samples in the same focus and measurement. For all deposition thicknesses the semi-continuous gold films have significantly larger TPL-signals than the bulk half of the sample, with the 5\,nm sample having the highest signal. We also see the expected FH-signal from the percolation films, with thicker depositions having higher reflectivity. In order to more thoroughly investigate the TPL-enhancement's dependence on deposition thickness, we fabricated a broad set of samples from $2$-$12$\,nm depositions.

\clearpage
\begin{figure}[h]
	\center
	\includegraphics[width=0.7\columnwidth]{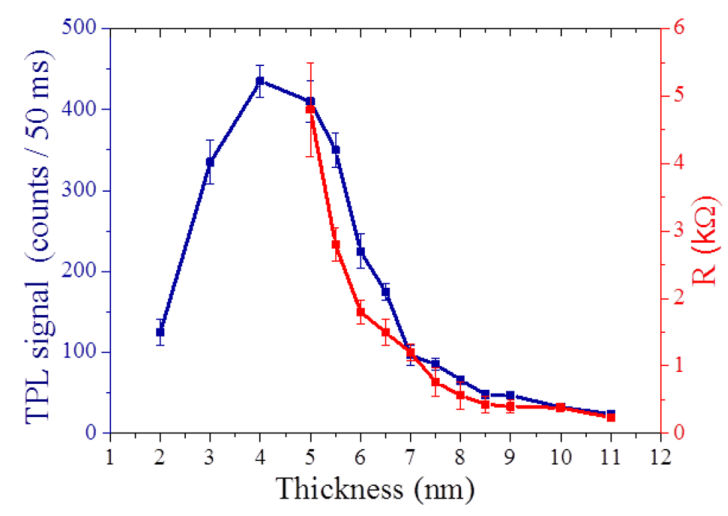}
    \caption{Electrical resistance and TPL-signal from various thicknesses of gold deposition on glass substrates. TPL-signal recorded at constant incident power and a wavelength of $740$\,nm}
    \label{fig:tpl_R}
\end{figure}

\noindent Using a photoresist and defining a simple narrow strip pattern with UV-lithography before metal deposition, it is possible to perform consistent electrical resistance measurements of the different film thicknesses. By restricting the percolation films into same size narrow strips, we can ensure that we are measuring the resistance on a comparable scale, even with a simple 2-point probe measurement. In Fig.~\ref{fig:tpl_R} we show that the percolation threshold for our fabrication is around 5\,nm deposition, as this is the point at which the film strips become conductive. In the same figure we have also plotted the averaged TPL-signal from the same sample thicknesses, showing that the TPL-signal has its largest enhancement around the film's percolation threshold.

We can understand this in terms of the plasmon distribution in percolation films based on their filling fraction. Generally, the largest field enhancements are observed for films close to the percolation threshold, while above and below the percolation threshold the predicted and observed enhancement is lower\cite{Shalaev:2007,Genov:2003b}. We can also understand this fact in terms of the geometry around the percolation threshold, as this point is uniquely defined to be when the system transitions to consist of a large connected cluster. That would automatically infer that just around the percolation threshold, the film will have the highest possible amount of small gaps between the metal clusters per area. Likewise, these gaps between the metal clusters are also likely to be only a few nanometres\cite{Greene:2010}, ensuring the highest degree of field localization in these gap modes.

\clearpage
\begin{figure}[h]
	\center
	\includegraphics[width=0.7\columnwidth]{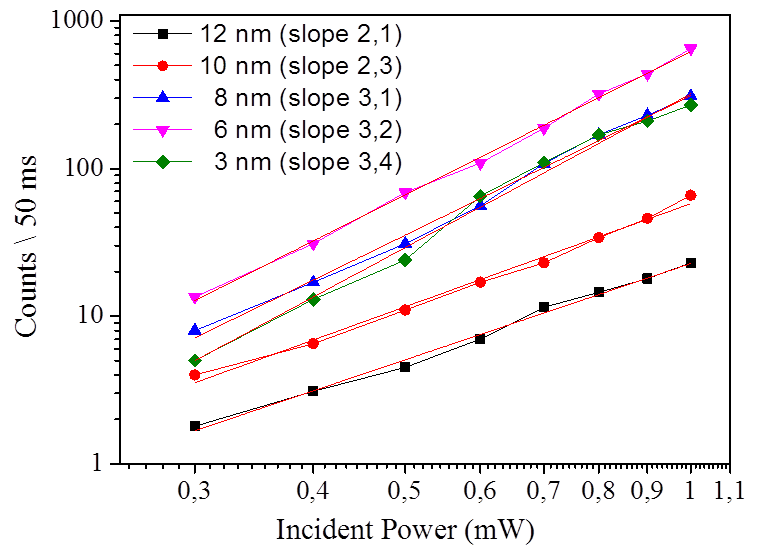}
    \caption{Measured TPL-signal from five different gold morphologies as a function of incident power. Red lines show linear fits to extract the slope. Excitation at 740\,nm.}
    \label{fig:tpl_enhancement}
\end{figure}

\noindent On Fig.~\ref{fig:tpl_enhancement} we see a log-log plot of the averaged TPL-intensities for 5 different films for different laser powers of excitation. For increasing power we would expect a quadratic dependence between the TPL-intensity and the incident power, as the TPL-signal is from a 2-photon process. However, for several of the samples we see slopes that are well above 2. We have attributed this increase in intensity to white light generation in the gold films\cite{Muhlschlegel:2005}. A low-pass filter is used to filter out the FH-signal from the photomultiplier tube collecting the TPL-signal (collecting TPL between 350-550\,nm). We could expect part of the white light signal to 'overlap' the TPL-measurement. From white-light generation in dielectrics, we would expect the white-light signal to have $I_{WL} \propto P_{in}^4$\cite{Muhlschlegel:2005}. This would then explain the larger than expected slopes (see more details in next section).

\begin{figure}[h]
	\center
	\includegraphics[width=0.7\columnwidth]{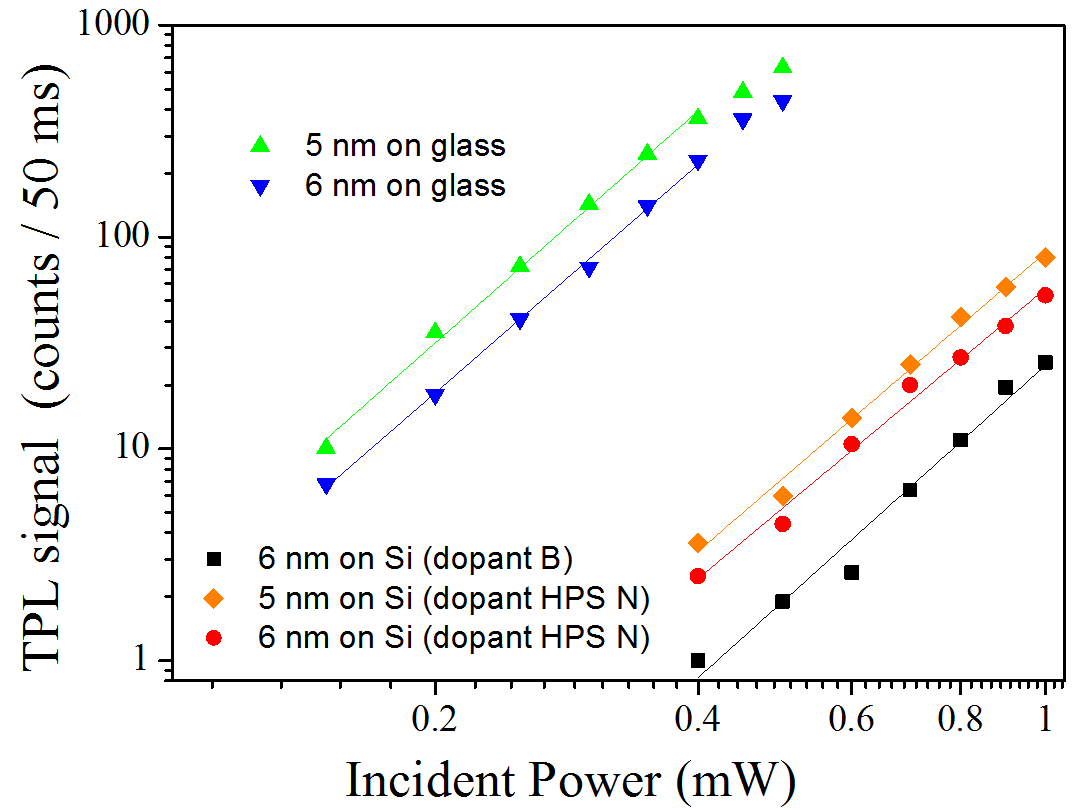}
    \caption{Comparison of TPL-signal from 5 and 6\,nm gold percolation films on glass and silicon substrates. Two types of silicon substrates have been used, boron doped (p-type) and high purity silicon (HPS)/undoped silicon. Excitation at 740\,nm.}
    \label{fig:tpl_si_sio2}
\end{figure}

\noindent Upon comparison between TPL-enhancement from percolation films on silicon and glass substrates we also notice a clear difference. For similar incident powers we record a significantly higher TPL-signal from glass substrates compared to silicon, almost by 2 orders of magnitude, see Fig.~\ref{fig:tpl_si_sio2}. Between the two different silicon substrates (p-doped and undoped) of Fig.~\ref{fig:tpl_si_sio2} we also see that the low conductivity undoped silicon substrate has TPL-signals 2-3 times larger than for the boron doped silicon.

We generally attribute the difference between glass and silicon substrates to the differences in heat conductivity between the two materials. Intrinsic silicon is about 2 orders of magnitude a better heat conductor than glass\cite{Weber:2002}. A thermally insulating substrate would allow the metallic structures where the plasmonic hotspots are hosted to heat up more (less heat carried away by substrate material). We know from the previous section above that phonons are an important aspect of gold photoluminescence. As a result, the temperature of the gold influences the luminescence spectrum\cite{Mooradian:1969}, with a hotter sample producing larger intensities for a broader wavelength range. Another explanation could be charge carrier excitation in the silicon, potentially screening the plasmon resonances in the gold, lowering field enhancement.

\clearpage
\section{White light generation}

\begin{figure}[h]
	\center
	\includegraphics[width=0.8\columnwidth]{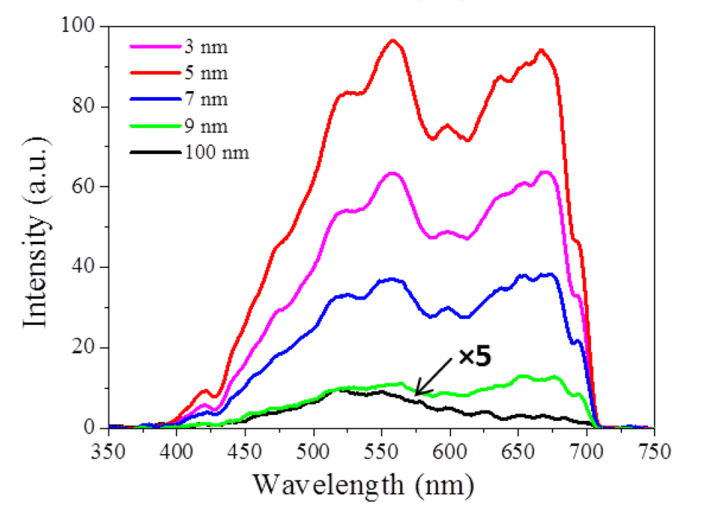}
    \caption{WL spectra recorded from several different gold percolation films on glass substrates. The sharp cut-off above 700\,nm is attributed to a filter in the measurement setup. Excitation at 780\,nm, and the average
incident power was \textasciitilde0.4 mW for thin gold films and \textasciitilde10 mW for the 100\,nm gold sample.}
    \label{fig:white_enh}
\end{figure}

\noindent White light (WL) generation from plasmonic gold nanostructures was first observed in 2005 by P.~M\"uhlschlegel \textit{et al.} \cite{Muhlschlegel:2005}\footnote{Although continuum generation from gold percolation films has been reported as early as 2000\cite{Ducourtieux:2000}}. In their experiment they investigate the optical response from plasmonic gap antennas by pulsed laser illumination. Besides seeing TPL coming from the gold making up the antennas, they also observe WL continuum generation in the gaps of their antennas. Likewise when plotting their signal strength compared to incident power, they find a higher than quadratic power dependence for WL generation. The exact mechanism for the WL generation is still not fully understood, but large field enhancement seems to be a requirement, and the effect is highly non-linear\cite{Haug:2015,Borys:2013,Zheng:2017}.

On Fig.~\ref{fig:white_enh} are several WL spectra from percolation films and a bulk gold sample. The spectra were recorded with a similar set-up as the one shown in Fig.~\ref{fig:tpl_setup}. The set-up has been modified to use a spectrometer as read-out for the 'TPL-signal' and a notch-filter has replaced the low-pass filter, to cut only the FH-signal. Like in reference \cite{Muhlschlegel:2005} we observe a noticeable peak around \textasciitilde550\,nm that could be attributed to the interband edge in gold\cite{Muhlschlegel:2005,Boyd:1986}. We also observe a noticeable peak around \textasciitilde675\,nm, however the notch filter to remove the 780\,nm excitation results in the hard cut-off around 720\,nm. The 5\,nm sample again proves to have the highest signal intensity, which we again attribute to the fact that films near the percolation threshold have the largest degree of field enhancement. Our measurements match the previously reported results from a very similar experiment in reference \cite{Ducourtieux:2000}, although they do not see a significant peak around 550\,nm. When we compare to the bulk (100\,nm) sample, we see most prominently a peak in this 500-550\,nm area, which is consistent with photoluminescence from gold's interband transistions. However, this signal is also significantly weaker than the ones measured from percolation films.

Ducourtieux \textit{et al.} attribute the continuum generation to optical non-linearities that are strongly enhanced from the field enhancement of the plasmon resonances in the gold films. The broad range of possible excitations thus provide a wide optical range of enhancement, leading to WL generation. They also identify the maximal WL generation to be from films at the percolation threshold, which in their fabrication is identified as 5.5\,nm deposition.

\begin{figure}[h]
	\center
	\includegraphics[width=0.75\columnwidth]{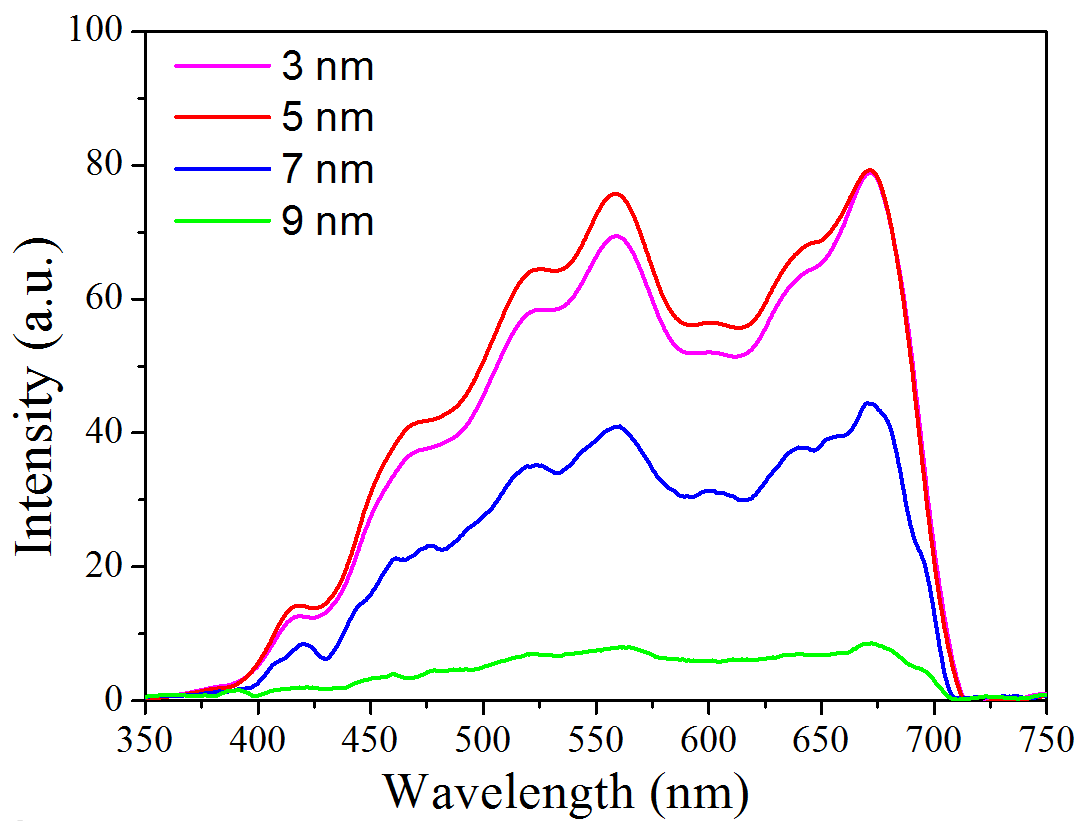}
    \caption{WL spectra from different percolation film thicknesses on intrinsic silicon substrates. Excitation was done at 780\,nm, and the average
incident power was \textasciitilde2.0 mW}
    \label{fig:white_si}
\end{figure}

\noindent Fig.~\ref{fig:white_si} shows WL spectra from percolation films on silicon substrates. The arbitrary units in both Fig.~\ref{fig:white_enh} and \ref{fig:white_si} corresponds to 200\,counts/s, but a much larger incident power of \textasciitilde2.0 mW was needed to achieve similar signal intensities from the percolation films. The spectra show a great deal of similarity to those from glass substrates, but do seem to show higher relative signal around 450\,nm than for the films on glass. The peak around \textasciitilde675\,nm also looks much more narrow.

To summarize our findings: We see broadband WL emission from our gold percolation film samples when illuminated with fs-laser pulses. The maximal amount of emission is seen from films close to the percolation threshold, which is around 5\,nm gold deposition. We also see a significant difference in efficiency of WL generation between gold films on glass or non-conductive silicon substrates, with almost 100 times higher signal from the films on glass substrates. Films on doped silicon substrates show even lower efficiency. We have attributed this difference due to the similar difference in heat conductivity between the two substrate materials, leading to a higher degree of plasmonic heating in the gold structures on glass substrates.

%% file: chapters/chap_snom.tex
\chapter{Near-field observation techniques}
\label{chap:snom}

A very commonly encountered limitation of conventional optical investigations, is that of the diffraction limit. The diffraction limit was first described by E.~Abbe in 1873\cite{Abbe:1873}. In practical terms, it says that a beam of light cannot be focused to a spot with a diameter smaller than half the wavelength of the light. This also means that using a microscope objective, it is not possible to resolve an image with greater resolution than what is set by this limit. For visible light, this results in a resolution limit around \textasciitilde200\,nm, which is very unsuited for any imaging in nanoscience. There are alternative imaging techniques, like electron microscopy (see chapter~\ref{chap:eels}) to get around this\footnote{The de Broglie wavelength of electrons at a similar energy as visible photons is still several orders of magnitude shorter, allowing perfectly for sub-diffraction imaging.}, however for some purposes it is still desired to stay in optics for sub-diffraction imaging/focusing. Plamonics offers a direct way to focus optical fields to spatial dimensions well below the diffraction limit\footnote{Far-field sub-diffraction imaging techniques like confocal microscopy\cite{Minsky:1988} and stimulated emission depletion (STED) microscopy also exists\cite{Hell:1994}.}, as we saw for both surface plasmon polaritons and localized surface plasmons in chapter~\ref{chap:plas}.

However, we are interested in direct imaging of sub-diffraction sized features. This kind of measurement for optical near-fields was first realized by the group of D.~W.~Pohl in 1984\cite{Pohl:1984,Novotny:2007}. Inspired by other scanning probe techniques like scanning tunnelling microscopy (STM), that also have their inception in the early 1980's, Pohl's group utilized a sharp metal coated quartz 'waveguide' with a roughly \textasciitilde20\,nm aperture defined at the tip. By then shining a laser through the small aperture, and scanning it across the sample in a few nanometres proximity, it was possible to record optical images with a resolution of about 10 times higher than that predicted for the diffraction limit\cite{Pohl:1984}. This technique has now come to be known as scanning near-field optical microscopy (SNOM, sometimes NSOM), and has been a popular method to study percolation films historically\cite{Shalaev:2007,Ducourtieux:2000}. Modern devices use a similar method as described above, although replacing the quartz waveguide with a tapered optical fiber, or in the case of scattering-type SNOM (s-SNOM) use a metal coated or uncoated atomic force microscope (AFM) cantilever (more on this later).

We will in this chapter briefly discuss the optical near-field, and in particular how it can be utilized for sub-diffraction imaging with a SNOM. We will also detail some practical aspects of SNOM experiments done here at DTU in the time of my PhD.

\section{The optical near-field}

As discussed above, the diffraction limit puts an upper threshold to the degree with which propagating optical fields can be focused and resolved. It can be shown that the origin of the diffraction limit comes from spatial filtering during propagation of the fields\cite{Sheppard:1982,Enderlein:2000}. If we regard a point-dipole source\footnote{The smallest electromagnetic source we can imagine. Any larger emitter could potentially be regarded as a superposition of such point dipoles.}, it will be characterized by a Dirac delta function, and such a function has an infinite range of frequencies associated with it. Upon propagation the high frequency components will be filtered out, and the imaging optics will also not be able to collect all components. Because of this loss of information, it is not possible to reconstruct the dipole source as a single point, and it will instead have a finite spatial size upon imaging\cite{Novotny:2012}.

\begin{figure}[h]
	\center
	\includegraphics[width=0.9\columnwidth]{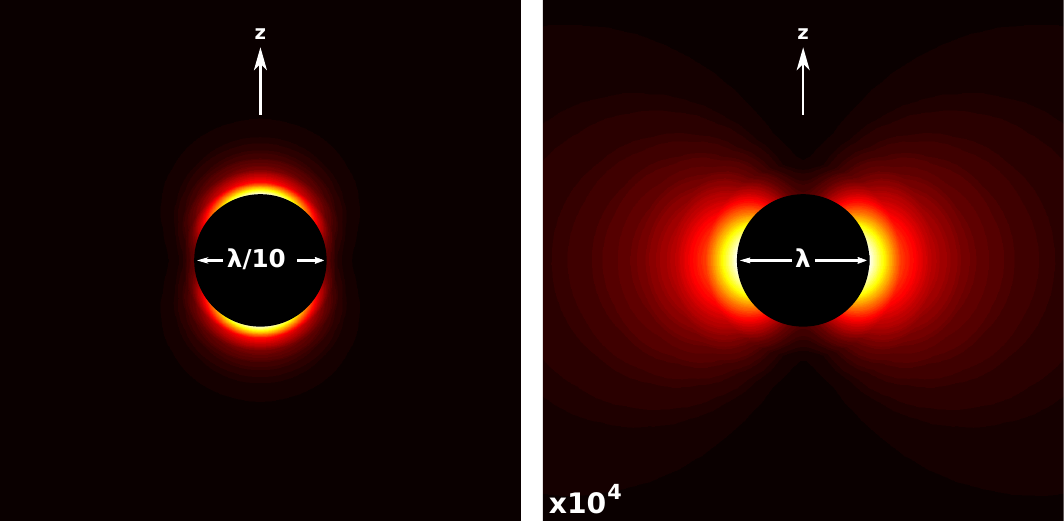}
    \caption{$\left|\mathbf{E}\right|^2$ plotted for a dipole orientated along the $z$-axis in different distances relative to the emitted wavelength\cite{Novotny:2007}. On the left image we see the near-field orientated along the dipole, on the right we see the far-field emitted transverse to the dipole. The colour scale for the far-field image is magnified by $10^4$ relative to the near-field image.}
    \label{fig:dipole_fields}
\end{figure}

\noindent However, this only applies to the propagating part of the field of a dipole. We have so far neglected the near-field of the source. The near-field manifests itself as an evanescent field-component that extends only a small fraction of a wavelength away from the dipole source, see Fig.~\ref{fig:dipole_fields}. We saw previously in chapter~\ref{chap:plas}, total internal reflection in a prism can produce evanescent fields. Likewise, an evanescent field can also be coupled into a prism or waveguide to result in a propagating mode (often called photon tunnelling\cite{Novotny:2012}). The intensity of this propagating field can then be collected with an ordinary photodetector. In other words, by bringing a small apertured waveguide probe close to a dipole-like source of radiation on a sample (or from other sources of evanescent fields, like plasmons), it is possible to measure the near-field intensity. Scanning this probe across the sample can then produce an image of the sample's near-field intensity with a spatial resolution limited to the aperture size of the probe (typically few tens of nanometres). Likewise, it is also possible to couple radiation into this probe and generate an evanescent field with which to excite a sample locally in the near-field.

These are the generalized operating principles of a SNOM. In the next section we will discuss the practical implementation of a SNOM in an optical set-up, as done here at DTU with the Danish Micro Engineering (DME) DS 45-40 SNOM.

\section{Scanning near-field optical microscopy}

In this section we will detail the operating principles of the DME DS 45-40 SNOM used during my PhD-studies (many of these principles will apply in general for similar SNOM systems). However, because of limitations of the system and lack of external technical support (the manufacturing company, Danish Micro Engineering (DME), underwent bankruptcy proceedings during my thesis studies), it has only been possible to acquire partial results from this system in the time limit of my studies.

\subsection{Fiber optic probes and shear-force feedback}

\begin{figure}[h]
	\center
	\includegraphics[width=0.45\columnwidth]{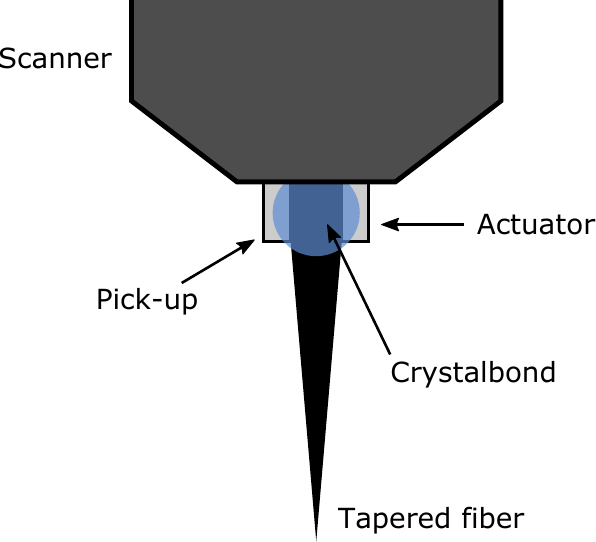}
    \caption{Overview of the mounting of the fiber optic probe.}
    \label{fig:snom_probe}
\end{figure}

\noindent We will start our discussion with the optical probes themselves, and the shear-force feedback system of the piezoelectric scanner used to track the probe/sample distance. A schematic of the system can be seen on Fig.~\ref{fig:snom_probe}.

\clearpage
The probes used have all been single mode optical fibers pulled to sharp tapers with an apex of \textasciitilde150\,nm. The fibers have not had any metal cladding or aperture defined, and were manufactured and purchased from the company LovaLite. The probes are mounted to the system by manually glueing\footnote{We use crystalbond 509 as a simple hot-melt glue. The hard resin-like texture of the solid glue seems ideal for the mechanical resonator purposes here.} them in-between two piezoelectric elements, which are then mounted on the rest of the 3D piezo scanner used for scanning the sample surface. One of the piezos in the mounting will act as an actuator to drive a vibration of the fibertip, while the second piezo will act as a pick-up to record the amplitude and frequency of the vibration. The fiber tip will oscillate like a stiff beam that is clamped in one end. By scanning across different actuation frequencies, it is possible to find the frequency for which the fibertip has its mechanical resonance frequency. Typical resonance frequencies for the mounted fibers are in the tens of kHz, with resonantor quality factors of 120-160 achievable with some care during the mounting process.

By driving the fiber at its resonance frequency, it is possible to detect damping of its oscillation with the pick-up piezo. By bringing the tip into close proximity of a sample surface (few to tens of nanometres), the tip of the fiber will experience van der Waals forces from the sample's surface\cite{Novotny:2007}. The result of this is a shear-force in the fiber that dampens its vibration. The degree of dampening depends on the distance between the probe and the sample surface. This kind of shear-force feedback is thus an ideal way to track the probe/surface distance.

\subsection{The DME DS 45-40 SNOM}

\begin{figure}[h]
	\center
	\includegraphics[width=0.95\columnwidth]{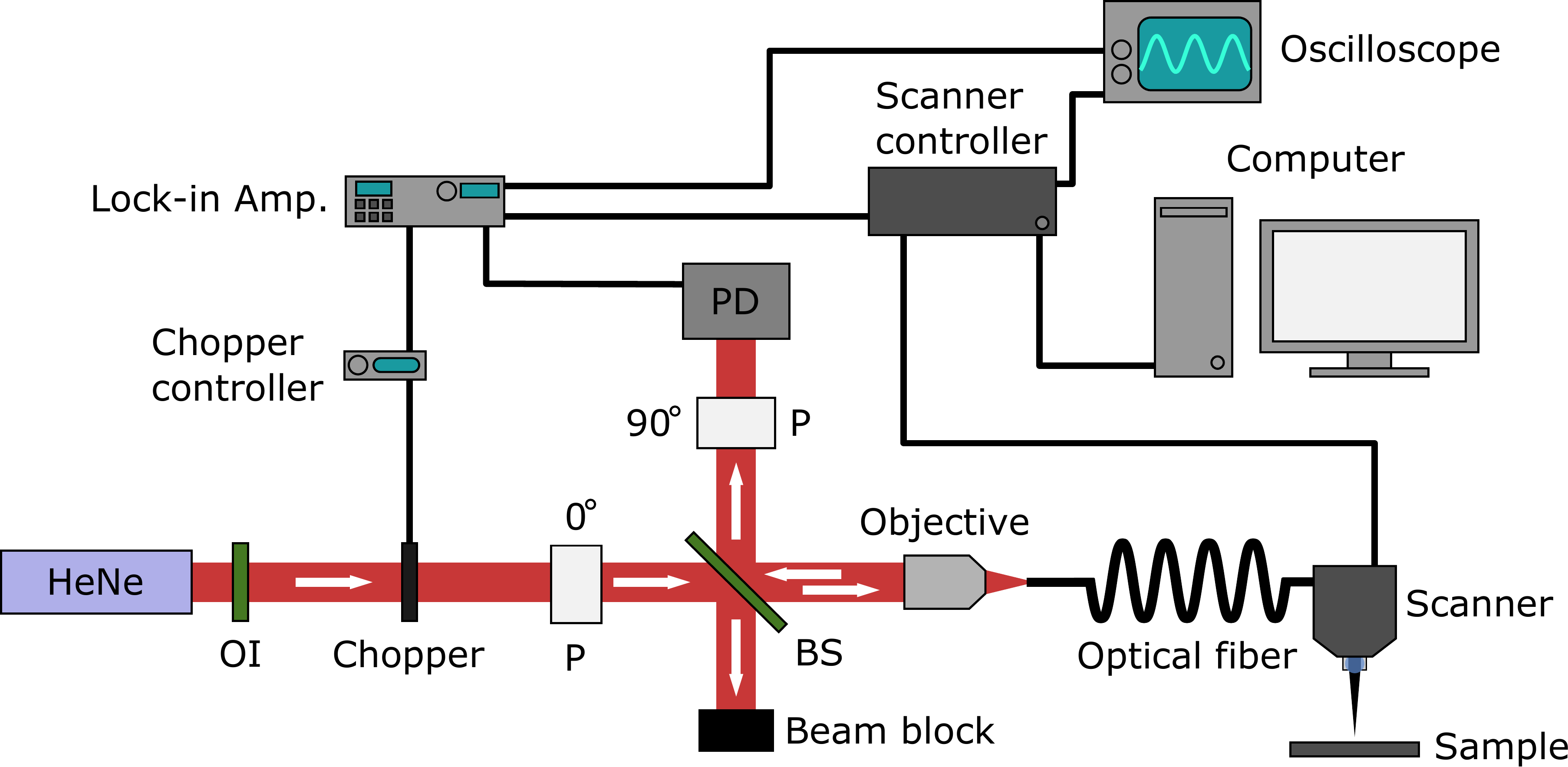}
    \caption{Schematic drawing of the optical set-up used for the DME DS 45-40 SNOM at DTU.}
    \label{fig:snom}
\end{figure}

On Fig.~\ref{fig:snom} is a schematic drawing of the optical set-up and supporting electronics for the SNOM experiments at DTU. The SNOM is run in a combined illumination and collection scheme. The light source is a 5.0\,mW helium neon laser (wavelength \textasciitilde633\,nm). The laser is first run through an optical isolator to cut any back-reflections into the laser cavity\footnote{The tapered end of the optical fiber dramatically limits the transmission of the fiber, resulting in a strong reflection back into the optical set-up.}. In order to utilize a lock-in amplifier for the collecting photodetector, we have to modulate the laser with an optical chopper. After the chopper, the laser is passed through a Glan-Thompson polarizer aligned with the polarization of the optical isolator. Passing through a 50/50 beamsplitter, the laser is now coupled into the flat end of the tapered optical fiber using a 3D stage and a microscope objective (20$\times$, $\text{NA}=0.5$). Some portion of the laser light will be transmitted from the fiber and excite the sample (the rest is reflected back into the set-up). The same fiber will also collect a signal from the sample, which is passed back and collected with the same microscope objective. The signal from the fiber will then be passed back to the beamsplitter, where it will be partially reflected. The signal is then passed through another Glan-Thompson polarizer that is rotated 90$^\circ$ in relation to the first. The purpose of this cross-polarization is to cut as much of the background signal from the rest of the set-up as possible\footnote{Any light that is reflected from the fiber will follow the same beam path as the signal we are interested in, but will be polarized to the first polarizer, thus cut in the cross polarization.}. Finally, the signal is collected by a photodetecter, which passes its voltage signal to the lock-in amplifier.

The lock-in amplifier is supplied with the chopper's modulation frequency from the chopper controller, and is able to amplify the signal from the detector by several orders of magnitude. The amplified signal is then passed on to the scanner controller, where it is correlated with the fibertip's position on the sample. The shear-force of the fiber is recorded simultaneously as the optical signal, and thus it is possible to record a near-field and shear-force topographic image at the same time.

An oscilloscope is also connected to both the lock-in amplifier and the scanner controller. From the signal of the lock-in it is possible to monitor that the chopper is accurately modulating the laser signal, and from the scanner controller it is possible to observe the shear-force feedback that is adjusting the height of the fiber tip during a scan.

Finally, the sample stage consists of a manual 2D translation stage. A simple 10$\times$ periscope mounted microscope is positioned perpendicular to the near-field probe. A camera is attached to the microscope and connected to the computer, allowing for real-space imaging of the probe's position on the sample, both before and while scanning.

In addition to the 'reflection' geometry shown in Fig.~\ref{fig:snom}, it is also possible to operate the system in 'transmission'. Here the laser source is instead shone through the sample from below, and the fibertip is scanned across the sample for collection of the near-field signal.

\subsection{Measurements}

\begin{figure}[h]
	\center
	\includegraphics[width=0.95\columnwidth]{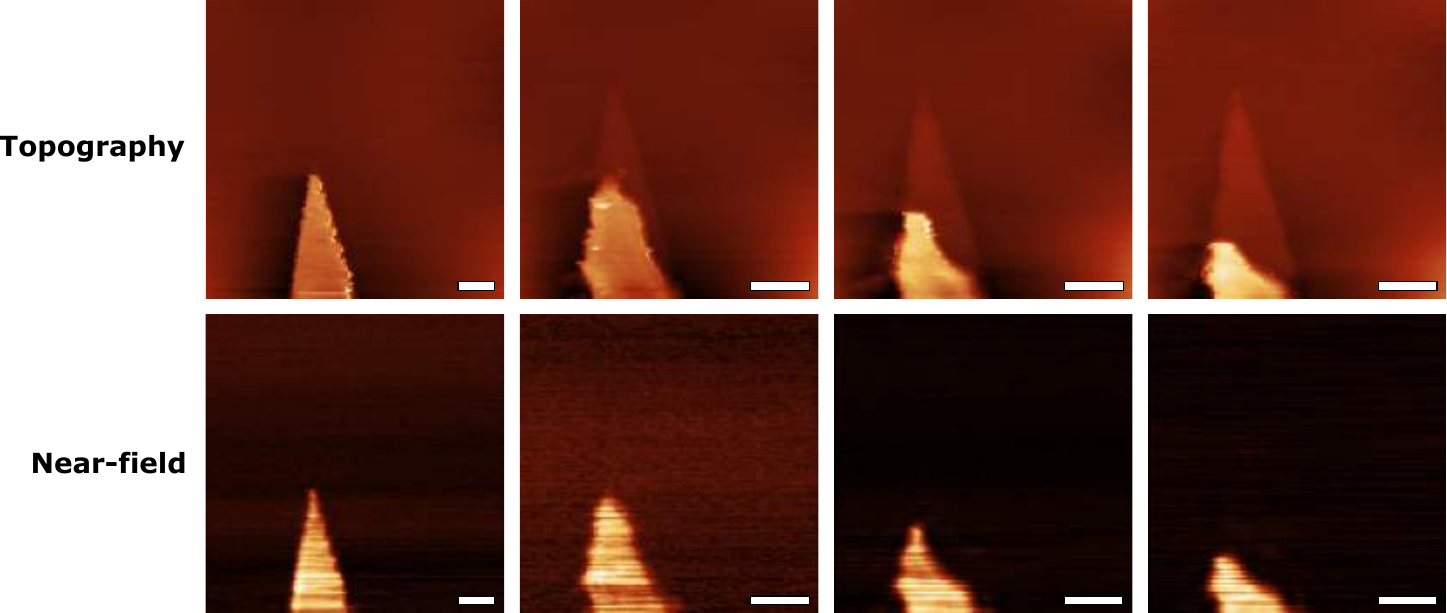}
    \caption{Consecutive scans of the same 40\,nm thick gold taper on SiO$_2$ substrate. The top row of images shows the recorded topography from the shear-force feedback, the bottom row the recorded near-field intensity. Scale bars are all 5\,$\upmu$m.}
    \label{fig:snom_peel}
\end{figure}

\noindent A constant limitation of the SNOM setup has been degradation of soft metal samples during scanning, likely caused by too large sample/probe interaction. Examples of recorded images of a gold taper on glass substrate can be seen on Fig.~\ref{fig:snom_peel}. With each consecutive scan we see how more and more of the gold taper is scratched off from the substrate.

\clearpage
\begin{figure}[h]
	\center
	\includegraphics[width=0.5\columnwidth]{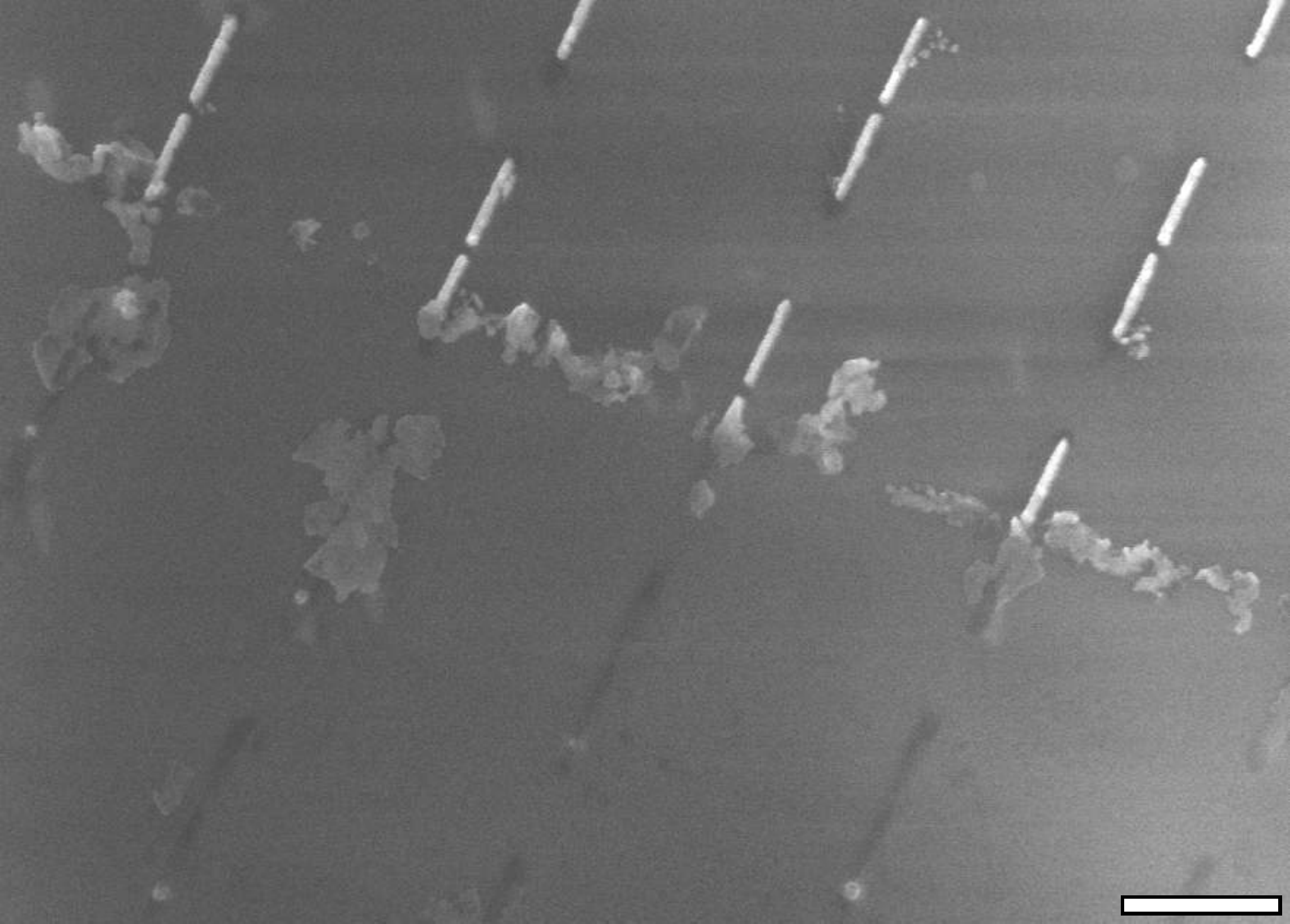}
    \caption{Edge of a region scanned with the SNOM system of gap nanoantenna on silicon substrate. The bottom half of the image shows the scanned region, where the antenna have been removed completely during scanning. The antenna are 40\,nm thick gold deposited on top 10\,nm titanium. Scale bar is 500\,nm.}
    \label{fig:snom_peel3}
\end{figure}

\noindent Fig.~\ref{fig:snom_peel3} shows a SEM-image of an array of nanoantenna that have been scanned with the SNOM. The scanned part of the antenna array very clearly shows damage to the nanostructures, with some having been removed even completely from the substrate.

\begin{figure}[h]
	\center
	\includegraphics[width=0.4\columnwidth]{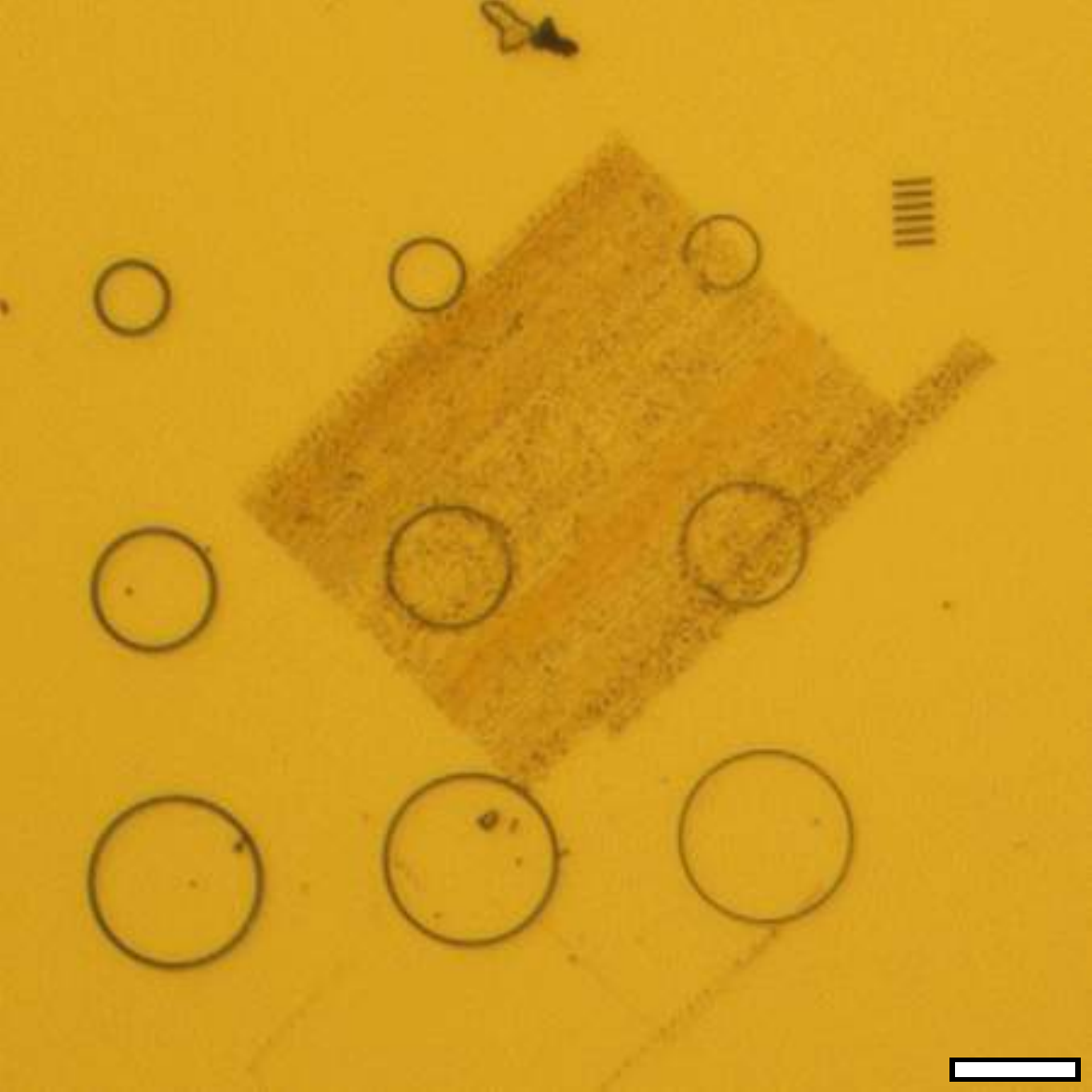}
    \caption{150\,nm gold film on glass substrate with circular trenches milled by FIB. A large square of high surface roughness can be seen left-over from a region scanned by the SNOM. Scale bar is 10\,$\upmu$m.}
    \label{fig:snom_peel2}
\end{figure}

\noindent The problem also extends to even thick gold films. Fig.~\ref{fig:snom_peel2} shows an optical microscope image of a 150\,nm thick gold film deposited on a glass substrate. Circular rings have been defined in the film using focused ion-beam (FIB) milling intended for plasmonic lenses\cite{Lerman:2009}. However, when attempting to scan the gold film, surface roughness can be seen to increase for each consecutive scan. When observing the regions scanned with the SNOM in a microscope, very clear and defined scratch marks can be observed.

\begin{figure}[h]
	\center
	\includegraphics[width=0.95\columnwidth]{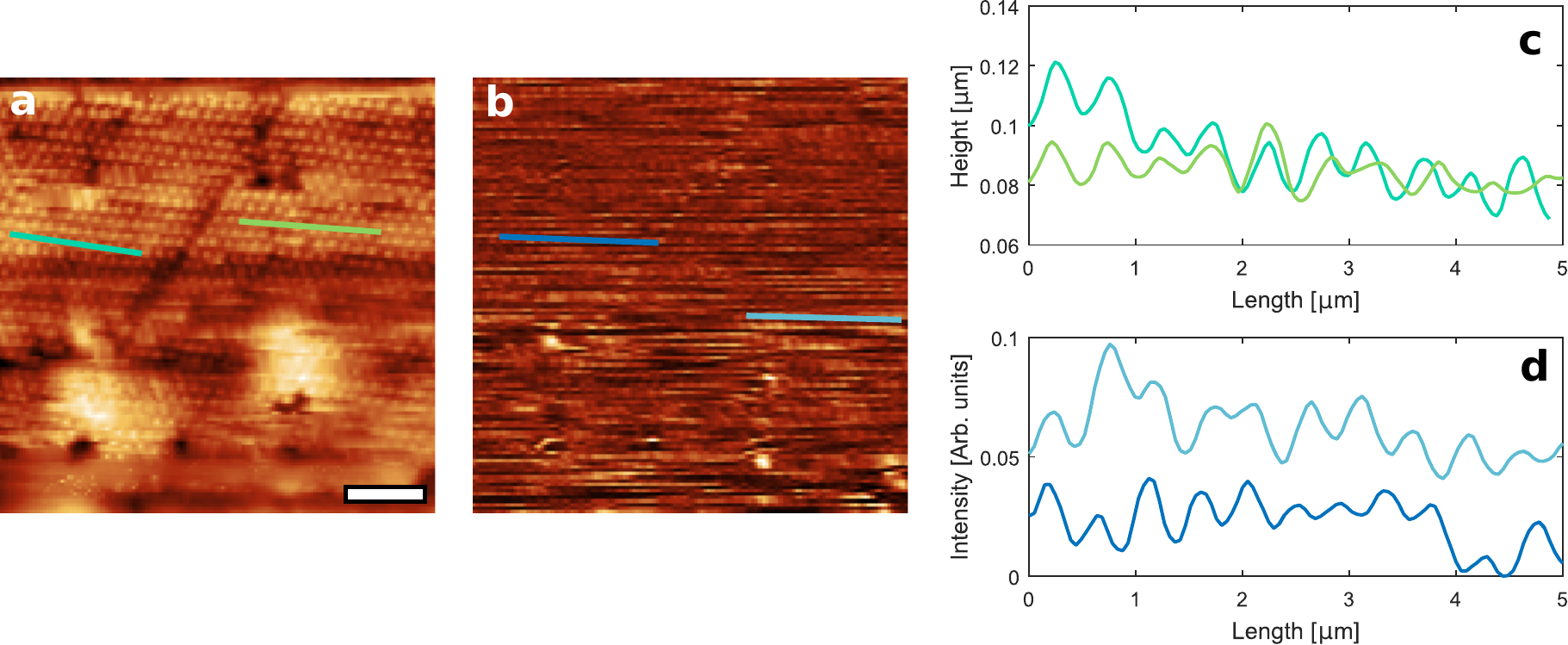}
    \caption{Profiles of latex spheres on SiO$_2$. \textbf{(a)} shows a topographical image recorded with the SNOM set-up, while \textbf{(b)} shows the simultaneously recorded optical signal. \textbf{(c)} shows profiles extracted from (a), and \textbf{(d)} shows profiles from (b). Scale bar for (a) and (b) is 3$\upmu$m.}
    \label{fig:snom_latex}
\end{figure}

Despite the issues, it has been possible to extract some information from the generally very noisy optical images. Fig.~\ref{fig:snom_latex} shows a scan over 500\,nm diameter latex spheres dispersed on a glass slide, where they have self-assembled into a lattice of closely packed spheres\cite{Fujimura:2000}. From both the topography and optical images, it is possible to discern the widths of the spheres from a profile extraction.

\subsection{Scattering-type SNOM}

So far all of our discussion has involved so-called apertured SNOMs, i.e. the near-field probes used have a clearly defined aperture that the evanescent fields are transmitted through. However, a second class of SNOM devices exists, where instead a sharp tip is brought close to alter the local optical properties of the sample. This tip can be metal coated to couple plasmonically to the sample, or uncoated to simply alter the local optical properties. By illuminating the sample and collecting a signal of scattered light in the far-field, it is possible to see how the same probe changes the sample's properties locally. Raster scanning the probe across the sample to record this in different positions can make an image. We refer to this kind of SNOM as aperture-less SNOM, or scattering-type SNOM (s-SNOM). 

\clearpage
\begin{figure}[h]
	\center
	\includegraphics[width=0.4\columnwidth]{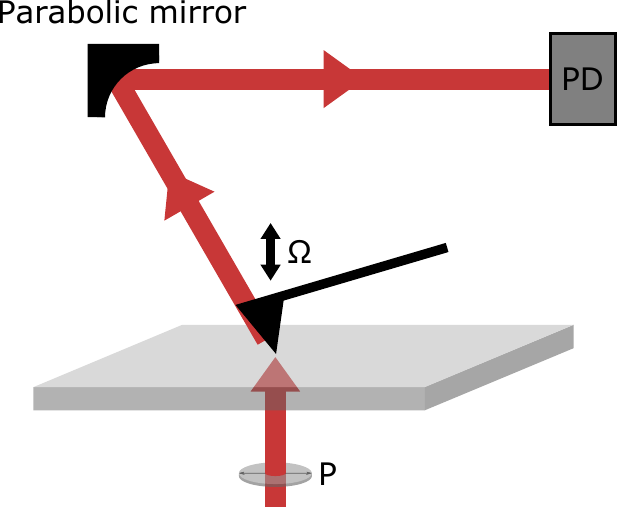}
    \caption{General beampath for a s-SNOM operated in transmission configuration. A polarized light source from below is shone at the sample, while the AFM cantilever is scanned across the sample. A parabolic mirror is used to collect and direct light scattered from the sample by the AFM tip toward a photodetector, where the intensity is measured.}
    \label{fig:ssnom}
\end{figure}

An example of the beampath for a s-SNOM in a transmission configuration can be see on Fig.~\ref{fig:ssnom}. Typically s-SNOM setups will have an additional parallel beampath for so-called pseudoheterodyne interferometry (not pictured in Fig.~\ref{fig:ssnom}). The modulated signal from the AFM cantilever's oscillation frequency, $\Omega$, is compared to a reflection from a mirror oscillating at a frequency $M$. By then demodulating the combined signal it is possible to measure the phase as well as the amplitude of the near-field signal\cite{Deutsch:2008}.

\begin{figure}[h]
	\center
	\includegraphics[width=\columnwidth]{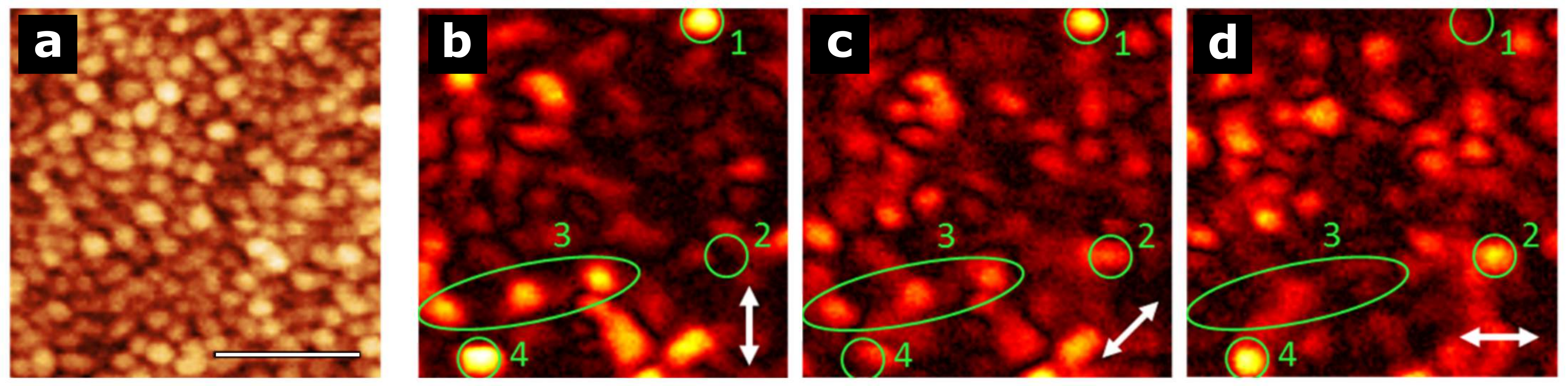}
    \caption{Topography, \textbf{(a)}, and near-field images, \textbf{(b-d)}, of a 5\,nm gold percolation film on glass substrate. All images were recorded from the same region of the sample. Excitation was done in the far-field with transmission from below of 1500\,nm light. The illumination source was set to three different polarizations. The polarization of the excitation source is marked with the white double arrows in (b-d). The scale bar is 200\,nm}
    \label{fig:snom_pol}
\end{figure}

\noindent Fig.~\ref{fig:snom_pol} shows images recorded of a 5\,nm gold film on glass substrate, recorded with a NeaSpec system in Prof. S.~I.~Bozhevolnyi's group by V.~A.~Zenin. The s-SNOM is operated in 'transmission' mode where the sample is illuminated from below with a linearly polarized 1500\,nm laser. The sample is then scanned consecutively in the same region, for three different polarizations of the illuminating laser. In Fig.~\ref{fig:snom_pol}.b-d, four regions have been highlighted. In these four regions we see distinct hotspots of near-field intensity for the three different polarizations. We also see that there is almost no overlap of the physical position of hotspots between images Fig.~\ref{fig:snom_pol}.b and d, where there is a 90$^\circ$ difference in excitation polarization. We can conclude from this that the plasmonic hotspots present in the films are highly polarized, and that the hotspots' spatial distribution for different excitation polarizations are also fairly unique (as also shown previously for general random clusters\cite{Stockman:1996,Stockman:1997}). We will show how this fact can be exploited for anisotropic optical modification in gold percolation films in chapter~\ref{chap:recon}.

%% file: chapters/chap_eels.tex
\chapter{Electron Energy-loss spectroscopy}
\label{chap:eels}

In chapter~\ref{chap:plas} we discussed how plasmons are able to couple and interact with external light. We also briefly touched on the fact that plasmons were initially understood experimentally, from exciting bulk plasmons in thin metal films with an electron beam in a transmission electron microscope (TEM)\cite{Watanabe:1956,Lang:1948,Ruthemann:1948}. The excitation of the plasmons were observed as energy losses in the electrons as they were passed through the samples. By using a magnetic field as a dispersive element\footnote{The strength of the Lorentz force from the magnetic field that deflects the electrons depends on their velocity, the magnitude of which is dependent on their kinetic energy.} for the electron beam after it has passed the sample, it is possible to observe the energy losses of the electrons over a wide energy spectrum. We refer to this method as electron energy-loss spectroscopy (EELS). EELS was first developed in 1944 by J.~Hillier and R.~F.~Baker\cite{Hillier:1944}. By filtering the electrons used to image a sample by energy, they were able to observe the high-energy core loss edges (hundreds to thousands of eV) in the probing electron beam. These edges are the result of the electron beam interacting with tightly bound inner shell electrons of the sample's lattice atoms. This is now a common method in EELS to make detailed maps of a sample's elemental composition, as the edges are unique identifiers for different elements\cite{Muller:2009}.

For plasmons, we are however interested in low energy losses (few eV). We will here briefly discuss how an EELS measurement is performed in modern microscopes, and what data processing procedures need to be done to extract the plasmon energies and their EELS amplitudes from the spectra.

\section{Transmission electron microscopy}

The field of an electron beam can be shown by explicit calculation with Maxwell's equations to be confined within close distance to the beam\cite{Abajo:2010}. This means that the electron beam can be regarded as a source of evanescent electric fields. As we saw in chapter~\ref{chap:plas} and \ref{chap:snom}, such fields can be used to great effect to excite plasmons. Furthermore, the beam's field can also be expressed in individual frequency (energy) components in terms of its Fourier transform. We can thus think of the electron beam as a supercontinuum evanescent light source\cite{Abajo:2010}, which is an ideal experimental probe for plasmon excitation. Optical experiments are typically restricted to one or few wavelengths of light, but in EELS we can probe multiple different energies, in one measurement, from the polychromatic source.

\clearpage
\begin{figure}[h]
	\center
	\includegraphics[width=0.6\columnwidth]{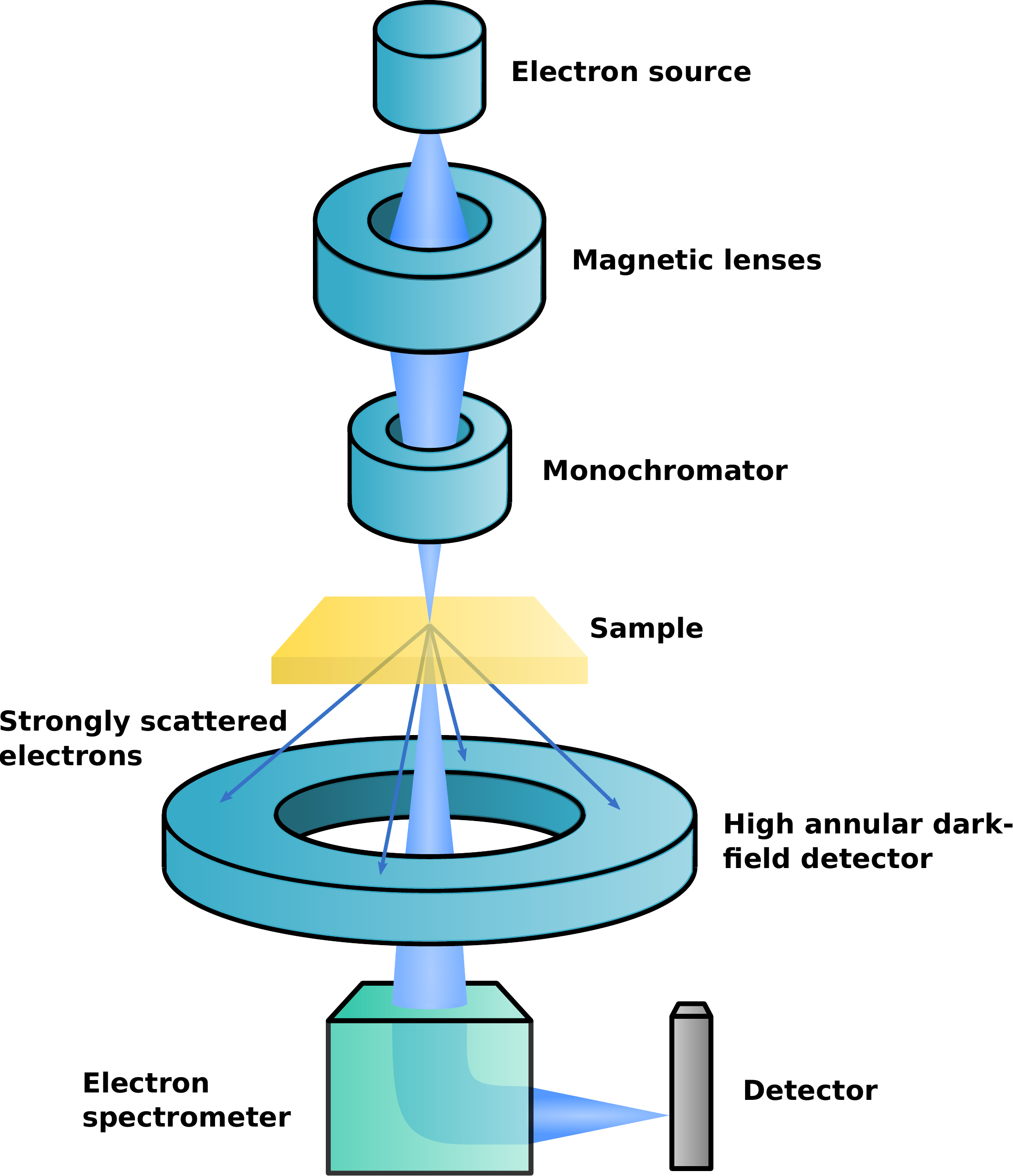}
    \caption{Schematic of the STEM EELS acquisition process. The electron beam is focused to a point in the sample and raster scanned across it. The electrons scattered at a high angle are collected by the ADF detector, while the electrons only weakly scattered are passed through and analysed in a spectrometer.}
    \label{fig:stem_eels}
\end{figure}

\noindent All EELS results presented in this thesis were acquired on a FEI Titan TEM with a monochromator and a probe aberration corrector, hosted at DTU's Center for Electron Nanoscopy (CEN). On Fig.~\ref{fig:stem_eels} is a simple schematic overview of how an EEL spectrum is recorded in a scanning transmission electron microscope (STEM). Electrons are emitted from a source (typically a field emission gun\cite{Egerton:2011}), and are then collected and focused onto the sample into a sub-nanometre spot through a series of magnetic lenses\cite{Egerton:2008}. The typical energy of the electrons used for such experiments is hundreds of keV, and to ensure enough signal electrons pass through the sample, its thickness should normally not exceed 100\,nm. In our case, the microscope is also fitted with a monochromator above the sample, to ensure as narrow a spread in electron energy as possible\cite{Keast:2007}. When passing through the sample the electrons can interact with it in different ways, although the majority of the electrons will not interact at all. These 'non-interacting' electrons make up a large peak in the final spectrum, that we will refer to as the zero-loss peak (ZLP). The width of the ZLP will also show the energy distribution of the electron beam, and it is from its full-width-half-maximum we can see the energy resolution\cite{Keast:2007,Keast:2012}.

Some of the electrons interacting with the sample will be scattered at very high angle. These electrons are collected by an annular dark-field detector (ADF) that sits like a ring below the sample. From this detector it is possible to record images of the sample's morphology in greyscale, where the contrast comes from the electron scattering rate. Heavier elements and thicker regions of the sample will appear brighter (more scattering), and different crystal orientations in a polycrystalline sample will appear with different contrasts as well, due to their differences in scattering rate\cite{Jesson:1995}.

Those electrons interacting only weakly with the sample (or not at all) will pass through to the electron spectrometer. Here by using a magnetic field they are dispersed onto an electron detector, their deflection angle and subsequent position on the detector revealing their kinetic energy.

\subsection{The energy-loss spectrum}
\begin{figure}[h]
	\center
	\includegraphics[width=0.85\columnwidth]{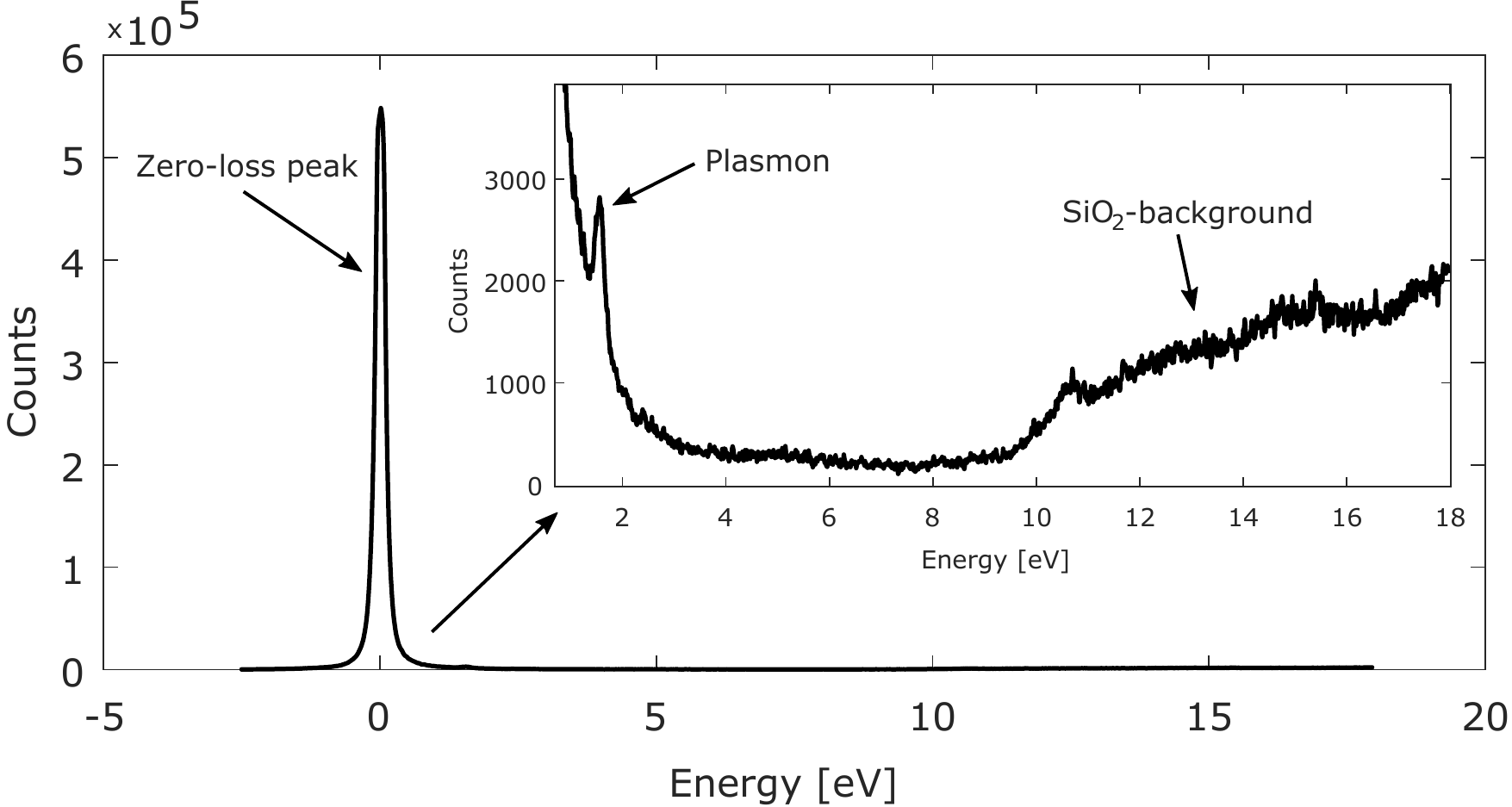}
    \caption{An example of an EEL spectrum taken from a gold percolation film on a 18\,nm SiO$_2$ membrane. The ZLP is clearly seen as the dominant spectral feature. An insert shows a magnified view of the rest of the spectrum, highlighting a plasmon peak around 1.5\,eV which is still partially inside the ZLP. The broad background signature of the SiO$_2$ membrane material is also seen from around 10\,eV.}
    \label{fig:zlp_spec}
\end{figure}
\noindent Fig.~\ref{fig:zlp_spec} shows an example a of a recorded EEL spectrum from near a gold cluster in a gold percolation film on a thin glass substrate (this type of sample is discussed in greater detail in chapter~\ref{chap:recon}). The ZLP is prominently visible in the spectrum, and in the magnified inset we can see how the ZLP overlaps to some degree with a peak that resembles a plasmon resonance. From around 10\,eV we also see a broad signal coming from the membrane structure acting as substrate for our sample. We will later use an integral of this to help normalize our data, as the intensity of this signal will be consistent across the sample. As we are interested in the plasmonic properties of our sample, and since we see a great deal of the resonances can overlap with the ZLP, we will have to perform a background subtraction on our data.

For all EELS results presented in this thesis, we have opted to fit a power-law to a subset of the data points that are to the left of the first plasmon resonance in a spectrum. The calculated fit is then subtracted from the entirety of the spectrum, and all data points below 0.5\,eV are discarded, as the background correction is not accurate for this range (anything from 0.0-0.5\,eV is also solidly inside the ZLP and difficult to interpret anyway).
\begin{figure}[h]
	\center
	\includegraphics[width=0.75\columnwidth]{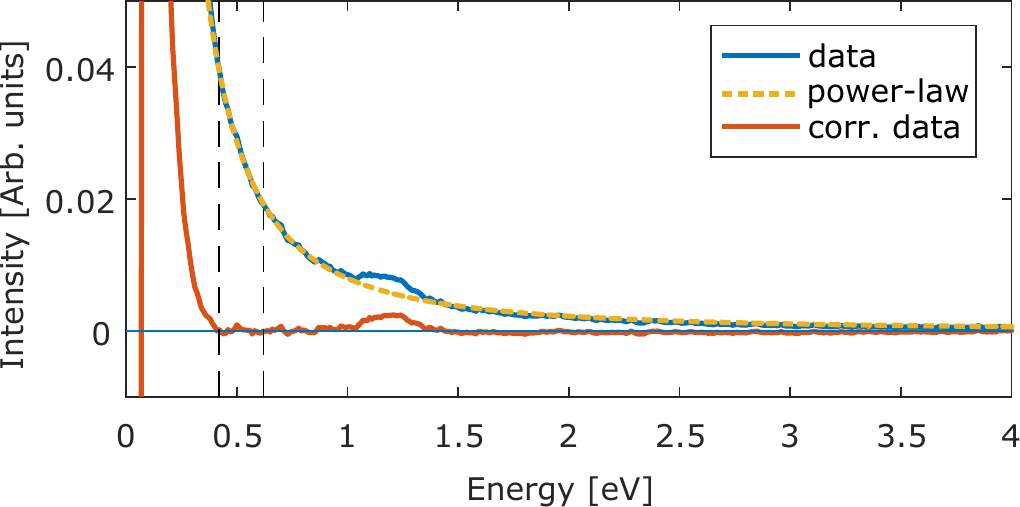}
    \caption{An example of how a small subset of the EELS data (marked in-between the dashed vertical lines) can be used to calculate a power-law fit for subtracting the ZLP.}
    \label{fig:zlp_remov}
\end{figure}
\clearpage

\section{Mapping plasmons with electrons}

A natural extension of being able to record EEL spectra from an electron beam passed through a sample, is to record such spectra in a 2D array from the sample. We refer to this kind of array as a hyperspectral image (sometimes also referred to as a data cube, 
as the EELS intensities are in $x$, $y$, and energy coordinates).

\begin{figure}[h]
	\center
	\includegraphics[width=0.75\columnwidth]{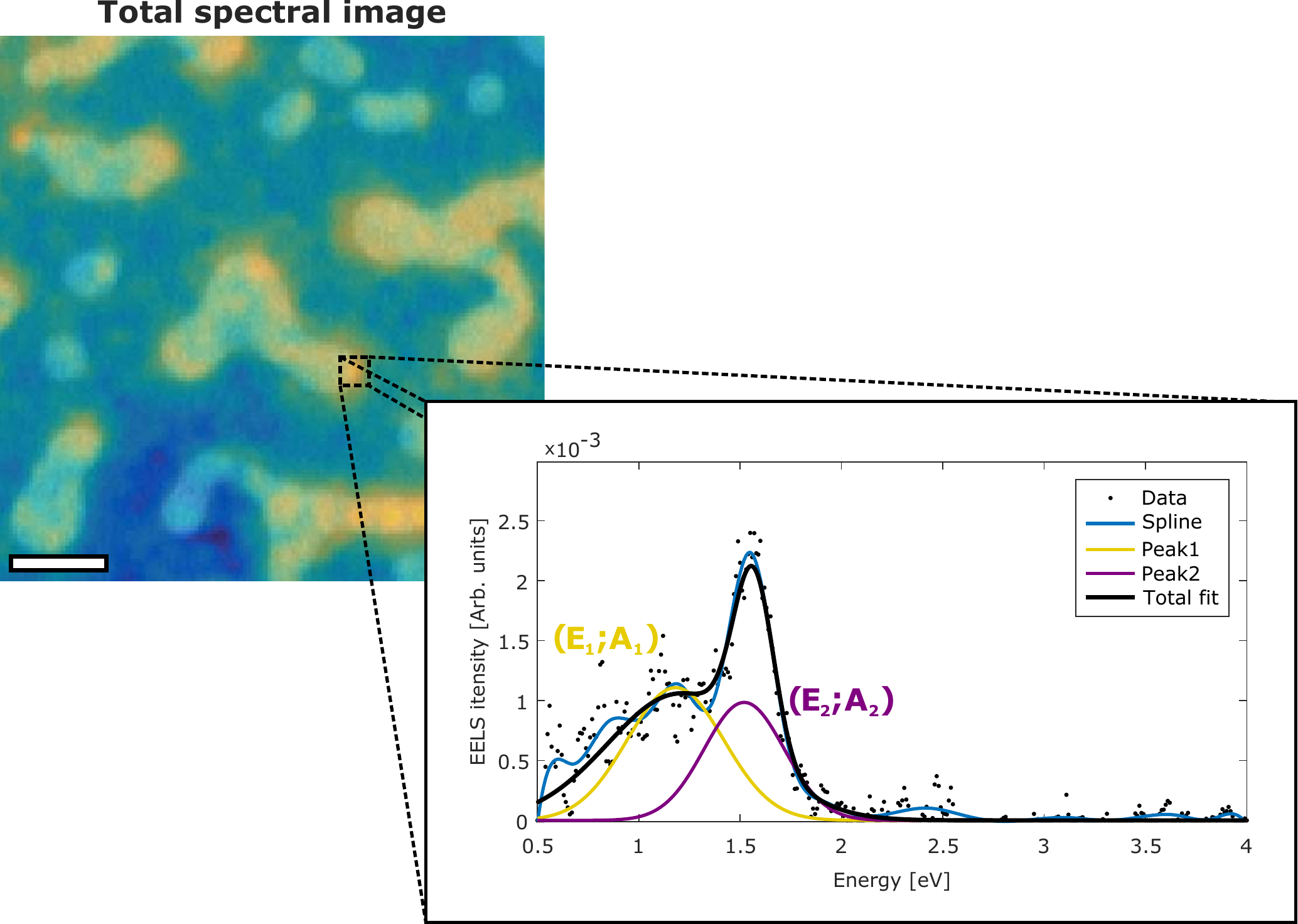}
    \caption{Example of a spectral image of a laser modified 5\,nm gold film on a thin glass substrate. An EEL spectrum from a point on one of the particles in the image has been highlighted, showing how successive Gaussian fitting can be used to identify plasmon peaks and extract their amplitudes and resonance energies. The scale bar for the spectral image is 75\,nm.}
    \label{fig:stat_meth}
\end{figure}

\noindent On Fig.~\ref{fig:stat_meth} we see an example of a summed EELS map overlaid on top of the sample morphology it was recorded from. Each spectrum in the image has had the ZLP subtracted as described above, and been normalized by the integral from 10-14\,eV of the SiO$_2$ background. An example EEL spectrum from a point near a gold particle has been highlighted. We can utilize the EELS data of a hyperspectral image in two main ways:
\begin{itemize}
	\item Integrate the EELS signal along the energy axis for a certain energy range to make a 2D colour intensity map that spatially maps the plasmons in this energy range.
    \item Extract the central position and height of the plasmon peaks in the spectra to categorize the resonance energies and intensities in the image.
\end{itemize}
\noindent In this thesis we will be utilizing both approaches with hyperspectral images of several different gold percolation film morphologies. A very big challenge however for extracting the plasmon resonance energies and amplitudes from percolation films, is the sheer volume of data from a typical spectral image. In our case, a typical image consists of \textasciitilde10,000 spectra. Each of those spectra can then contain vastly different amounts of plasmon resonances, at widely different energies (some even overlapping considerably). As a result, we have had to develop a very computationally intensive automatic fitting routine. In a sort of 'first sorting', for a given spectral image we first average all spectra together and plot the result. In this summed set of spectra a set of Gaussian functions are fitted by hand to the resulting 'primary modes' of the system - those modes either so intense or frequent in the individual spectra that they end up dominating the average of all spectra.

\begin{figure}[h]
	\center
	\includegraphics[width=0.95\columnwidth]{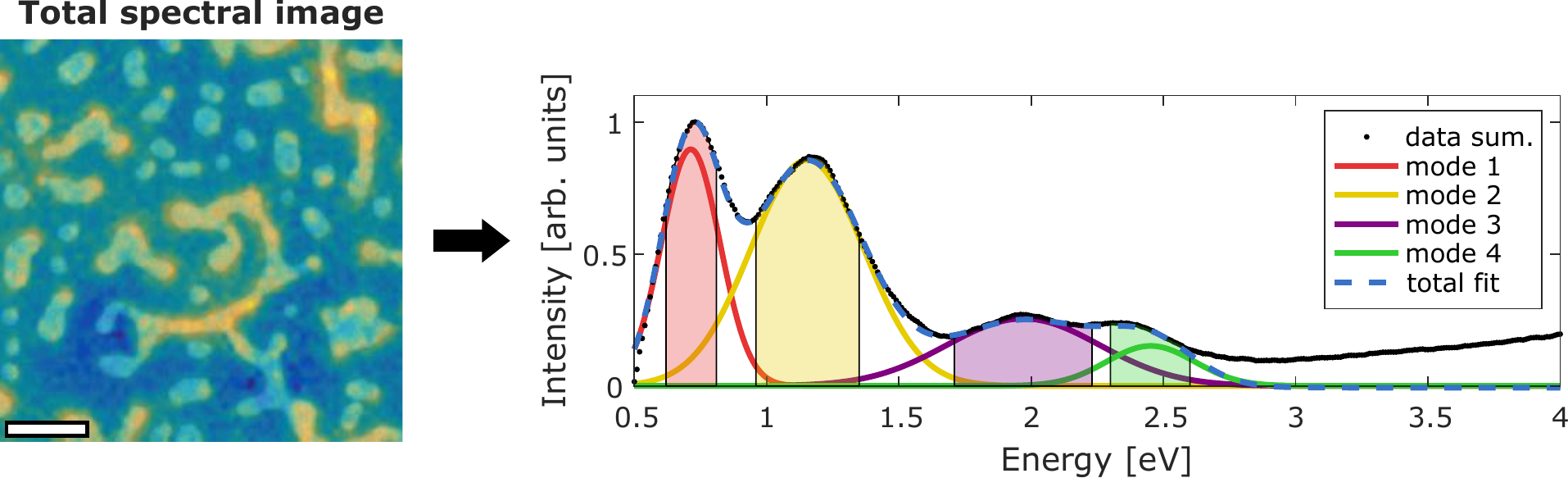}
    \caption{Summed spectra from a spectral image, showing four primary modes present in the image. The scale bar for the spectral image is 150\,nm.}
    \label{fig:prim_modes}
\end{figure}

\noindent An example of such a summed image can be seen on Fig,~\ref{fig:prim_modes}. The data are from the same spectral image as the excerpt in Fig.~\ref{fig:stat_meth}. A fit of the sum of four Gaussian functions is also highlighted, with the individual function terms also plotted. From the four Gaussian functions we can designate regions of interest within a portion of their widths (in this case within 50\% and above of their amplitude, i.e. full-width-half-maximum). For each of the individual spectra that make up the spectral image, we now make a smoothing spline (to reduce the random oscillations from noise in the data), and we look primarily for peaks in the spectra within the energy ranges defined by the widths from the primary modes. After noting the amount of peaks identified, a series of independent Gaussian fits are carried out for each peak in only the data from the vicinity of the peak. The quality of this fit (in this vicinity) is then evaluated from the residual, and if it is found to be too poor, it is ignored for the further data treatment. 

\clearpage
After double-checking the validity of the independent Gaussian fits, the width, amplitude, and central position of each of the remaining fits are then used as initial guesses for fitting of a sum of the same number of Gaussian functions across the full spectrum data. From this total fit we can now extract the amplitude and resonance energy of all of the identified (and verified) plasmon resonances in the spectrum. An example of this series of fits is shown in Fig.~\ref{fig:stat_meth}. This procedure is then repeated for each spectrum of the image, and the collected data can later be described statistically.

This method is of course not perfect, and will not catch all the different varieties of plasmon resonances present in the image. We justify the use of the primary modes to initially 'target' our search for plasmons, by the fact that our images are of 500$\times$500\,nm$^2$. This size fits quite well with the diffraction limited spot in an optical experiment. So for any optical experiments we could compare our EELS data with, this kind of averaging of the modes in the image would already have occurred in the optical experiment as well. In other words, an optical experiment would also be most sensitive to the primary modes of the system, as everything else would be averaged out by the large spotsize.

%% file: chapters/chap_recon.tex
\chapter{Optical modification of semi-continuous gold films}
\label{chap:recon}

A general energy loss process for plasmons in metals is due to Ohmic resistive losses. Basically, the electron scattering processes as we understand them in the Drude picture (see chapter~\ref{chap:opt}) is the transfer of energy from the electrons to the material's crystal lattice by phonons. This transfer of energy to the lattice is also what we traditionally understand as heating of a material. The more general case of Ohmic heating is that of an electrical resistor with a large current passed through it. Due to resistive losses in the resistor, energy from the electron current will be deposited as heat, causing it to increase its temperature. Because of gold's relatively low melting point ($1060$\,$^\circ$C), it is possible to melt gold locally from the Ohmic heating caused by the decay of strong plasmon excitations. In this chapter we will discuss some of the general processes for this reshaping, and also how it is possible to alter the general properties of gold percolation films by such local reshaping from plasmon excitation. The exact mechanisms at play and parameters needed to attain certain reshaping modes has been beyond the scope of the work in this thesis, but we will refer to results of prior literature on the topic to frame our own work.

\section{Photothermal processes in metal nanoparticles}

\begin{figure}[h]
	\center
	\includegraphics[width=0.7\columnwidth]{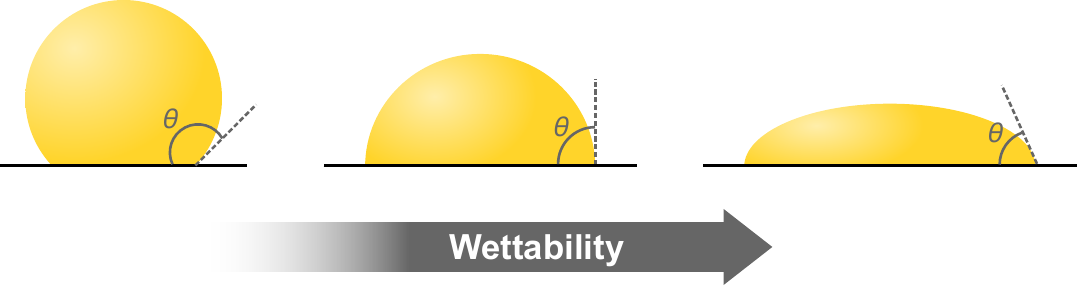}
    \caption{Three different substrates with increasing levels of wettability for a liquid drop on top of the substrate. The wetting angle $\theta$ is indicated for each case.}
    \label{fig:wet}
\end{figure}

\noindent There are several ways in which gold can be reshaped by plasmonic heating, but the most pedagogical case is probably that of a gold nanorod melting into a sphere. For now we simply deal with the case of a metal nanorod being heated, and neglect the complications of the heat energy being delivered by plasmon decay\footnote{Several processes happen at different timescales when a metallic particle is plasmonically excited\cite{Brongersma:2015,Hu:2012,Link:1999,Link:2000}.}. When the metal rod reaches a temperature matching its melting point, which may well be lower than the melting point of the bulk metal\footnote{Due to the high relative surface energy associated with nanoparticles, as compared to bulk materials, their melting points can often be considerably lower\cite{Jiang:2003,Nayebi:2009,Lee:2009}.}, the now liquid metal will try to contract into a sphere because of surface tension forces\cite{Link:2000}. A very general trend observed for liquids (and plasmas) is that they will contract to spherical shapes to limit their surface energy (the atoms at the surface have higher energy than the bulk atoms because of the lack of neighbour atoms to bond with). We can understand why a spherical shape is preferred from the simple fact that a sphere is the 3D Euclidean shape with the least amount of surface area per volume\cite{Link:2000}. However, when dealing with liquids on a substrate things become a little more complex, as the spherical droplet will be deformed based on the surface's wettability\cite{Moore:1959,Sangiorgi:1988,Kojima:2008}. For a surface with high wettability (strong liquid-surface interaction) the liquid droplet will be flattened and spread out across the substrate, while for a low wettability substrate (low liquid-surface interaction) it will remain almost spherical. Fig.~\ref{fig:wet} shows examples of different wetting angles. For the specific case of water we refer to surfaces with low wettability as hydrophobic, and high wetability as hydrophilic. We will discuss the influence of the substrate material when melting metal particles in more detail in a later section.

As stated above, we would expect a general shortening and 'rounding' of isolated particles when they are melted. We can also consider the case of a coupled system of plasmonic nanoparticles (see chapter~\ref{chap:plas}). Here the situation is a little more complicated. In general two processes can occur: Separation, or merging/welding of the particles into one. For weakly coupled particles with a relatively large separation between them, we would expect the case of them decoupling. However, for the case of two very closely situated particles they can actually exchange material and start to melt together with the right excitation\cite{Hu:2012,Garnett:2012,Mertens:2016}. To make matters more complicated, a kind of intermediate process can also occur, where two particles are joined by a small 'neck' extending between them, instead of fully merging. This kind of 'necking' has been studied for joining together self-assembled nanosphere chains\cite{Herrmann:2014,Hu:2012,Huang:2012}.

\begin{figure}[h]
	\center
	\includegraphics[width=0.95\columnwidth]{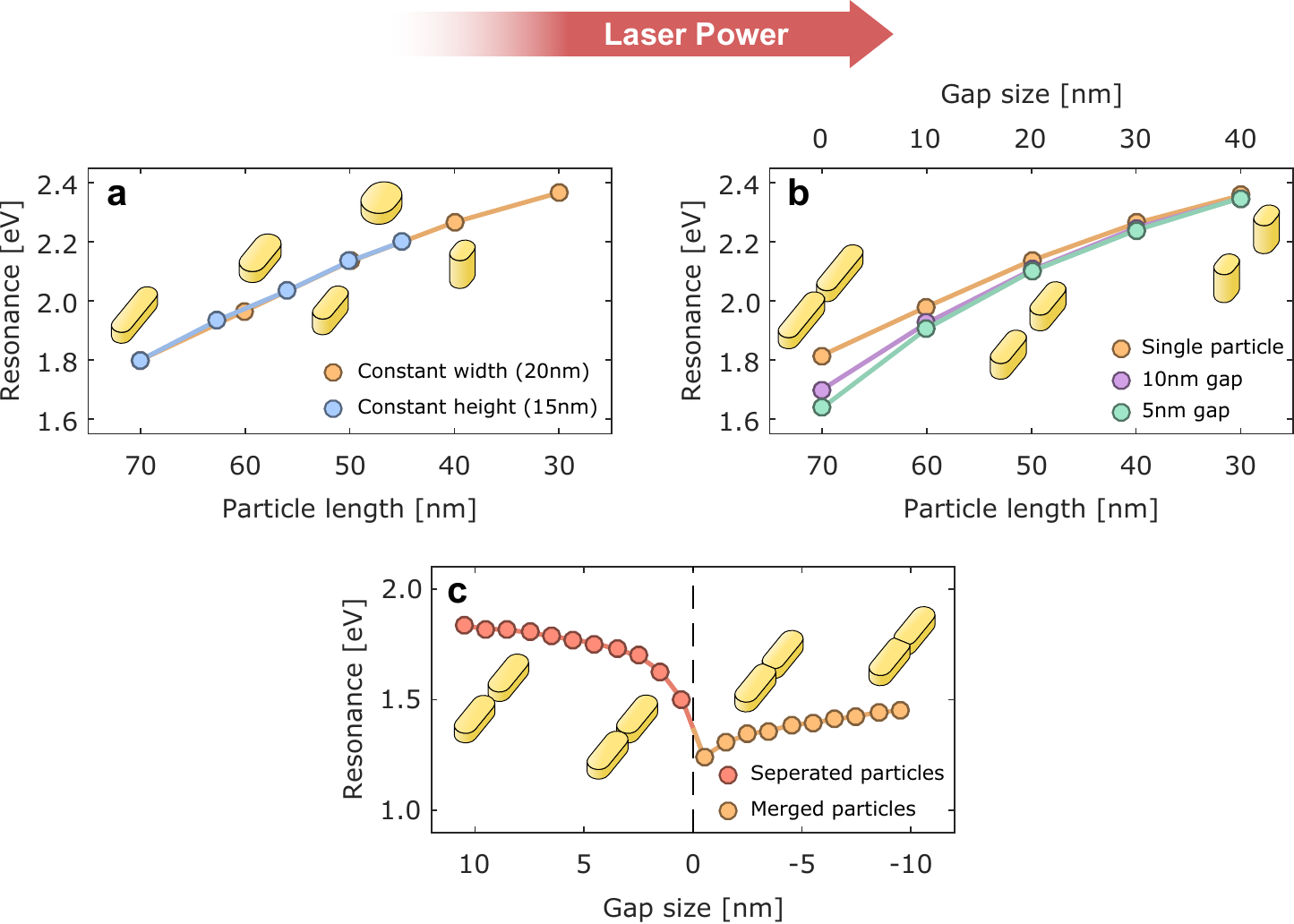}
    \caption{Simulations of the longitudinal plasmon resonance for gold rods being reshaped as if melted. In all plots reading from left to right corresponds to an increase in deposited power into the system. \textbf{(a)} The case of a single gold nanorod being contracted, while its volume is preserved. To compensate for the shortening, volume is either added to the height or the width of the particle. \textbf{(b)} The case of two coupled gold rods being shortened, so that they eventually decouple. Lost volume from contraction added back as height. \textbf{(c)} Two gold nanorods moved towards each other, to emulate merging. Lost volume in the particles' overlapping region is added to the height of the connected particle.}
    \label{fig:melt}
\end{figure}

On Fig.~\ref{fig:melt} we have highlighted how the longitudinal plasmon resonance for elongated particles are influenced by their reshaping, as if being partially melted/con\-tracted towards more spherical shapes, while preserving the volume of the particles. The curves presented here are based on simulations done in collaboration with T.~Rep\"an.

\clearpage

\noindent On Fig.~\ref{fig:melt}.a is the initial case of an elongated particle being contracted, and to maintain its volume it is either made taller or wider. We see the expected outcome that its plasmon resonance blueshifts as it contracts in length. We also see that the blueshift is only affected by the change in dimension parallel to the plasmon excitation, i.e. both the widening and heightening of the particle corresponds to similar curves. In Fig.~\ref{fig:melt}.b we study the case for a set of coupled particles. By shortening both of the two particles we effectively shorten the gap between them, which as expected for the bonding mode of a plasmonic dimer causes a blueshift in resonance energy. We have also plotted the resonance of a single particle of the same shape, to see at which point the particle coupling is broken and they effectively become two separated particles. Finally, in Fig.~\ref{fig:melt}.c we study the case of two particles merging. Here, instead of shortening the particles, their centre coordinates are moved towards each other. As this is initially somewhat the opposite case as in Fig.~\ref{fig:melt}.b, we expect to see a redshift before the two particles merge. As soon as the two particles merge, we return somewhat to the case of Fig.~\ref{fig:melt}.a, as we have one new very long particle. As the two particles overlap more and more, any lost volume from the overlap is added to the joint particle's height. Effectively, the continued merging of the particles acts as a shortening of the joint particle, and we return to the case of Fig.~\ref{fig:melt}.a with a blueshift.

Another photothermal process worth briefly mentioning is that of laser ablation, i.e. directly evaporating or sublimating material by transfer of large amounts of energy from high laser intensity. This technique is however typically not reliant on plasmonic excitation (although it can be), opting instead for simply delivering the energy by high intensity photon absorption. Laser ablation has found uses for the synthesis/fabrication of nanoparticles by ablating material into an aqueous solution\cite{Amendola:2006,Amendola:2009}. Because the process mostly relies on intensity of the laser itself, it is also compatible with a wide variety of materials\cite{Amendola:2006,Amendola:2009}. As the processes studied in this thesis are not believed to involve ablation, we will not discuss it in further detail here.

\subsection{Substrate influence}

In our own experiments of melting and reshaping gold percolation films with a fs-laser (the same laser as in the TPL set-up discussed in chapter~\ref{chap:tpl}), we have noticed a significant difference in the morphology changes between percolation films on glass and silicon substrates.

\begin{figure}[h]
	\center
	\includegraphics[width=0.65\columnwidth]{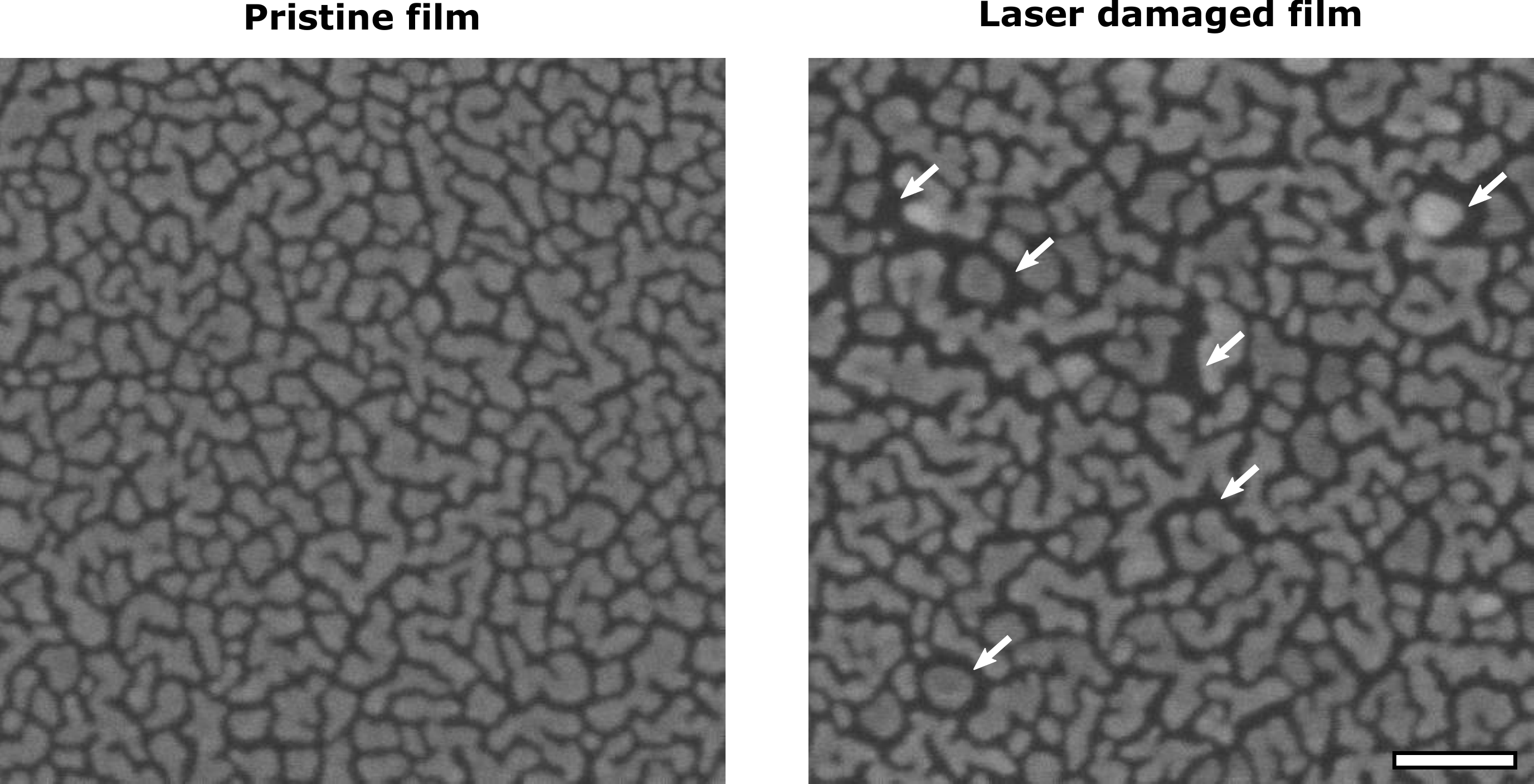}
    \caption{SEM-image of how the fs-laser affects a 5\,nm gold percolation film on an intrinsic silicon substrate (undoped). An average laser power of \textasciitilde3\,mW was used to damage the film, at a wavelength of 740\,nm. Arrows added to highlight regions of large morphology change. The scale bar is 100\,nm.}
    \label{fig:silicon_dam}
\end{figure}

\clearpage
\noindent On Fig.~\ref{fig:silicon_dam} are examples of a 5\,nm gold deposition on an intrinsic silicon substrate. One part of the film has been left as deposited, and another part has been illuminated with a 740\,nm \textasciitilde3\,mW fs-laser pulse to induce morphology changes. The overall changes in the film's morphology are quite subtle, but manifests themselves in two ways. First, several areas of the laser illuminated film shows abnormally wide gaps between neighbouring clusters when compared with the pristine film. Secondly, several rounded or spherical looking particles show up in the damaged image. These particles are also always situated near larger gaps to neighbour clusters. Both of these things would indicate that a gold cluster has been melted and turned into a gold droplet from surface tension. Lastly, some regions of the film appear virtually unchanged, indicating that the injection of heat into the system happens very locally, and it also does not propagate a great deal from its injection point. The local introduction of heat is very consistent with the idea that the heating happens from plasmon Ohmic losses, as heat can only be introduced in the specific plasmonic hotspots corresponding to the laser wavelength used.

\begin{figure}[h]
	\center
	\includegraphics[width=0.65\columnwidth]{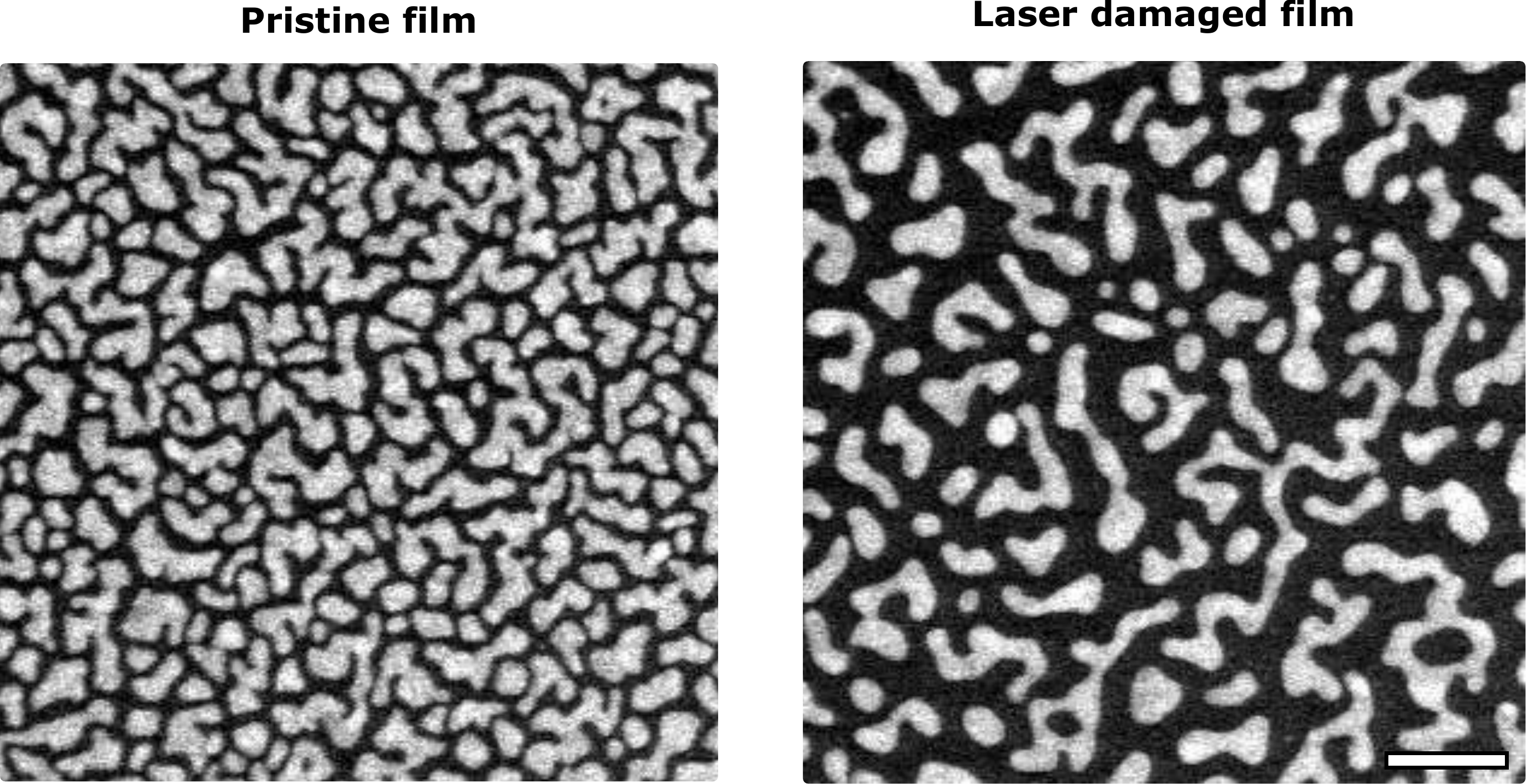}
    \caption{STEM dark-field image of laser damage done in a 5\,nm gold percolation film on 18\,nm glass membrane substrate. A laser power of \textasciitilde0.5\,mW was used for the damage at 740\,nm wavelength. The scale bar is 100\,nm.}
    \label{fig:SiO2_dam}
\end{figure}

\noindent If we now consider the case of a 5\,nm percolation film on a 18\,nm thin SiO$_2$-substrate as shown on Fig.~\ref{fig:SiO2_dam}, we see that the morphology changes from the laser are significantly different. Despite having been illuminated at a much lower average laser power of \textasciitilde0.5\,mW compared to the \textasciitilde3\,mW of Fig.~\ref{fig:silicon_dam}, the morphology for the glass substrate seems to have been affected much more. Not only do the changes on individual clusters seem more significant (more rounded and spherized), the changes also seem to have affected every cluster in the region, and is no longer isolated like for the silicon case.

Gold has a similar wetting angle of about 135$^\circ$ on both polished glass and silicon\cite{Moore:1959,Sangiorgi:1988,Kojima:2008}. However, despite having similar specific heat capacities, the heat conductivity of silicon is almost 100 times higher than that of glass\cite{Weber:2002}. We attribute the massive differences in morphology changes to be due this discrepancy between the two substrate materials. Because the glass substrate is incapable of acting like a proper heat sink to divert the heat away from the gold structures where injected, the heat instead dissipates into the substrate plane. The result is that the effects of the laser are much more equally felt by all of the percolation film.

This analysis is of course not conclusive as a direct comparison between glass and silicon, as the glass samples are also placed on thin 18\,nm membranes. As a result, the mass of the substrate is much lower in the glass samples. This will give it a lower heat capacity and make it function worse as a heat sink/temperature reservoir, in addition to being a worse heat conductor. However, when trying to perform polarization sensitive laser modifications to the films by laser writing, samples on thick glass substrates were always found to be less consistent and reliable compared with silicon\cite{Novikov:2017}. This fact also leads us to believe that the silicon substrate is much better suited for isolating the melting to just specific sites of plasmonic hotspots, as this would give better polarized behaviour. We will discuss this kind of polarization manipulation in the next section.

\section{Polarization inscription}

We have studied how it is possible to 'inscribe' specific polarization sensitivity into the optical properties gold percolation films, both on silicon and on 18\,nm thin glass TEM membranes. All fs-laser writing and TPL-measurements were like in chapter~\ref{chap:tpl} done in collaboration with Prof. S.~I.~Bozhevolnyi's group at University of Southern Denmark. We will start our discussion here by looking at the case of gold films on silicon substrates, as it is the most straightforward.

\subsection{Silicon substrates}

As we discussed in chapter~\ref{chap:metal}, and saw in chapter~\ref{chap:snom}, the plasmonic hotspots in percolation films are strongly polarized. We can use this fact along with photothermal processes to remove hotspots for a given polarization from a sample. In fact, this specific removal of hotspots with a given polarization is what is visible on Fig.~\ref{fig:silicon_dam}, as the laser has only damaged the local part of the film that hosted a plasmon resonance matching the polarization of the laser. We can see the consequence of this removal of polarization when we study TPL-enhancement, like in chapter~\ref{chap:tpl}, from the altered films.

\clearpage

\begin{figure}[h]
	\center
	\includegraphics[width=0.85\columnwidth]{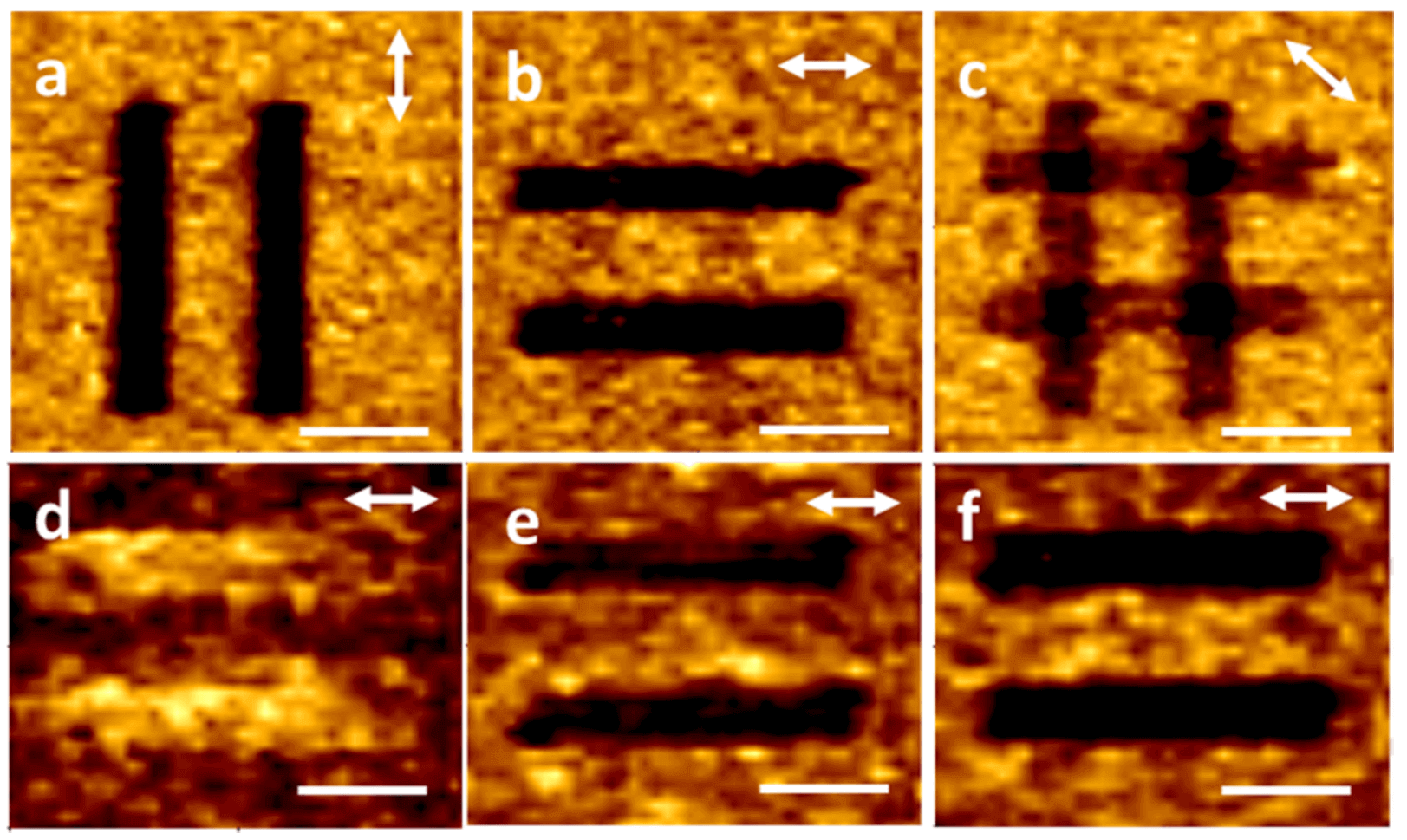}
    \caption{TPL-images from a 5\,nm gold percolation film on silicon substrate with laser lines written in it. \textbf{(a-c)} shows polarization and \textbf{(d-f)} shows wavelength dependencies. Two pairs of lines in \textbf{(a-c)} were written with the laser polarization aligned along the lines. Both writing and reading were done at the wavelength of 740\,nm. A pair of lines in \textbf{(d-f)} were written at the wavelength of 780\,nm with the laser polarization along the lines again. Reading was done at the wavelength of \textbf{(d)} 740, \textbf{(e)} 780, and \textbf{(f)} 820\,nm. The white arrows indicates the laser polarization during the read-out. Laser powers of \textasciitilde3 and \textasciitilde1\,mW were used for writing and reading, respectively. The scale bars are 5\,$\upmu$m.}
    \label{fig:tpl_pol}
\end{figure}

\noindent On Fig.~\ref{fig:tpl_pol} we see a set of TPL-images recorded from two different regions of a 5\,nm gold film. In two different areas lines have been 'written' with the fs-laser. In Fig.~\ref{fig:tpl_pol}.a-c two pairs of lines have been written before recording TPL-images, with the polarization of the laser aligned along each pair of lines. TPL-images have then been recorded in the same laser set-up, using a lower laser power than for the writing. The result is that when recording a TPL-image with the polarization that is parallel to the polarization of the writing, the laser written areas show up in the TPL-images as a clear dark contrast with almost no TPL-signal. For the case of Fig.~\ref{fig:tpl_pol}.c, we see that when aligning the read-out polarization to 45$^\circ$, we can see all lines at the same time. What we see here, is that it is possible to selectively 'extinguish' the TPL-signal from a region of the film for a given polarization. 

The case of Fig.~\ref{fig:tpl_pol}.d-e demonstrates the effects of different wavelengths. Here again a pair of lines have been written with a laser polarization along the lines, at a wavelength of 780\,nm. Then the lines have been scanned again, but now with a shorter wavelength of 740\,nm in Fig.~\ref{fig:tpl_pol}.d, the same wavelength of 780\,nm in Fig.~\ref{fig:tpl_pol}.e, and a longer wavelength of 820\,nm on Fig.~\ref{fig:tpl_pol}.f. For the case of Fig.~\ref{fig:tpl_pol}.d we see a larger enhancement of TPL-signal from the laser written region, while for the cases of Fig.~\ref{fig:tpl_pol}.e and f we see the expected reduction of TPL-signal. We can understand this 'contrast inversion' from our discussion in the previous section above: When resonant elongated plasmonic particles are shortened due to photothermal processes their resonance frequency will blueshift. In other words, the clusters that were affected by the longer wavelength will be shifted towards a shorter wavelength. Hotspots were however already present at the shorter wavelength, and were themselves unaffected by the writing process at the longer wavelength. The net effect is that now at the shorter wavelength a larger concentration of resonances for the given polarization has been achieved, and we thus see an even larger enhancement of TPL when recording images at this shorter wavelength.

\begin{figure}[h]
	\center
	\includegraphics[width=0.95\columnwidth]{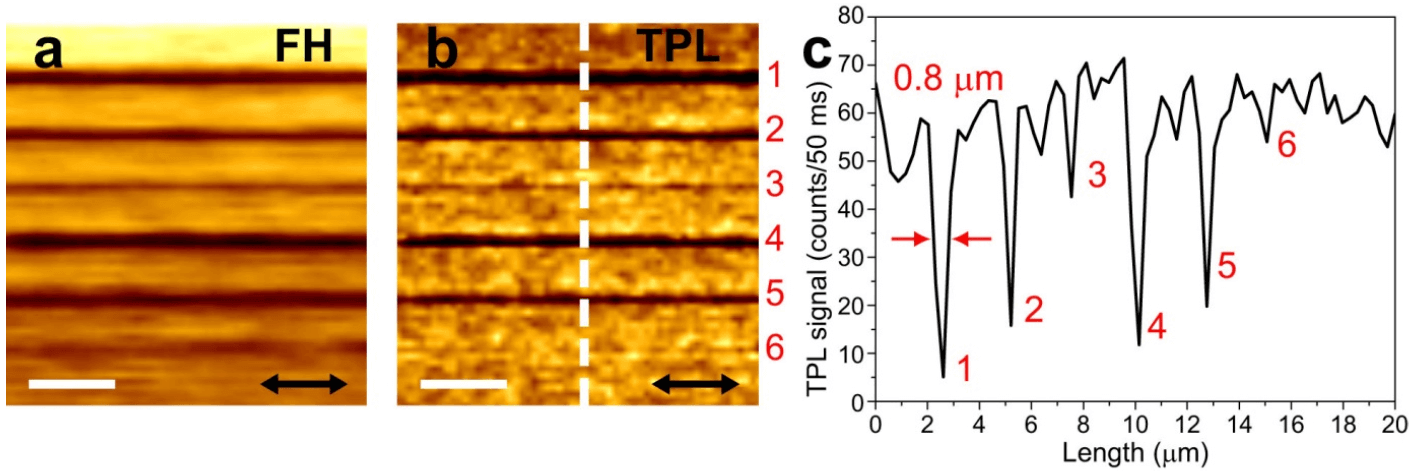}
    \caption{\textbf{(a)} FH and \textbf{(b)} TPL-images of a 5\,nm thin gold film on a silicon substrate, obtained with different laser powers during the writing procedure (exposure time of 50\,ms, scan speed of 20\,$\upmu$m/s, spot size of \textasciitilde750\,nm) and different scanning step sizes. Read-out and writing performed with the same polarization. For the scanning step size of 360\,nm, the FH writing powers are: P$_1$\textasciitilde5\,mW, P$_2$\textasciitilde4\,mW, and P$_3$\textasciitilde3\,mW. For the step size of 800\,nm powers are: P$_4$\textasciitilde5\,mW, P$_5$\,\textasciitilde4\,mW, and P$_6$\textasciitilde3\,mW. \textbf{(c)} Cross sections taken from the dashed line in (b). The scale bars are 5\,$\upmu$m.}
    \label{fig:tpl_lines}
\end{figure}

\noindent On Fig.~\ref{fig:tpl_lines} some examples of the degree of signal reduction achievable for different step sizes and laser powers can be seen. Here three different laser powers and two step sizes for the laser scan have been compared. For the case of 5\,mW and a step size of 360\,nm (peak 1 in Fig.~\ref{fig:tpl_lines}.c) we see it is even possible to almost completely extinguish the TPL-signal for a given polarization (when reading back at same or longer wavelengths). A very similar writing process has also been reported recently, highlighting writing with three different polarizations in the same spot, and wavelength separated images as well, see reference \cite{Zheng:2017}.

\subsection{Glass substrates}

We have now seen how the localized reshaping of plasmonic hotspots in gold films on silicon substrates can be used for polarization control of the film's optical properties. Now we will move on to the more complex case of reshaping gold films on glass substrates. As discussed above, here the heat seems to dissipate much more into the gold films than when compared with silicon samples. However, we will show that it is still possible to achieve polarization effects
in these structures, despite the fact that the morphology changes are much less isolated.

\begin{figure}[h]
	\center
	\includegraphics[width=\columnwidth]{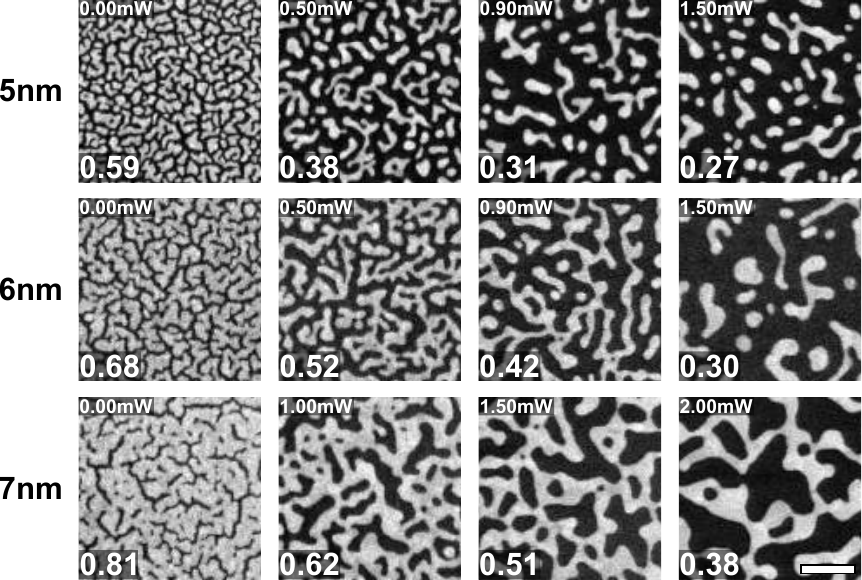}
    \caption{STEM dark-field images of different gold percolation film morphologies on 18\,nm SiO$_2$ TEM membranes. The films have been illuminated with the same scan parameters, but different laser powers. Numbers in the upper left corners are the laser power used for illumination, and for all samples a wavelength of 740\,nm was used. The numbers in the lower left corners of the images indicate found filling fractions of gold from image analysis. All images are at the same scale, the scale bar is 150\,nm.}
    \label{fig:SiO2_dam2}
\end{figure}

\noindent Fig.~\ref{fig:SiO2_dam2} shows a small 'catalogue' of how different deposition thicknesses give different film morphologies, and how these morphologies are influenced by various levels of fs-laser power. The films were illuminated with the laser in separate positions with the different laser powers. The filling fraction of gold in each image has also been calculated. For the 5\,nm sample we see that it initially consists of smaller isolated clusters, and the effect of the laser seems to be to melt and merge many of these clusters. For increasing laser powers we also consistently see the filling fraction drop, indicating that more and more of the gold in the sample is being collected into the new clusters (assuming minimal evaporation/dissipation into the substrate).

The 6\,nm sample appears to be closer to the percolation threshold than the 5\,nm sample. It consists of bigger gold clusters that are however still isolated. The effects of the laser also seem somewhat different, as the resulting morphologies for the two lower laser powers seems to become merged together in long tortuous filaments, with a few almost spherical particles mixed in-between. The case of 1.50\,mW is however different. It seems this laser power is above the threshold to nearly melt the clusters completely, as they do not seem able to form the large windy clusters.

The 7\,nm sample appears to be well above the percolation threshold, looking like a solid single cluster, with many small gaps in it. The effects of laser illumination seems mainly to be either fusing or opening up of the gaps in the film. For the 2.00\,mW case the result appears to be a lattice of thick gold clusters that are interconnected through thin windy wires.

\begin{figure}[h]
	\center
	\includegraphics[width=0.75\columnwidth]{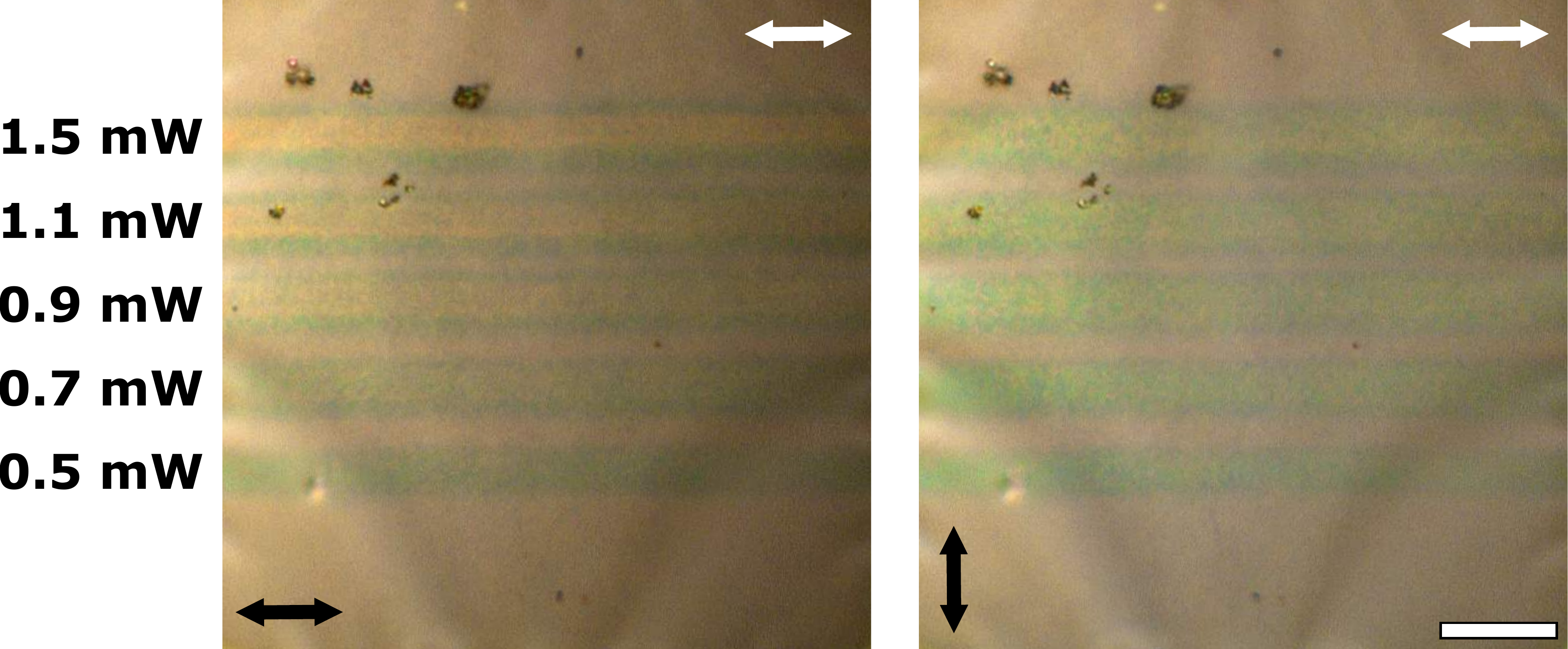}
    \caption{Microscope images of five lines written with the fs-laser at different powers in a 5\,nm gold percolation film on a 18\,nm SiO$_2$ membrane. The white double arrows indicate the polarization of the writing laser, the black arrows show the orientation of a polarizer inserted in the microscope before the camera. All lines were written with a wavelength of 740\,nm. The scale bar is 10\,$\upmu$m.}
    \label{fig:tem_sample}
\end{figure}

\noindent From just viewing the laser illuminated regions afterwards in an ordinary microscope, it is very clear that the regions have been changed compared to the surrounding gold on the glass membrane. An example of several lines written next to each other on the 5\,nm gold film sample can be seen on Fig.~\ref{fig:tem_sample}. When observing the sample through an optical polarizer, it also becomes apparent that there is a difference between aligning this parallel or perpendicular to the polarization of the laser used for the writing process. When aligning the polarizer parallel to the writing laser's polarization, a reddish hue becomes visible in the centre of the written lines (most apparent for the 1.5\,mW line.). This colour is however absent when viewing in the perpendicular orientation. The 6 and 7\,nm samples show similar behaviour, but it is less pronounced than for the 5\,nm sample. We will focus our attention on this sample as a result.

\begin{figure}[h]
	\center
	\includegraphics[width=0.80\columnwidth]{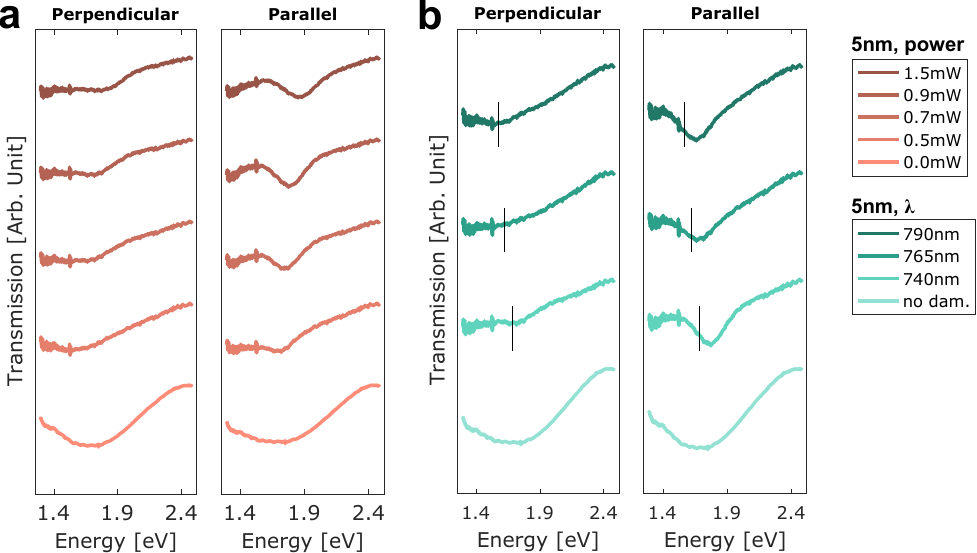}
    \caption{Transmission spectra of a 5\,nm percolation film that has been altered with different laser powers and wavelengths. The microscope light source for illumination has been polarized either parallel or perpendicular to the polarization of the laser used to alter the films. \textbf{(a)} are spectra from regions altered with different laser powers, at a wavelength of 740\,nm. \textbf{(b)} shows the influence of the wavelength of the laser used for the damage. The vertical black bars corresponds to the photon energy of the laser during the writing process.}
    \label{fig:perco_trans}
\end{figure}

Because the laser modified samples are on a transparent substrate, we can perform optical transmission spectroscopy on them. The results of such measurements done in collaboration with M.~Geisler can be seen on Fig.~\ref{fig:perco_trans}. 

\noindent In the transmission microscope used, we are able to change the polarization of the white light source used for sample illumination. We have recorded transmission spectra for when the microscope light source is aligned either parallel or perpendicular to the polarization of the laser used to perform the morphology changes in the samples. From these spectra we see very clearly that a polarized extinction feature is present after the laser writing. When aligning the microscope light source to parallel with the laser used for the writing process, we see a sharp dip in transmission around 1.80-2.00\,eV. For increasing laser powers this dip tends to blueshift, which could indicate that the gold clusters hosting the polarized mode are further shortened along this axis in the samples made with higher laser powers. On Fig.~\ref{fig:perco_trans}.b we also show spectra from sample regions illuminated with the same laser power, but at different wavelengths. From these spectra we see that the wavelength of the laser used for the photothermal process is also important, as the dip in transmission seems to be just slightly blueshifted relative to the laser wavelength used (reminiscent of what happens in Fig.~\ref{fig:tpl_pol}.d).

\begin{figure}[h]
	\center
	\includegraphics[width=0.95\columnwidth]{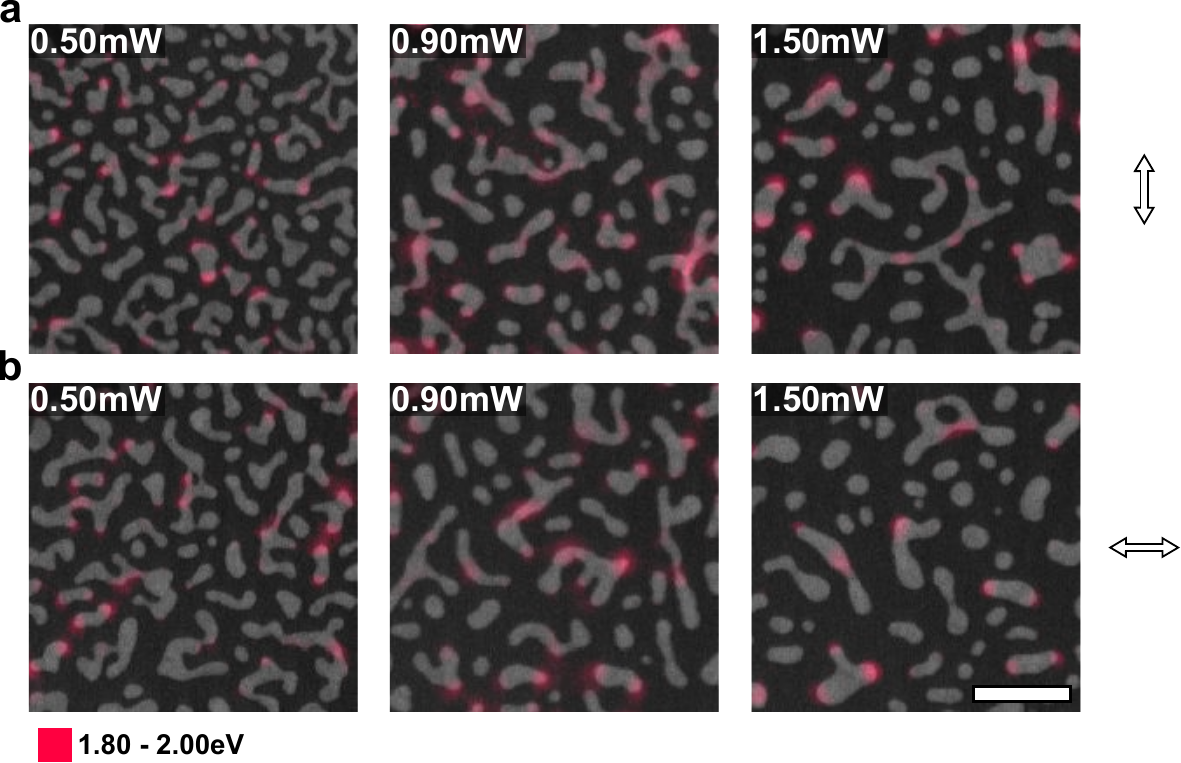}
    \caption{Colour maps of the recorded EEL spectra integrated in the energy range of 1.80--2.00\,eV and overlaid on top of their respective STEM dark-field images of the 5\,nm samples. Each row of images consists of samples with the same polarization used for laser writing (white arrows), (\textbf{a}) being the samples with $y$-polarized laser writing, and (\textbf{b}) the samples with $x$-polarized laser writing. The numbers in the top left corners of the images indicate the amount of power used for laser writing. The scale bar is 150\,nm.}
    \label{fig:eels_5nm}
\end{figure}

To study the near-field cause of this emergent polarization sensitive colour, we have recorded detailed hyperspectral images of the different 5\,nm film morphologies using EELS. By integrating the EELS data in the range of 1.80-2.00\,eV, we can now construct a colour map showing the regions with high electron-energy loss in the energy ranges where we see the transmission dip in Fig.~\ref{fig:perco_trans}. The result can be seen on Fig.~\ref{fig:eels_5nm} for three different powers of laser writing, and for two sets of samples that have been laser written with polarizations 90$^\circ$ apart.

\begin{figure}[h]
	\center
	\includegraphics[width=0.8\columnwidth]{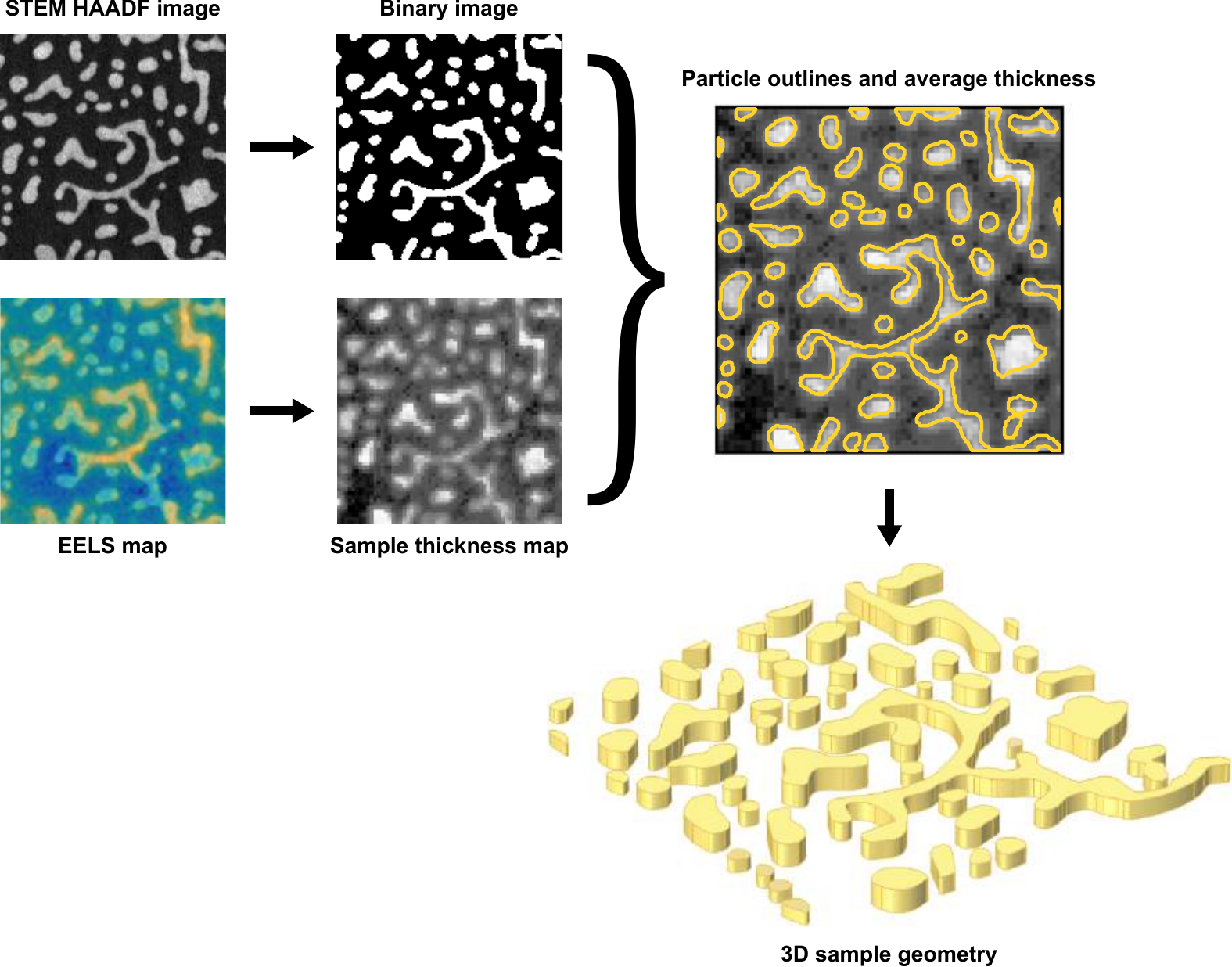}
    \caption{Overview of the process to generate the thin film geometries used in simulations. The dark-field STEM images are converted into black/white binaries to recover the particle outlines, and the EELS data is integrated in order to construct relative thickness maps of the samples. Using the thickness map and the particle outlines, an average thickness of each gold particle is found, and the sample geometry can be reconstructed with the clusters as relatively scaled prisms.}
    \label{fig:sim_geo}
\end{figure}

\clearpage
\noindent In these images we can identify several elongated particles that are about \textasciitilde50-100\,nm long and show very strong EELS signals at their endpoints. Many of these elongated particles also appear to be aligned along the polarization used when damaging the sample with the laser (especially in the 1.5\,mW sample images), although many also appear to be oriented more or less randomly. Elongated gold particles with dipolar plasmon resonances could explain the polarization sensitivity that we observe in Figs.~\ref{fig:tem_sample} and \ref{fig:perco_trans}. However, from EELS measurements alone we cannot resolve the polarization sensitivity of the structures, as the excitation with an electron beam is not polarized. The other problem, as discussed in chapter~\ref{chap:eels}, is that the excitation with electrons can also excite plasmonic dark modes, resonances normally inaccessible with just light\cite{Koh:2009}. As the polarization sensitivity observed in Fig.~\ref{fig:perco_trans} is from a polarized far-field light source, we will have to correlate our near-field EELS data to how a far-field excitation would look like. In collaboration with T.~Rep\"an we have reconstructed the 3D geometry of the gold morphologies, by using our STEM dark-field images and EELS data. The process is outlined in Fig.~\ref{fig:sim_geo}.

\clearpage
\noindent After reconstructing the geometries for the two regions of the 5\,nm gold film damaged with 1.5\,mW and different polarizations (seen in the right column of Fig.~\ref{fig:eels_5nm}), we performed finite element simulations of far-field plane wave excitations on them. From these simulations it is possible to align the polarization of the exciting wave to parallel or perpendicular to the polarization of the laser used in the writing process. In other words, we can try to now realize how the plasmons would be excited on the gold clusters like in the transmission spectroscopy measurements of Fig.~\ref{fig:perco_trans}.

\begin{figure}[h]
	\center
	\includegraphics[width=\columnwidth]{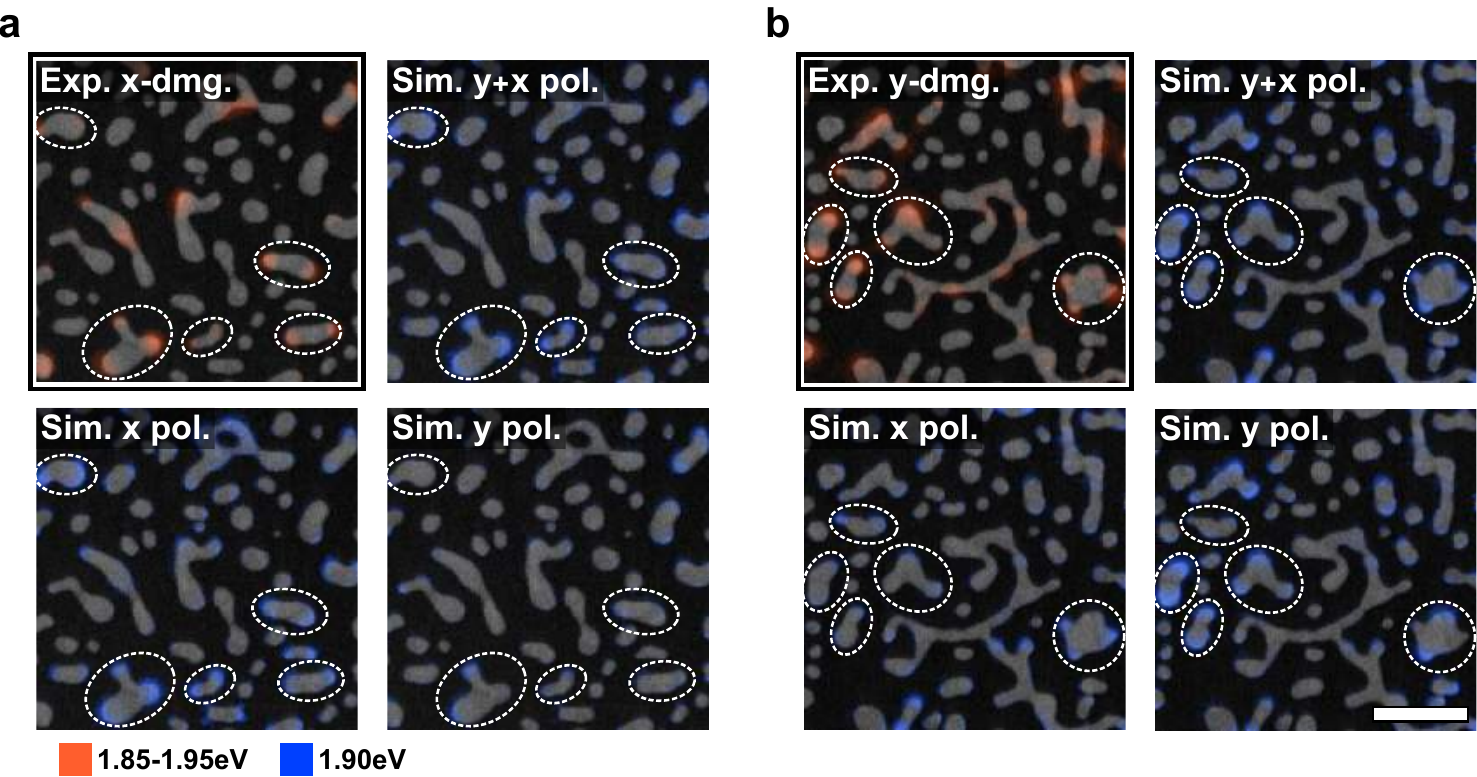}
    \caption{Comparison between EELS data and the norms of electric field $z$-components from simulated polarized far-field excitation of the 1.5\,mW laser written regions in the 5\,nm sample. (\textbf{a}) is for a writing laser with polarization along the images' $x$-axis, and (\textbf{b}) for the sample with laser writing polarized along the $y$-axis. The EELS data was integrated in the region of 1.85--1.95\,eV and the excitation energy for simulations was 1.90\,eV. The electric field norms from the simulation were taken right below the gold particles. The scale bar is 150\,nm.}
    \label{fig:eels_5nm2}
\end{figure}

\noindent The results of the simulations can be seen on Fig.~\ref{fig:eels_5nm2}. We have chosen to highlight the norm of the electric field $z$-component in the images, as it is the field component the electrons in EELS are most sensitive to\cite{Abajo:2010}. Because we have the two orthogonal polarizations for the far-field as separate simulations, we can compare the two individually, or look at their sum. We expect the sum of the two polarizations to give the greatest agreement with the EELS data, as it would most closely represent the uniform non-polarized case. We find generally good agreement between the summed simulations' field distributions and the EELS intensities. The two individual polarization simulations also seem to confirm our suspicion that indeed the rod-like particles host dipolar resonances, and they are aligned parallel to the polarization of the laser writing. For the simulations with a perpendicular polarization to the laser writing, we see very little plasmonic excitation in the gold clusters. For the simulations polarized parallel to the laser writing, we see large excitation with field distributions very similar to the EELS data.

To summarize: The effects of the fs-laser on gold percolation films on a glass substrate are very different to films on a silicon substrate. For the silicon samples the photothermal damage of the laser is very localized to the specific regions of the plasmon hotspots (likely due to increased heat conductivity of the substrate). The net effect of this is that hotspots of a certain polarization are removed from the sample after laser writing. For the glass case however, it appears that there is no big loss of plasmon resonances at the laser wavelength, as we see no strong indication of this in Fig.~\ref{fig:perco_trans}.b. However, similar to the silicon samples, at a slightly blueshifted wavelength (with respect to the laser used for writing), we see a large aggregation of plasmonic excitation. This effect is also strongly polarized to be aligned with the laser polarization used in the writing process, and is likely caused by gold nanorod-like particles that have formed in the sample.

\section{Reconfiguration of modes}

\begin{figure}[h]
	\center
	\includegraphics[width=\columnwidth]{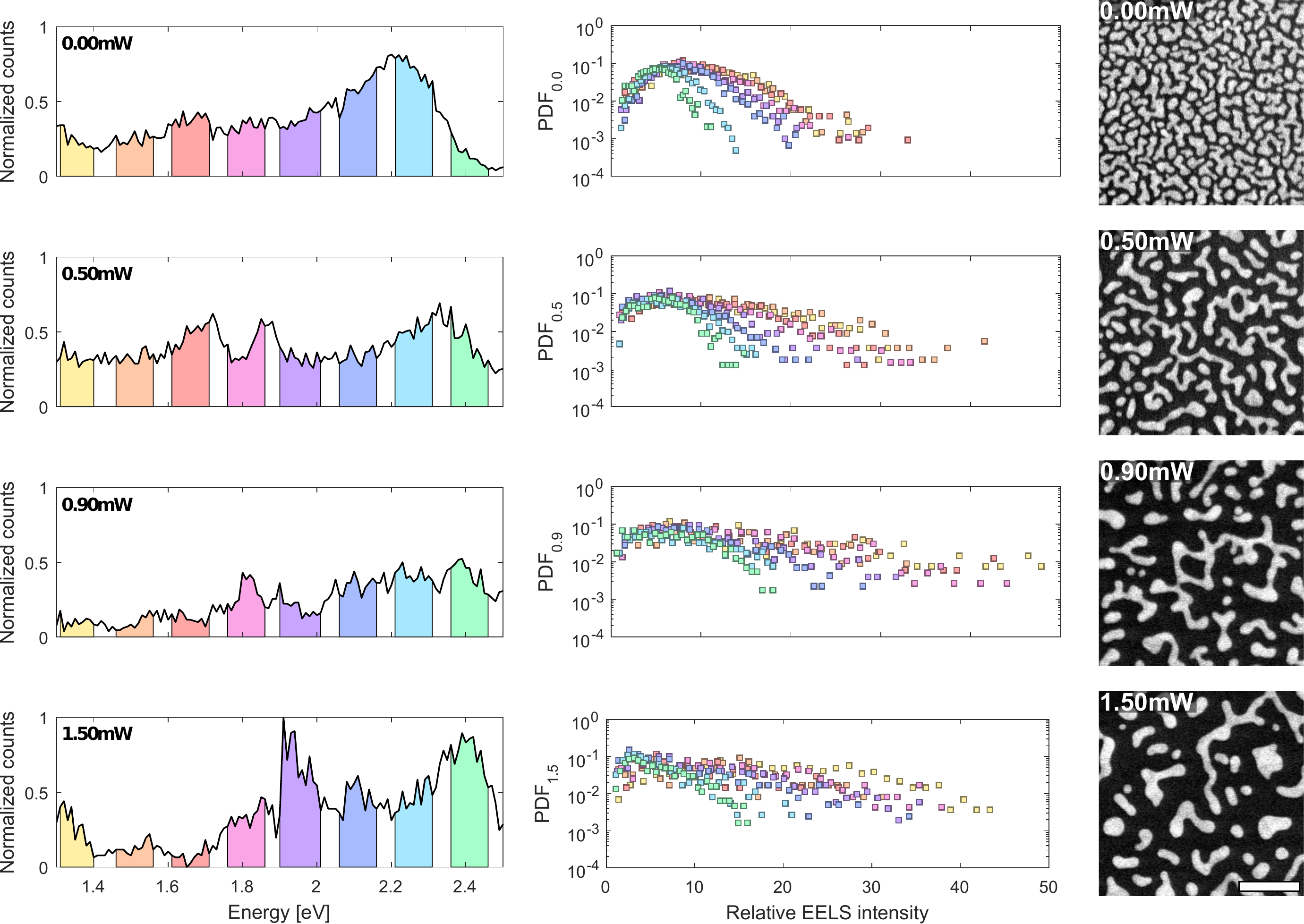}
    \caption{Plasmon modes for an intrinsic 5\,nm gold film, and three 5\,nm gold film samples that have been modified with different laser powers. The left column shows histograms of the identified central energies of plasmon resonances in the spectral images. The centre column consists of PDFs showing the EELS intensity distribution across different energy ranges matching the highlighted colours in the histograms on the left. On the right are shown STEM dark-field images of example film morphologies. The scale bar is 150\,nm.}
    \label{fig:recon_5nm}
\end{figure}

We detailed in chapter~\ref{chap:eels} how it is possible to extract the resonance energy and EELS signal amplitude of plasmon resonances in an EELS map. By recording hyperspectral images of intrinsic percolation films and films that have been modified with a fs-laser we can also study how the plasmon resonances in the films evolve with the morphology changes. The results of such an analysis can be seen here on Fig.~\ref{fig:recon_5nm}. We have mapped 500$\times$500\,nm$^2$ regions of the four different film morphologies. Like discussed for the similar analysis in chapter~\ref{chap:metal}, such an area is large enough that we can consider the sample as isotropic at this scale, and can use this area to make general statements about the film as a whole.

\clearpage

\noindent By identifying the plasmon resonances in the EELS data we can construct histograms of how many modes are present at which energy, like in the left of Fig.~\ref{fig:recon_5nm}. If we then take all the resonances within a narrow range of these energies (highlighted colours on the histograms), we can sort them by their magnitude in EELS intensity in the spectra (amplitude), and bin them into 30 values based on their intensity. We then divide this distribution function of intensities by the total count of intensities, and we get a probability distribution function (PDF) of the EELS intensity for that energy range. Taking the 50 minimum intensities from each of the four spectral images, and calculating the median value, we can normalize the PDFs to this value. This gives us the PDFs for the relative EELS intensities, which let's us compare the PDFs between the four samples. These PDFs are highlighted in the centre column of Fig.~\ref{fig:recon_5nm}. The shape of these PDFs now offers us insight into the system that would have been very difficult to attain by just EELS intensity colour maps. First of all, the shape of the PDF can tell us about the degree to which the gold clusters are coupled together. In prior work it has been well documented that metal percolation films exhibit Gaussian function-like PDFs for their electric field intensities\cite{Genov:2003a,Genov:2003b}, while isolated plasmonic particles show power-law-like PDFs\cite{Bozhevolnyi:2001,Stockman:1994}\footnote{A single isolated dipole would simply have its field intensity scale with the distance as $I\propto1/r^4$ in space, whereas many interacting and interfering dipoles would have a much more complex spatial scaling of intensity.}. In our case, we see the PDFs for the intrinsic films start out with quite Gaussian shapes, but eventually get more and more twisted into a scaling power-law shape for the samples modified with larger laser powers. When compared with the large drops in filling fraction observed in Fig.~\ref{fig:SiO2_dam2}, we can conclude that we are indeed seeing the plasmonic coupling of the percolation films being lifted from the photothermal reshaping by laser illumination. Secondly, by comparing the right-most points of the PDFs (i.e. largest relative EELS intensity) between the different energy ranges, we see that for the intrinsic film we get the expected behaviour of the lower energy modes also supporting the most intense hotspots\cite{Genov:2003a}. However, we also see that we generally find more intense hotspots from the laser modified films, and some of the higher energy modes especially (1.90-2.00\,eV and 2.10-2.20\,eV) experience almost a factor of 2 increase in their highest relative EELS intensities.

We can understand the effects of the laser as a decoupling of the clusters in the percolation films. This also causes a redistribution of the initially broad ranges of plasmon resonances into more narrow bands of resonances (seen in the histograms in Fig.~\ref{fig:recon_5nm}).

%% file: chapters/chap_con.tex
\chapter{Conclusion}

In conclusion, we have reviewed and studied the general physics of metal percolation films. We have studied how such films can be fabricated using e-beam metal deposition systems, and explained the metal growth process responsible for the self-assembly of the deposited metal into percolation structures. We have characterized the morphologies of different deposition recipes by SEM- and TEM-imaging. We have found only minor differences between glass and silicon substrates in terms of achieved morphologies from similar gold depositions, while the later studied properties of the films have deviated dramatically based on their substrate materials.

We have shown how two-photon luminescence (TPL) of gold can be plasmonically enhanced in gold percolation films. Different deposition thicknesses of metal will produce different degrees of enhancement, but we have shown that the highest degree of enhancement is found for films near the percolation threshold. In addition, we have found that percolation films on glass substrates produce TPL-enhancement by about 2 orders of magnitude higher than when compared to silicon substrates. We preliminarily attribute this to the differences between the substrates to act as heat sinks, i.e. to remove heat from the Ohmic losses in the metal structures. We have also observed white light generation from the gold percolation films, when excited with a fs-laser. The exact mechanism(s) of the white light generation in gold is still not understood. Additionally, we have also studied how it is possible to perform polarization sensitive writing with the TPL-setup in gold percolation films. We found that it is possible to locally extinguish the TPL-signal for a given polarization in the percolation films. This is achieved through photothermal reshaping processes of the gold, from scanning the region of the film with high power fs-laser pulses. We found this effect is not as easily reproducible for percolation films on glass substrates, when compared to silicon substrates. We again attribute this difference to the substrates' ability to channel heat away from the gold clusters, ensuring much more localized reshaping of clusters on the silicon substrates.

We have measured the distribution of plasmonic hotspots in our gold percolation film samples, by using s-SNOM and EELS. Using polarized excitation in the s-SNOM experiment, we have reconfirmed that the hotspots are strongly linearly polarized, and that the spatial distribution of hotspots is unique between a 90$^\circ$ difference in excitation polarization. Using EELS we have mapped the spatial distribution of plasmonic hotspots in percolation films, down to few nanometre resolution. This reveals the incredible complexity of how the plasmonic hotspots are formed, and how the field distributions in these films are not linked to the geometry in a simple, or predictable, way\cite{Losquin:2013}.

Using a quantitative/statistical analysis method, we have studied the distributions of plasmon excitations in various percolation film morphologies on 18\,nm thin glass membranes. Many of these morphologies have been achieved by photothermal reshaping of the metal clusters present in intrinsic films by fs-laser illumination. We have studied how the plasmon modes are gradually more and more reshuffled for increasing laser powers used to reshape the morphologies. We have also shown how the plasmon modes in the film become less and less coupled to each other, as the system transitions into morphologies more akin to isolated gold particles for the higher laser powers.

We have also studied how this kind of laser writing leaves behind a strongly polarized plasmon excitation in the films, observable in the far-field. We have carefully reconstructed our film morphologies observed in the EELS experiments, and performed finite-element simulations of plane wave excitations on them. From these simulations we have highlighted how the polarization dependence observed in the far-field is caused by nanorod-like particles formed along the polarization of the fs-laser used for the laser writing process.

\section{Outlook}

Besides the many already interesting optical applications for metal percolation films, the possibility to modify the local optical properties of the films by fs-laser writing opens up a new set of interesting applications.

For one, the polarization sensitive writing effect in TPL-signal from percolation films on silicon substrates allows for writing of two distinct images on top of each other, with minimal cross-talk between them on read-out. This, along with the fact that the percolation films are very easy to fabricate with just simple metal evaporation systems (a method that is also easily scalable to very large surface areas), makes high density information storage devices seem like an obvious application. In addition, the materials used would also just be gold, silicon, and possibly glass. The potential lifetime of such storage devices could be hundreds (if indeed not thousands) of years, with minimal to no loss of information. Additional work would need to be done to determine the smallest writing spotsize needed for writing and readout of information. Further experiments are also needed to determine if it would be possible to encapsulate the percolation films in silicon/SiO$_2$ to create a sandwiched structure of multiple layers for multiplexing, like previously demonstrated with gold nanorods\cite{Zijlstra:2009}.

Secondly, the ability to manipulate wavelength and polarization sensitivity in the percolation films both on silicon and glass substrates by fs-laser writing could open new sensing applications and devices for percolation films.

The ability of the gold percolation films to also generate white light at fairly low pump powers (few tenths of mW) could also have novel applications. One aspect that also remains unstudied is how photothermal reshaping of the metal films could influence parts of the white light spectrum, by removing or adding additional plasmon resonances for certain wavelengths.

Finally, from just observing the percolation films on glass substrates after fs-laser writing, a clear red colour becomes apparent. Further investigations are needed, but potentially metal percolation films could act as cheap, large area, substrates for emerging plasmonic colour laser printing technologies.